\documentstyle[12pt]{article}
\makeatletter

\@addtoreset{equation}{section}
\makeatother

\begin{document}

%%%%%%%%%%%%%%%%%%%%%%%%%%%%%%%%%%%%%%%%%%%%%%%%%%%%%%%%%%%%%%%%%%
%%%%%%%%%%%%%%%%%%%%%%%% Title %%%%%%%%%%%%%%%%%%%%%%%%%%%%%%%%%%%
%%%%%%%%%%%%%%%%%%%%%%%%%%%%%%%%%%%%%%%%%%%%%%%%%%%%%%%%%%%%%%%%%%
\begin{flushright}
DPUR/TH/1\\
DFPD07/TH/06\\
April, 2007\\
%hep-th/070****\\
\end{flushright}
\vspace{20pt}

%\magnification=\magstep1
\pagestyle{empty}
\baselineskip15pt
%\font\cmssB=cmss17
%\font\cmssS=cmss10

\begin{center}
{\large\bf Y-formalism and $b$ ghost in the Non-minimal Pure Spinor 
Formalism of Superstrings
\vskip 1mm }

\vspace{20mm}
Ichiro Oda
          \footnote{
          E-mail address:\ ioda@phys.u-ryukyu.ac.jp} \\

\vspace{5mm}
           Department of Physics, Faculty of Science, University of the 
           Ryukyus,\\
           Nishihara, Okinawa 903-0213, Japan.\\

\vspace{5mm}

and

\vspace{5mm}

Mario Tonin
          \footnote{
          E-mail address:\ mario.tonin@pd.infn.it}\\
\vspace{5mm}
          Dipartimento di Fisica, Universita degli Studi di Padova,\\
          Instituto Nazionale di Fisica Nucleare, Sezione di Padova,\\
          Via F. Marzolo 8, 35131 Padova, Italy\\

\end{center}

%\maketitle

\vspace{5mm}
\begin{abstract}
We present the Y-formalism for the non-minimal pure spinor quantization of 
superstrings. In the framework of this formalism we compute, at the quantum 
level, the explicit form of the compound operators involved in the 
construction of the $b$ ghost, their normal-ordering contributions and the
relevant relations among them. We use these results to  construct the 
quantum-mechanical $b$ ghost in the non-minimal pure spinor formalism. 
Moreover we show that this non-minimal $b$ ghost is cohomologically equivalent
to the non-covariant $b$ ghost. 

\end{abstract}

\newpage
\pagestyle{plain}
\pagenumbering{arabic}
%\setcounter{page}{1}

%%%%%%%%%%%%%%%%%%%%%%%%%%%%%%%%%%%%%%%%%%%%%%%%%%%%%%%%%%%%%%%%%%%%%
%%%%%%%%%%%%%%%%%%%%%%%%%%%%%%   SEC  1    %%%%%%%%%%%%%%%%%%%%%%%%%%
%%%%%%%%%%%%%%%%%%%%%%%%%%%%%%%%%%%%%%%%%%%%%%%%%%%%%%%%%%%%%%%%%%%%%
\section{Introduction}
Several years ago, a new formalism for the covariant quantization of
superstrings was proposed by Berkovits \cite{Ber1}. Afterward, it has been 
recognized that this new formalism not only solves the longstanding problem 
of covariant quantization of the Green-Schwarz (GS) superstring, but also 
it is suitable to deal with problems that appear almost intractable in the 
Neveu-Schwarz-Ramond (NSR) approach, such as those involving space-time 
fermions and/or backgrounds with R-R fields.

In this approach, the GS superstring action (let us say in the left-moving 
sector) is replaced with a free action 
for the bosonic coordinates $X^a$ and their fermionic partners $\theta^\alpha$
with their conjugate momenta $p_\alpha$,
plus an action for the bosonic ghosts $\lambda^\alpha$ and their conjugate 
momenta $\omega_\alpha$, where $\lambda^\alpha$ satisfy the "pure spinor 
constraint"  $\lambda \Gamma^a \lambda = 0$.
The $\omega-\lambda$ action looks like a free action but is not really free
owing to the pure spinor constraint, which is necessary to have 
vanishing  central charge and correct level of the Lorentz algebra. This 
 formulation is nowadays called "pure spinor formulation of 
superstrings" and many studies \cite{Ber2}-\cite{Adam} were devoted to it in 
the recent years. $\footnote{Alternative formalisms to remove the constraint 
were proposed in \cite{Grassi2, Kazama1}.}$

Another key ingredient in the pure spinor formulation is provided by the BRST
charge
$Q = \oint \lambda^\alpha d_\alpha$ where $d_\alpha \approx 0$ contains the
constraints generating a fermionic $\kappa$ symmetry in the GS superstring 
and has the role of a spinorial derivative in superspace.
The peculiar feature associated with this BRST charge is that $Q$ is 
nilpotent only when the bosonic spinor $\lambda^\alpha$ satisfies the pure 
spinor condition.
This peculiar feature is in fact expected since the constraint $d_\alpha 
\approx 0$ in the GS approach involves both the first-class and the 
second-class constraints. Roughly speaking,
the pure spinor condition is needed to handle the second-class constraint of 
the GS superstring, keeping the Lorentz covariance manifest. 

Since the BRST charge $Q$ is nilpotent, one can define the cohomology and 
examine 
its physical content. Indeed, it has been shown that the BRST cohomology 
determines 
the physical spectrum which is equivalent to that of the RNS formalism and 
that of the GS formalism in the light-cone gauge \cite{Ber3}. 
Moreover, the BRST charge $Q$ of the pure spinor formalism was found to be 
transformed to that of the NSR superstring \cite{Ber4}
as well as that of the GS superstring in the light-cone gauge \cite{Ber11, 
Kazama4}.

Even if the pure spinor formalism provides a Lorentz-covariant superstring
theory with manifest space-time supersymmetry even at the quantum level,
there are some hidden sources of possible violation of Lorentz covariance.

One of such sources is related to the $b$ field defined by $T = \{Q, b \}$ 
with $T$ being the stress-energy tensor, which is necessary to compute 
higher loop amplitudes. Since the pure spinor formulation is not derived 
from a diffeomorphism-invariant action and does not contain the $b-c$ ghosts 
of diffeomorphisms, the usual antighost $b$ is not present in this approach.
In \cite{Ber3} a compound $b$ field whose BRST variation gives the 
stress energy tensor, was obtained. However this $b$ field is not
Lorentz-covariant.

The same $b$ field follows from an attempt \cite{Matone} to derive,
at the classical level, the pure spinor formulation from a (suitably 
gauge-fixed and twisted) $N=2$ superembedding approach. In this approach 
the $b$ field is the twisted current of one of the two world-sheet (w.s.) 
supersymmetries whereas the integrand of the BRST charge $Q$ is the twisted 
current of the other supersymmetry, suggesting an $N=2$ topological origin 
of the pure spinor approach.

This $b$ field turns out to be proportional to the quantity $Y_\alpha
= {{v_\alpha}\over{v\lambda}}$ where $v_\alpha$ is a constant pure spinor,
such that $b_{Y} = Y_\alpha G^\alpha$ where $G^\alpha$ is a covariant, 
spinor-like compound field, so that $b_{Y}$ is not only Lorentz non-covariant 
but also singular at $v\lambda = 0$.

A way to overcome the problem of the non-covariance and singular nature  of 
$b_{Y}$ was given
in \cite{Ber10} where a recipe to compute higher loop amplitudes was 
proposed, in terms of a picture-raised $b$ field constructed with the
help of suitable covariant fields $G^\alpha$, $H^{\alpha\beta}$, 
$K^{\alpha\beta\gamma}$ and $L^{\alpha\beta\gamma\delta}$ and 
some picture-changing operators $Z's$ and $Y's$. \footnote{The picture-lowering 
operators $Y_C$, which are needed to absorb the zero modes of the 
ghost $\lambda^\alpha$, break the Lorentz-covariance but this breaking is
BRST trivial and then harmless.}

Recently, a very interesting formalism called "non-minimal pure spinor 
formalism" has been put forward \cite{Ber12}.
In this formalism, a non-minimal set of variables are added to that of the 
(minimal) pure spinor formulation. These non-minimal variables form a BRST 
quartet and have the role of changing the ghost-number anomaly from $-8$ to
 $+3$ 
without changing the central charge and the physical mass spectrum. 
A remarkable thing is that, in this formalism, one can define a 
Lorentz-covariant $b$ ghost without the need of picture-changing operators.
With the help of a suitable regulator, a recipe has been given to compute 
scattering amplitudes up to two-loop amplitudes.
The OPE's between the relevant operators that result in this approach 
show that the 
(non-minimal) pure spinor formulation is indeed a hidden, critical, $N=2$ 
topological string theory.
A significant improvement was obtained in \cite{BerNek}. Here a gauge 
invariant,  BRST trivial regularization of the $b$ field is proposed,
that allows for a consistent prescription to compute amplitudes at any loop.

A further source of possible non-covariance arises at intermediate steps
of calculations, since the solution of the pure spinor constraint in terms 
of independent fields implies the breaking of $SO(10)$ to $U(5)$.
\footnote{In the extended pure spinor formalism \cite{Kazama1}, 
the same non-covariance can be found in the ghost sector where the ghosts 
are invariant under only $U(5)$ group, but not $S(10)$ group.}   To be more 
precise, the space of (Euclidean) pure spinors in ten dimensions has 
the geometrical structure of a complex cone $Q = \frac{SO(10)}{U(5)}$ 
\cite{Ber13}. This space has been studied by Nekrasov \cite{Nekrasov}
and the obstructions to its global definition are analyzed. 
It was shown that the obstructions are absent if the tip of the cone  is 
removed. Then this complex cone is covered by 16 charts, $U^{(\alpha)},(\alpha)
= 1, \cdots, 16$ and in each chart the local parametrization of the pure spinor,
which breaks $SO(10)$ to $U(5)$, is taken such that the parameter that describes
the generatrix of the cone is non-vanishing. This parametrization can be used 
to compute the relevant OPE's \cite{Ber1, Ber3} (U(5)-formalism).   

In a  previous work \cite{Oda4}, we have proposed a new formalism named 
"Y-formalism" for purposes of handling this unavoidable non-covariance 
stemming from the pure spinor condition. This Y-formalism is closely related 
to the $U(5)$-formalism, but has an advantage of treating
all operators in a unified way without going back to the $U(5)$-decomposition.
It is based on writing the fundamental OPE between $\omega$ and $\lambda$ in 
a form that involves   $Y_\alpha = \frac{v_\alpha}{v\lambda}$.
Strictly speaking, one needs 16, orthogonal, constant pure spinors $v^
{(\alpha)}$ (and 16 $Y^{(\alpha)}$) for each chart, such that 
$U^{(\alpha)}$ $(v^{(\alpha)}\lambda)\neq 0$ in each chart. 
However, for our puposes it is sufficient to work in a given chart.

Actually, it  turned out that the Y-formalism is quite useful to find the 
full expression of $b$ ghost \cite{Oda4}. More recently, the Y-formalism 
was also utilized to construct a four-dimensional pure spinor superstring
\cite{Chandia}. The $Y$-field also arises in the regularization prescription 
proposed in \cite{BerNek}.

The aim of the present paper is to extend the Y-formalism to the non-minimal
case and  to discuss in the framework of this formalism the 
non-minimal, covariant $b$ field in addition to the fields  $G^\alpha$, 
$ H^{\alpha
\beta}$, $ K^{\alpha\beta\gamma}$ and $L^{\alpha\beta\gamma\delta}$, which 
are the building blocks of the $b$ field. This will be done not only at the 
classical but also 
at the quantum level, by taking into account the subtleties of normal 
ordering.
The consistent results which we will get in this article, could be regarded 
as a good check of the consistency of the Y-formalism.
Moreover we shall show that the non-minimal, covariant $b$ field is 
cohomologically equivalent to the non-covariant $b$ field $b_{Y}$,
improved by the term coming from the non-minimal sector.

In section 2, we will review the Y-formalism for the minimal pure spinor 
formalism.
In section 3, the operators $G^\alpha$, $H^{\alpha\beta}$, $K^{\alpha\beta
\gamma}$ 
and $L^{\alpha\beta\gamma\delta}$, and their (anti-)commutation relations 
with the BRST charge, 
will be examined from the quantum-mechanical viewpoint. In section 4, we will
construct the Y-formalism for the non-minimal pure spinor formalism.
In section 5, based on the Y-formalism at hand, we will construct 
the Lorentz-covariant quantum $b$ ghost, which satisfies the defining 
equation
 $\{Q, b \} = T$. We shall also show that it is cohomologically equivalent to
the
non-covariant $b$ ghost $b_{Y}$ (improved by the term coming from the 
non-minimal sector).
Section 6 is devoted to conclusion and discussions. 
Some appendices are added.
Appendix A contains our notation, conventions and useful identities. 
In  Appendix B, we will review the normal-ordering prescriptions, the 
generalized Wick theorem and the rearrangement theorem which we will use 
many times in this article. Finally in Appendix C  we give some details of
the main calculations.

%%%%%%%%%%%%%%%%%%%%%%%%%%%%%%%%%%%%%%%%%%%%%%%%%%%%%%%%%%%%%%%%%%%%%
%%%%%%%%%%%%%%%%%%%%%%%%%%%%%%   SEC  2    %%%%%%%%%%%%%%%%%%%%%%%%%%
%%%%%%%%%%%%%%%%%%%%%%%%%%%%%%%%%%%%%%%%%%%%%%%%%%%%%%%%%%%%%%%%%%%%%
\section{Review of the Y-formalism}

In this section, we start with a brief review of the (minimal) pure spinor 
formalism of superstrings \cite{Ber1}, and then explain the Y-formalism 
\cite{Oda4}.
For simplicity, we shall confine ourselves to only the left-moving 
(holomorphic) sector of a closed superstring theory. The generalization
to the right-moving (anti-holomorphic) sector is straightforward. 

The pure spinor approach is based on the BRST charge
%**   2.1 %%%%%%%%%%%%%%%%%%%%%%%%%%%%%%%%%%%%%%%%%%%%%%%%%%%%%%%%%
\begin{eqnarray}
Q = \oint dz \lambda^\alpha d_\alpha,
\label{2.1}
\end{eqnarray}
%%%%%%%%%%%%%%%%%%%%%%%%%%%%%%%%%%%%%%%%%%%%%%%%%%%%%%%%%%%%%%%%%%%
and the action 
%**   2.2 %%%%%%%%%%%%%%%%%%%%%%%%%%%%%%%%%%%%%%%%%%%%%%%%%%%%%%%%%
\begin{eqnarray}
I = \int d^2z (\frac{1}{2} \partial X^a \bar \partial X_a + p_\alpha
\bar \partial \theta^\alpha - \omega_\alpha \bar \partial \lambda^\alpha),
\label{2.2}
\end{eqnarray}
%%%%%%%%%%%%%%%%%%%%%%%%%%%%%%%%%%%%%%%%%%%%%%%%%%%%%%%%%%%%%%%%%%%
where $\lambda$ is a pure spinor
%**   2.3 %%%%%%%%%%%%%%%%%%%%%%%%%%%%%%%%%%%%%%%%%%%%%%%%%%%%%%%%%
\begin{eqnarray}
\lambda \Gamma^a \lambda = 0.
\label{2.3}
\end{eqnarray}
%%%%%%%%%%%%%%%%%%%%%%%%%%%%%%%%%%%%%%%%%%%%%%%%%%%%%%%%%%%%%%%%%%%
This action is manifestly invariant under (global) super-Poincar\'e 
transformations. It is easily shown that the action $I$ is also invariant 
under 
the BRST transformation generated by the BRST charge $Q$ which is nilpotent 
owing to the pure spinor condition (\ref{2.3}).  
Notice that in order to use $Q$ as BRST charge it is implicit that 
the pure spinor condition is required to vanish in a strong sense.

Moreover, the action $I$ is invariant under the $\omega$-symmetry
%**   2.4 %%%%%%%%%%%%%%%%%%%%%%%%%%%%%%%%%%%%%%%%%%%%%%%%%%%%%%%%%
\begin{eqnarray}
\delta \omega_\alpha = \Lambda_a (\Gamma^a \lambda)_\alpha,
\label{2.4}
\end{eqnarray}
%%%%%%%%%%%%%%%%%%%%%%%%%%%%%%%%%%%%%%%%%%%%%%%%%%%%%%%%%%%%%%%%%%%
where $\Lambda^a$ are local gauge parameters.
At the classical level the ghost current is 
%**   2.5 %%%%%%%%%%%%%%%%%%%%%%%%%%%%%%%%%%%%%%%%%%%%%%%%%%%%%%%%%
\begin{eqnarray}
J_0 = \omega \lambda,
\label{2.5}
\end{eqnarray}
%%%%%%%%%%%%%%%%%%%%%%%%%%%%%%%%%%%%%%%%%%%%%%%%%%%%%%%%%%%%%%%%%%%
and the Lorentz current for the ghost sector is given by
%**   2.6 %%%%%%%%%%%%%%%%%%%%%%%%%%%%%%%%%%%%%%%%%%%%%%%%%%%%%%%%%
\begin{eqnarray}
N_0^{ab} = {1 \over 2} \omega \Gamma^{ab} \lambda,
\label{2.6}
\end{eqnarray}
%%%%%%%%%%%%%%%%%%%%%%%%%%%%%%%%%%%%%%%%%%%%%%%%%%%%%%%%%%%%%%%%%%%
which together with $T_{0\lambda} = \omega \partial \lambda$ are the
only super-Poincar\'e covariant bilinear fields involving $\omega$ and
gauge invariant under the $\omega$-symmetry.
{}From the field equations it follows that $p$, $\theta$, $\omega$ and 
$\lambda$
are holomorphic fields. At the quantum level, one obtains 
the following OPE's  \footnote{According to Appendix B, we should call 
them
not the OPE's but the contractions, but we have called "OPE's" since 
the terminology is usually used in the references of the pure spinor 
formulation.} 
involving the superspace coordinates $Z^M = (X^a, \theta^\alpha)$
and their super-Poincar\'e covariant momenta $P_M = (\Pi_a , p_\alpha)$: 
%**   2.7 %%%%%%%%%%%%%%%%%%%%%%%%%%%%%%%%%%%%%%%%%%%%%%%%%%%%%%%%%
\begin{eqnarray}
<X^a(y) X^b(z)> &=& - \eta^{ab} \log(y - z), \nonumber \\
<p_\alpha(y) \theta^\beta(z)> &=& \frac{1}{y - z} \delta_\alpha^\beta,
 \label{2.7}
\end{eqnarray}
%%%%%%%%%%%%%%%%%%%%%%%%%%%%%%%%%%%%%%%%%%%%%%%%%%%%%%%%%%%%%%%%%%%
so that
%**   2.8 %%%%%%%%%%%%%%%%%%%%%%%%%%%%%%%%%%%%%%%%%%%%%%%%%%%%%%%%%
\begin{eqnarray}
<d_\alpha(y) d_\beta(z)> &=& - \frac{1}{y - z} \Gamma^a_{\alpha\beta} 
\Pi_a(z), \nonumber\\
<d_\alpha(y) \Pi^a(z)> &=& \frac{1}{y - z} (\Gamma^a 
\partial \theta)_\alpha(z),
\label{2.8}
\end{eqnarray}
%%%%%%%%%%%%%%%%%%%%%%%%%%%%%%%%%%%%%%%%%%%%%%%%%%%%%%%%%%%%%%%%%%% 
where
%**   2.9 %%%%%%%%%%%%%%%%%%%%%%%%%%%%%%%%%%%%%%%%%%%%%%%%%%%%%%%%%
\begin{eqnarray}
d_\alpha &=& p_\alpha - \frac{1}{2}(\partial X^a + \frac{1}{4} \theta
\Gamma^a \partial \theta) (\Gamma_a \theta)_\alpha, \nonumber\\
\Pi^a &=& \partial X^a + \frac{1}{2} \theta \Gamma^a \partial \theta, 
\nonumber\\
\bar \Pi^a &=& \bar \partial X^a + \frac{1}{2} \theta \Gamma^a \bar 
\partial \theta.
\label{2.9}
\end{eqnarray}
%%%%%%%%%%%%%%%%%%%%%%%%%%%%%%%%%%%%%%%%%%%%%%%%%%%%%%%%%%%%%
As for the ghost sector, the situation is a bit more complicated owing to the
pure spinor condition (\ref{2.3}). Namely, it would be inconsistent to assume
a free field OPE between $\omega$ and $\lambda$.  The reason is that since 
the pure spinor condition 
must vanish identically, not all the components of $\lambda$ are independent:
solving the condition, five of them are expressed nonlinearly in terms of 
the others. Accordingly, five components of $\omega$ are pure gauge.
 
This problem is nicely resolved by introducing the Y-formalism.
Let us first define the non-covariant object 
%**   2.10 %%%%%%%%%%%%%%%%%%%%%%%%%%%%%%%%%%%%%%%%%%%%%%%%%%%%%%%%%
\begin{eqnarray}
Y_\alpha = \frac{v_\alpha}{v \lambda},
\label{2.10}
\end{eqnarray}
%%%%%%%%%%%%%%%%%%%%%%%%%%%%%%%%%%%%%%%%%%%%%%%%%%%%%%%%%%%%%%%%%%%
such that
%**   2.11 %%%%%%%%%%%%%%%%%%%%%%%%%%%%%%%%%%%%%%%%%%%%%%%%%%%%%%%%%
\begin{eqnarray}
Y_\alpha \lambda^\alpha = 1,
\label{2.11}
\end{eqnarray}
%%%%%%%%%%%%%%%%%%%%%%%%%%%%%%%%%%%%%%%%%%%%%%%%%%%%%%%%%%%%%%%%%%%
where $v_\alpha$ is a constant pure spinor $Y \Gamma^a Y = 0$.
Then it is useful to define the projector
%**   2.12 %%%%%%%%%%%%%%%%%%%%%%%%%%%%%%%%%%%%%%%%%%%%%%%%%%%%%%%%%
\begin{eqnarray}
K_\alpha \ ^\beta= \frac{1}{2}(\Gamma^a \lambda)_\alpha
(Y \Gamma_a)^\beta,
\label{2.12}
\end{eqnarray}
%%%%%%%%%%%%%%%%%%%%%%%%%%%%%%%%%%%%%%%%%%%%%%%%%%%%%%%%%%%%%%%%%%%
which, since $Tr K = 5$, projects on a 5 dimensional subspace of the 16 
dimensional spinor space in ten dimensions. The orthogonal 
projector is $(1 - K)_\alpha \ ^\beta$. 
Now the pure spinor condition implies
%**   2.13 %%%%%%%%%%%%%%%%%%%%%%%%%%%%%%%%%%%%%%%%%%%%%%%%%%%%%%%%%
\begin{eqnarray}
\lambda^\alpha K_\alpha \ ^\beta = 0.
\label{2.13}
\end{eqnarray}
%%%%%%%%%%%%%%%%%%%%%%%%%%%%%%%%%%%%%%%%%%%%%%%%%%%%%%%%%%%%%%%%%%%
Since $K$ projects on a 5 dimensional subspace, Eq. (\ref{2.13}) is a simple 
way to understand why a pure spinor has eleven independent components.

Then we postulate the following OPE between $\omega$ and $\lambda$:
%**   2.14 %%%%%%%%%%%%%%%%%%%%%%%%%%%%%%%%%%%%%%%%%%%%%%%%%%%%%%%%%
\begin{eqnarray}
<\omega_\alpha (y) \lambda^\beta (z)> = \frac{1}{y-z} 
(\delta_\alpha^\beta - K_\alpha \ ^\beta (z)).
\label{2.14}
\end{eqnarray}
%%%%%%%%%%%%%%%%%%%%%%%%%%%%%%%%%%%%%%%%%%%%%%%%%%%%%%%%%%%%%%%%%%%
It follows from Eq. (\ref{2.14}) that the OPE between $\omega$ and
the pure spinor condition vanishes identically. Moreover, the BRST charge
$Q$  is then strictly nilpotent even acting on $\omega$.
It is useful to notice that, with the help of the projector $K$, one can 
obtain a non-covariant but gauge-invariant antighost $\tilde\omega$ defined
as
%**   2.15 %%%%%%%%%%%%%%%%%%%%%%%%%%%%%%%%%%%%%%%%%%%%%%%%%%%%%%%%%
\begin{eqnarray}
\tilde \omega_\alpha = ( 1 - K )_\alpha \ ^\beta \omega_\beta. 
\label{2.15}
\end{eqnarray}
%%%%%%%%%%%%%%%%%%%%%%%%%%%%%%%%%%%%%%%%%%%%%%%%%%%%%%%%%%%%%%%%%%%

In the framework of this formalism one can compute \cite{Oda4} the OPE's among 
the ghost current, Lorentz current and stress energy tensor and one can
obtain  the  quantum version of these operators. Indeed, it has been shown
in \cite{Oda4} that all the non-covariant, Y-dependent contributions in the
r.h.s. of the OPE's among these operators disappear if the stress energy 
tensor, the Lorentz
current for the ghost sector, and the ghost current  at the quantum level, are 
improved by  $Y$-dependent correction terms, those are
%**   2.16%%%%%%%%%%%%%%%%%%%%%%%%%%%%%%%%%%%%%%%%%%%%%%%%%%%%%%%%%
\begin{eqnarray}
T &=& - \frac{1}{2} \partial X^a \partial X_a - p_\alpha \partial 
\theta^\alpha + T_{\lambda}  \nonumber\\ 
&=& - \frac{1}{2} \Pi^a \Pi_a - d_\alpha \partial \theta^\alpha
+ \omega_\alpha \partial \lambda^\alpha + \frac{3}{2} \partial(Y \partial 
\lambda),
\label{2.16}
\end{eqnarray}
%%%%%%%%%%%%%%%%%%%%%%%%%%%%%%%%%%%%%%%%%%%%%%%%%%%%%%%%%%%%%%%%%%%
%**   2.17 %%%%%%%%%%%%%%%%%%%%%%%%%%%%%%%%%%%%%%%%%%%%%%%%%%%%%%%%%
\begin{eqnarray}
N^{ab} = \frac{1}{2} \Big[ \omega \Gamma^{ab} \lambda 
- \frac{3}{2} \partial Y \Gamma^{ab} \lambda - 2 Y \Gamma^{ab} 
\partial \lambda \Big],
\label{2.17}
\end{eqnarray}
%%%%%%%%%%%%%%%%%%%%%%%%%%%%%%%%%%%%%%%%%%%%%%%%%%%%%%%%%%%%%%%%%%%
%**   2.18 %%%%%%%%%%%%%%%%%%%%%%%%%%%%%%%%%%%%%%%%%%%%%%%%%%%%%%%%%
\begin{eqnarray}
J = \omega \lambda + \frac{7}{2}  Y \partial \lambda.
\label{2.18}
\end{eqnarray}
%%%%%%%%%%%%%%%%%%%%%%%%%%%%%%%%%%%%%%%%%%%%%%%%%%%%%%%%%%%%%%%%%%%

Then the OPE's among $T$, $N^{ab}$ and $J$ read
%**   2.19 %%%%%%%%%%%%%%%%%%%%%%%%%%%%%%%%%%%%%%%%%%%%%%%%%%%%%%%%%
\begin{eqnarray}
<T(y) T(z)> = \frac{2}{(y-z)^2} T(z) + \frac{1}{y-z} \partial T(z),
\label{2.19}
\end{eqnarray}
%%%%%%%%%%%%%%%%%%%%%%%%%%%%%%%%%%%%%%%%%%%%%%%%%%%%%%%%%%%%%%%%%%%
%**   2.20 %%%%%%%%%%%%%%%%%%%%%%%%%%%%%%%%%%%%%%%%%%%%%%%%%%%%%%%%%
\begin{eqnarray}
<T(y) J(z)> = \frac{8}{(y-z)^3} + \frac{1}{(y-z)^2} J(z) 
+ \frac{1}{y-z} \partial J(z),
\label{2.20}
\end{eqnarray}
%%%%%%%%%%%%%%%%%%%%%%%%%%%%%%%%%%%%%%%%%%%%%%%%%%%%%%%%%%%%%%
%**   2.21 %%%%%%%%%%%%%%%%%%%%%%%%%%%%%%%%%%%%%%%%%%%%%%%%%%%%%%%%%
\begin{eqnarray}
<T(y) N^{ab}(z)> = \frac{1}{(y-z)^2}N^{ab}(z) + \frac{1}{y-z}
\partial N^{ab}(z),
\label{2.21}
\end{eqnarray}
%%%%%%%%%%%%%%%%%%%%%%%%%%%%%%%%%%%%%%%%%%%%%%%%%%%%%%%%%%%%%%%%%%%%
%**   2.22 %%%%%%%%%%%%%%%%%%%%%%%%%%%%%%%%%%%%%%%%%%%%%%%%%%%%%%%%%
\begin{eqnarray}
<J(y) J(z) >= - \frac{4}{(y-z)^2},
\label{2.22}
\end{eqnarray}
%%%%%%%%%%%%%%%%%%%%%%%%%%%%%%%%%%%%%%%%%%%%%%%%%%%%%%%%%%%%%%%%%%%
%**   2.23 %%%%%%%%%%%%%%%%%%%%%%%%%%%%%%%%%%%%%%%%%%%%%%%%%%%%%%%%%
\begin{eqnarray}
<J(y) N^{ab}(z)> = 0,
\label{2.23}
\end{eqnarray}
%%%%%%%%%%%%%%%%%%%%%%%%%%%%%%%%%%%%%%%%%%%%%%%%%%%%%%%%%%%%%%%%%%%
%**   2.24 %%%%%%%%%%%%%%%%%%%%%%%%%%%%%%%%%%%%%%%%%%%%%%%%%%%%%%%%%
\begin{eqnarray}
<N^{ab}(y) N^{cd}(z)> = - \frac{3}{(y-z)^2} \eta^{d[a}\eta^{b]c}
- \frac{1}{y-z} (\eta^{a[c}N^{d]b} - \eta^{b[c}N^{d]a}),
\label{2.24}
\end{eqnarray}
%%%%%%%%%%%%%%%%%%%%%%%%%%%%%%%%%%%%%%%%%%%%%%%%%%%%%%%%%%%%%%%%%%%
%%%%%%%%%%%%%%%%%%%%%%%%%%%%%%%%%%%%%%%%%%%%%%%%%%%%%%%%%%%%%%%%%%%
which are in full agreement with \cite{Ber1, Ber3}.
Note that although the correction terms in the currents depend on the 
non-covariant Y-field explicitly, these can be rewritten as BRST-exact 
terms.

Now a remark is in order. It appears at first sight that, due to the 
correction terms, the operators $J$, $N^{ab}$ and $T$ are singular at 
$v\lambda = 0$ but the opposite is in fact true: it is clear from 
Eqs. (\ref{2.19})-(\ref{2.24})
that the $Y$-dependent correction terms have just the r\^ole of cancelling 
the singularites which are present in the operators  $T_0$, $N_0^{ab}$ and 
$J_0$, owing to the singular nature of the OPE (\ref{2.14}) 
between $\omega$ and $\lambda$.
\footnote {As anticipated in the notation , we will append a suffix 
$"0"$ when we refer to compound fields at the classical level, that is, given 
in terms of $T_0$, $N_0^{ab}$ and $J_0$, and we will reserve the notation 
without suffix $"0"$ in denoting the corresponding quantities at the
quantum level, given in terms of $T$, $N^{ab}$ and $J$.}

It will be convenient to rewrite (\ref{2.17}), (\ref{2.18}) and $T_{\lambda}$ 
as
%**   2.25%%%%%%%%%%%%%%%%%%%%%%%%%%%%%%%%%%%%%%%%%%%%%%%%%%%%%%%%%%%%%
\begin{eqnarray}
N^{ab} = \frac{1}{2} [ \Omega \Gamma^{ab} \lambda 
 - 2 Y \Gamma^{ab} \partial \lambda ],
\label{2.25}
\end{eqnarray}
%%%%%%%%%%%%%%%%%%%%%%%%%%%%%%%%%%%%%%%%%%%%%%%%%%%%%%%%%%%%%%%%%%%%%%
%**   2.26 %%%%%%%%%%%%%%%%%%%%%%%%%%%%%%%%%%%%%%%%%%%%%%%%%%%%%%%%%
\begin{eqnarray}
J = \Omega \lambda +  2 Y \partial \lambda,
\label{2.26}
\end{eqnarray}
%%%%%%%%%%%%%%%%%%%%%%%%%%%%%%%%%%%%%%%%%%%%%%%%%%%%%%%%%%%%%%%%%%%%%%%   
%**   2.27%%%%%%%%%%%%%%%%%%%%%%%%%%%%%%%%%%%%%%%%%%%%%%%%%%%%%%%%%%%%%%
\begin{eqnarray}
T_ {\lambda} = \Omega \partial \lambda + 3 \partial Y \partial \lambda 
+ \frac{3}{2} Y \partial^2 \lambda,
\label{2.27}
\end{eqnarray}
%%%%%%%%%%%%%%%%%%%%%%%%%%%%%%%%%%%%%%%%%%%%%%%%%%%%%%%%%%%%%%%%%%%%%%%
where we have introduced the quantity
%%%%%%%%2.28%%%%%%%%%%%%%%%%%%%%%%%%%%%%%%%%%%%%%%%%%%%%%%%%%%%%%%%%%%%
\begin{eqnarray}
\Omega_\alpha = \omega_\alpha - \frac{3}{2} \partial Y_\alpha.
\label{2.28}
\end{eqnarray} 
%%%%%%%%%%%%%%%%%%%%%%%%%%%%%%%%%%%%%%%%%%%%%%%%%%%%%%%%%%%%%%%%%%%%%%%

The Y-formalism explained thus far is also useful to deal with the $b$ field
which plays an important role in computing higher loop 
amplitudes.  Its main property is
%**   2.29 %%%%%%%%%%%%%%%%%%%%%%%%%%%%%%%%%%%%%%%%%%%%%%%%%%%%%%%%%%%%
\begin{eqnarray}
\{ Q, b(z) \} = T(z),
\label{2.29}
\end{eqnarray}
%%%%%%%%%%%%%%%%%%%%%%%%%%%%%%%%%%%%%%%%%%%%%%%%%%%%%%%%%%%%%%%%%%%
where $T$ is the stress energy tensor. 
%%%%%%%%%%%%%%%%%%%%%%%%%%%%%%%%%%%%%%%%%%%%%%%%%%%%%%%%%%%%%%%%%%%
Since in the pure spinor formulation the reparametrization ghosts do not
exist, $b$ must be a composite field. Moreover, since the $b$ ghost has ghost 
number $-1$ and the covariant fields, which include $\omega_\alpha$ and are 
gauge invariant under the $\omega$-symmetry, always have ghost number zero or positive, one 
must use $Y_\alpha$ (which also has ghost number $-1$) to construct the $b$ 
ghost. 
Therefore $b$ is not super-Poincar\'e invariant.  The $b$ ghost has been 
constructed for the first time in \cite{Ber3} in the U(5)-formalism in such 
a way 
that it satisfies Eq. (\ref{2.29}). In the Y-formalism at hand, at the 
classical level it takes the form
%**   2.30 %%%%%%%%%%%%%%%%%%%%%%%%%%%%%%%%%%%%%%%%%%%%%%%%%%%%%%%%%
\begin{eqnarray}
b_{0 Y} = \frac{1}{2} \Pi^a Y \Gamma_a d + \omega ( 1 - K ) \partial \theta
= Y_\alpha G_0^\alpha,
\label{2.30}
\end{eqnarray}
%%%%%%%%%%%%%%%%%%%%%%%%%%%%%%%%%%%%%%%%%%%%%%%%%%%%%%%%%%%%%%%%%%%
where
%**   2.31  %%%%%%%%%%%%%%%%%%%%%%%%%%%%%%%%%%%%%%%%%%%%%%%%%%%%%%%%%%%%%
\begin{eqnarray}
G_0^{\alpha} = \frac{1}{2}\Pi_a (\Gamma^a d)^\alpha - \frac{1}{4} 
N_0^{ab} (\Gamma_{ab}\partial \theta)^{\alpha} - \frac{1}{4} J_0 \partial 
\theta^\alpha.
\label{2.31}
\end{eqnarray}
%%%%%%%%%%%%%%%%%%%%%%%%%%%%%%%%%%%%%%%%%%%%%%%%%%%%%%%%%%%%%%%%%%%%%%%%%%
The last equality in (\ref{2.30}) follows from the identity (\ref{A.3}).
The expression of $b_{Y}$ at the quantum level will be derived in section 5.

The non-covariance of $b_Y$ is not dangerous since, as we shall show in section
5, the Lorentz variation of $b_Y$ (or of its improvement at the non-minimal 
level) is BRST-exact.
However, this operator  cannot be accepted as  insertion to compute
higher loop amplitudes. Indeed, contrary to the operators 
$T$, $N^{ab}$ and $J$, it has a true singularity at $v\lambda = 0$ of the form
$(v\lambda)^{-1}$. The point is that there exists an operator $\xi = 
Y\theta$, singular with a pole at $v\lambda \rightarrow 0 $, such that
$ \{ Q,\xi \} = 1 $ and the cohomology would become trivial if this 
operator is allowed in the Hilbert space, since for any closed operator $V$, 
 $V = \{Q,\xi V\} $. Then, for consistency, operators 
singular at  $v\lambda \rightarrow 0 $ must be excluded from the Hilbert 
space.

%%%%%%%%%%%%%%%%%%%%%%%%%%%%%%%%%%%%%%%%%%%%%%%%%%%%%%%%%%%%%%%%%%%%%
%%%%%%%%%%%%%%%%%%%%%%%%%%%%%%   SEC  3    %%%%%%%%%%%%%%%%%%%%%%%%%%
%%%%%%%%%%%%%%%%%%%%%%%%%%%%%%%%%%%%%%%%%%%%%%%%%%%%%%%%%%%%%%%%%%%%%
\section{Fundamental operators and normal-ordering effects}

When we attempt to construct a $b$ ghost covariantly, either a 
picture-raised 
$b$ ghost \cite{Ber10, Oda4} or a covariant $b$ ghost in the framework
of the non-minimal approach \cite{Ber12}, we encounter several fundamental 
operators, $G^\alpha$, $H^{\alpha\beta}$, $K^{\alpha\beta\gamma}$
and $L^{\alpha\beta\gamma\delta}$ \cite{Ber10, Oda4}, which are a 
generalization 
of the constraints introduced by Siegel some time ago in \cite{Siegel}. 
Thus, in this section, we will consider those operators in order. 
We will pay a special attention to a consistent treatment of the 
normal-ordering effects.

Let us notice that in addition to $G^\alpha$, the totally antisymmetrized 
operators 
$H^{[\alpha\beta]}$, $K^{[\alpha\beta\gamma]}$ and $L^{[\alpha\beta\gamma
\delta]}$ are the more fundamental objects and are of particular interest 
since they are involved in the construction of the $b$ field in the 
non-minimal formulation. 
At the classical level, $G^\alpha$ is defined in (\ref{2.31}) and  
$H^{[\alpha\beta]}$, $ K^{[\alpha\beta\gamma]}$ and $L^{[\alpha\beta
\gamma\delta]}$ 
are given by
%**   3.1 %%%%%%%%%%%%%%%%%%%%%%%%%%%%%%%%%%%%%%%%%%%%%%%%%%%%%%%%%
\begin{eqnarray}
H_0^{[\alpha\beta]} &=& \frac{1}{384} \Gamma^{\alpha\beta}_{abc} 
( d \Gamma^{abc} d + 24 N_0^{ab} \Pi^c ), \nonumber\\
K_0^{[\alpha\beta\gamma]} &=& - \frac{1}{96} \Gamma^{[\alpha\beta}_{abc} 
( \Gamma^a d )^{\gamma]} N_0^{bc}, \nonumber\\
L_0^{[\alpha\beta\gamma\delta]} &=& - \frac{1}{3072} (\Gamma_{abc})^{[\alpha
\beta} (\Gamma^{ade})^{\gamma\delta]} N_0^{bc} N_{0de}.
\label{3.1}
\end{eqnarray}
%%%%%%%%%%%%%%%%%%%%%%%%%%%%%%%%%%%%%%%%%%%%%%%%%%%%%%%%%%%%%%%%%%%%
They satisfy the following recursive relations:
%**   3.2  %%%%%%%%%%%%%%%%%%%%%%%%%%%%%%%%%%%%%%%%%%%%%%%%%%%%%%%%%
\begin{eqnarray}
\{ Q, G_0^\alpha \} &=& \lambda^\alpha T_0, \nonumber\\  
\big[ Q, H_0^{[\alpha\beta]} \big] &=& \lambda^{[\alpha} G_0^{\beta]}, 
\nonumber \\
\{ Q, K_0^{[\alpha\beta\gamma]} \} &=& \lambda^{[\alpha} H_0^{\beta\gamma]},  
\nonumber \\
\big[ Q, L_0^{[\alpha\beta\gamma\delta]} \big] &=& \lambda^{[\alpha}
K_0^{\beta\gamma\delta]}, \nonumber \\
\lambda^{[\alpha} L_0^{\beta\gamma\delta\rho]} &=& 0,
\label{3.2}
\end{eqnarray}       
%%%%%%%%%%%%%%%%%%%%%%%%%%%%%%%%%%%%%%%%%%%%%%%%%%%%%%%%%%%%%%%%%%%%%
which one can verify easily.
The full fields $H_0^{\alpha\beta}$, $ K_0^{\alpha\beta\gamma}$ and 
$L_0^{\alpha\beta\gamma\delta}$, which are involved in the construction of 
the picture-raised $b$ ghost, can be obtained by adding new terms symmetric 
with respect to at least a couple of adjacent indices, and they satisfy the
recursive relations
%**   3.3  %%%%%%%%%%%%%%%%%%%%%%%%%%%%%%%%%%%%%%%%%%%%%%%%%%%%%%%%%%%
\begin{eqnarray}
\big[ Q, H_0^{\alpha\beta} \big] &=& \lambda^{\alpha} G_0^{\beta} + \cdots, 
\nonumber \\
\{ Q, K_0^{\alpha\beta\gamma} \} &=& \lambda^{\alpha} H_0^{\beta\gamma} +
\cdots,  \nonumber \\
\big[ Q, L_0^{\alpha\beta\gamma\delta} \big] &=& \lambda^{\alpha}
K_0^{\beta\gamma\delta} + \cdots, \nonumber \\
\lambda^{\alpha} L_0^{\beta\gamma\delta\rho} &=& 0 + \cdots,
\label{3.3}
\end{eqnarray}       
%%%%%%%%%%%%%%%%%%%%%%%%%%%%%%%%%%%%%%%%%%%%%%%%%%%%%%%%%%%%%%%%%%%%%%%% 
where the dots denote "$\Gamma_1$-traceless terms", i.e. terms that vanish if
saturated with a $\Gamma_a^{\alpha_{i}\alpha_{i+1}}$ between two adjacent 
indices. The fields $H_0^{\alpha\beta}$, $ K_0^{\alpha\beta\gamma}$ and 
$L_0^{\alpha\beta\gamma\delta}$ are defined modulo $\Gamma_1$-traceless terms.

In this section we wish to discuss these operators and their recursive relations
at the quantum level.
A remark is in order. At the quantum level, in dealing with holomorphic
operators composed of fields with singular OPE's, a normal-ordering prescription 
is needed for their definition. 
As a rule, for the normal ordering of two operators $A$ and $B$ we shall adopt in 
this paper the generalized normal-ordering prescription, denoted by $(A B)$ in 
\cite{Francesco} since it is convenient in carrying out explicit calculations. 
As explained in Appendix B, this prescription consists in subtracting the singular 
poles, evaluated at the point of the second entry and it is given by the contour integration 
%**  3.4 %%%%%%%%%%%%%%%%%%%%%%%%%%%%%%%%%%%%%%%%%%%%%%%%%%%%%%%%%
\begin{eqnarray}
(A B)(z) = \oint_z \frac{dw}{w-z} A(w) B(z).
\label{3.4}
\end{eqnarray}
%%%%%%%%%%%%%%%%%%%%%%%%%%%%%%%%%%%%%%%%%%%%%%%%%%%%%%%%%%%%%%%%%%%
Often, for simplicity, in dealing with this prescription  the outermost 
parenthesis is suppressed and the normal ordering is taken from the right so that, 
in general, 
$A_1 A_2 A_3...A_n$  means $(A_1(A_2(A_3( \cdots A_n) \cdots )))$.

A different prescription denoted as $:A B:$, that we shall call "improved",
consists in subtracting the full contraction $<A(y)B(z)>$, included a possible finite 
term, as computed from the canonical OPE's (\ref{2.7}) and (\ref{2.14}).
In many cases the two prescriptions coincide but when they are different, 
it happens, as we shall see, that the final  results look more natural 
if expressed  in the improved prescription.  
             
\subsection{$G^\alpha$}

$G^\alpha$ is obtained from  (\ref{2.31}) by replacing $N_0^{ab}$ and
$J_0$ with $ N^{ab}$ and $J$ as defined in Eqs. (\ref{2.17}) and (\ref{2.18}) 
and adding a normal-ordering term parametrized by a constant  $c_1$ 
%**   3.5 %%%%%%%%%%%%%%%%%%%%%%%%%%%%%%%%%%%%%%%%%%%%%%%%%%%%%%%%%%%%%%
\begin{eqnarray}
G^\alpha &=&  \frac{1}{2} \Pi^a (\Gamma_a d)^\alpha 
- \frac{1}{4} N_{ab} (\Gamma^{ab} \partial \theta)^\alpha  
- \frac{1}{4} J \partial \theta^\alpha + c_1 \partial^2 \theta^\alpha 
\nonumber\\
&\equiv& G_1^\alpha + G_2^\alpha + G_3^\alpha + G_4^\alpha.
\label{3.5}
\end{eqnarray}
%%%%%%%%%%%%%%%%%%%%%%%%%%%%%%%%%%%%%%%%%%%%%%%%%%%%%%%%%%%%%%%%%%%%%%%%%
The constant $c_1$ will be determined from the requirement that
 $G^\alpha$ should be a 
primary field of conformal weight $2$. Then we have to compute the OPE 
$<T(y) G^\alpha(z)>$. The three terms $G_1^\alpha \equiv  \frac{1}{2} \Pi^a
(\Gamma_a d)^\alpha$, $G_2^\alpha \equiv - \frac{1}{4} N_{ab} (\Gamma^{ab} 
\partial \theta)^\alpha$ and $G_3^\alpha \equiv - \frac{1}{4} J 
\partial \theta^\alpha$ 
are all products of two operators of conformal weight $1$ so that 
their OPE's with the stress energy tensor can be easily calculated. 
One finds that only $G_2^\alpha$ 
is a primary field. $G_1^\alpha$ has a triple pole with residuum $- 5 \partial 
\theta^\alpha$ and $G_3^\alpha$ has a triple pole with residuum  $-2 
\partial \theta^\alpha$.
Moreover, the normal-ordering term $G_4^\alpha \equiv  c_1 \partial^2 
\theta^\alpha$ also has a triple pole with residuum $2 c_1 \partial 
\theta^\alpha$. Therefore, putting them together, one has
%**   3.6 %%%%%%%%%%%%%%%%%%%%%%%%%%%%%%%%%%%%%%%%%%%%%%%%%%%%%%%%%%%%%%%%
\begin{eqnarray}
<T(y) G^\alpha(z)> = \frac{- 5 - 2 + 2 c_1}{(y-z)^3} \partial \theta^\alpha(z)
+ \frac{2}{(y-z)^2} G^\alpha(z) +  \frac{1}{y-z} \partial G^\alpha(z).
\label{3.6}
\end{eqnarray}
%%%%%%%%%%%%%%%%%%%%%%%%%%%%%%%%%%%%%%%%%%%%%%%%%%%%%%%%%%%%%%%%%%%%%%%%%%
Hence, the requirement that $G^\alpha$ must be a primary field of conformal 
weight $2$ is satisfied  when we select the constant $c_1$ to be 
$\frac{7}{2}$. 

In spite of the appearance, it turns out that this figure is in agreement 
with the result of \cite{Ber10} where the value $- \frac{1}{4}$ is indicated 
as the coefficient in front of the normal-ordering term $ \partial^2 
\theta^\alpha $ in $ G^\alpha $. 
The difference is an artifact of the different normal-ordering 
prescriptions, the generalized normal-ordering prescription in (\ref{3.5}) 
and the improved one. 
Whereas the two prescriptions coincide for $G_2^\alpha$ 
and $G_3^\alpha$, there appears a difference in $G_1^\alpha$. Indeed, since
%**   3.7 %%%%%%%%%%%%%%%%%%%%%%%%%%%%%%%%%%%%%%%%%%%%%%%%%%%%%%%%%%%%%%%%%%
\begin{eqnarray}
\Pi^a(x) d_\alpha(z) 
&=& \frac{1}{2} \frac{1}{(x-z)^2} [ (\Gamma^a \theta)_\alpha(z) 
- (\Gamma^a \theta)_\alpha(x) ]  \nonumber\\
&-& \frac{1}{2} \frac{1}{x-z} (\Gamma^a \partial \theta)_\alpha(x)
+ : \Pi^a(z) d_\alpha(z) : + \cdots,
\label{3.7}
\end{eqnarray}
%%%%%%%%%%%%%%%%%%%%%%%%%%%%%%%%%%%%%%%%%%%%%%%%%%%%%%%%%%%%%%%%%%%
we obtain
%**   3.8 %%%%%%%%%%%%%%%%%%%%%%%%%%%%%%%%%%%%%%%%%%%%%%%%%%%%%%%%%%%%%%
\begin{eqnarray}
\frac{1}{2} ( \Pi^a (\Gamma_a d)^\alpha ) 
= - \frac{15}{4} \partial^2 \theta^\alpha
+ \frac{1}{2} : \Pi^a d_\alpha :.
\label{3.8}
\end{eqnarray}
%%%%%%%%%%%%%%%%%%%%%%%%%%%%%%%%%%%%%%%%%%%%%%%%%%%%%%%%%%%%%%%%%%%%%%%%%%
Substituting this result into Eq. (\ref{3.5}), setting $c_1 = \frac{7}{2}$,
we have
%**   3.9 %%%%%%%%%%%%%%%%%%%%%%%%%%%%%%%%%%%%%%%%%%%%%%%%%%%%%%%%%%%%%%
\begin{eqnarray}
G^\alpha = : \frac{1}{2} \Pi^a (\Gamma_a d)^\alpha :
- \frac{1}{4} N_{ab} (\Gamma^{ab} \partial \theta)^\alpha  
- \frac{1}{4} J \partial \theta^\alpha - \frac{1}{4}\partial^2 \theta^\alpha,
\label{3.9}
\end{eqnarray}
%%%%%%%%%%%%%%%%%%%%%%%%%%%%%%%%%%%%%%%%%%%%%%%%%%%%%%%%%%%%%%%%%%
which precisely coincides with the expression given in  \cite{Ber10}.

Next, we want to derive the quantum counterpart of the first 
(classical) recursive relations in (\ref{3.2}) and, for that, we need to 
compute  $\{ Q, G^\alpha \}$.
In doing this calculation, one must be careful to deal with the order of the 
factors in the 
terms coming from the (anti)commutator among $Q$ and $G^\alpha$ and 
use repeatedly the 
rearrangement theorem, reviewed in Appendix B, in order to recover the 
operator $\lambda^{\alpha}T $. The details of this calculation are presented 
in Appendix C. As expected from 
the covariance of  $\{ Q, G^\alpha \}$, the terms involving $Y$, coming from 
the rearrangement procedure, cancel exactly those coming from the 
$Y$-dependent correction terms
of the operators $N^{ab}$ and $J$ (see (\ref{2.17}) and (\ref{2.18})) present 
in the definition of $G^{\alpha}$. The final result is  

%**   3.10 %%%%%%%%%%%%%%%%%%%%%%%%%%%%%%%%%%%%%%%%%%%%%%%%%%%%
\begin{eqnarray}
\{ Q, G^{\alpha} \} = \lambda^{\alpha} T - \frac{1}{2} \partial^2 \lambda^
\alpha.
\label{3.10}
\end{eqnarray}
%%%%%%%%%%%%%%%%%%%%%%%%%%%%%%%%%%%%%%%%%%%%%%%%%%%%%%%%%%%%
The normal-ordering term $- {\frac{1}{2}} \partial^2 \lambda^\alpha$ 
in (\ref{3.10}) might appear to be strange 
at first sight, but it is indeed quite reasonable. The point is  that
it is not $\lambda^\alpha T$ but $\lambda^\alpha T - \frac{1}{2} 
\partial^2 \lambda^\alpha$ 
that is a primary field of conformal weight 2 when we take account of the 
normal-ordering effects. In fact, since
%**   3.11 %%%%%%%%%%%%%%%%%%%%%%%%%%%%%%%%%%%%%%%%%%%%%%%%%%%%%%%%%%%%%%
\begin{eqnarray}
< \lambda^\alpha(y) T(z) > \equiv \frac{R_1^\alpha(z)}{y-z}
= - \frac{\partial \lambda^\alpha(z)}{y-z},
\label{3.11}
\end{eqnarray}
%%%%%%%%%%%%%%%%%%%%%%%%%%%%%%%%%%%%%%%%%%%%%%%%%%%%%%%%%%%%%%
$<T(y) (\lambda^\alpha T)(z)>$ has a triple pole with residuum
 $+ \partial^2\lambda$,
and $\frac{1}{2} \partial^2 \lambda$ has the same triple pole, it follows 
 that
%**   3.12 %%%%%%%%%%%%%%%%%%%%%%%%%%%%%%%%%%%%%%%%%%%%%%%%%%%%%
\begin{eqnarray}
B^\alpha_1 = \lambda^\alpha T - \frac{1}{2} \partial^2 \lambda^\alpha,
\label{3.12}
\end {eqnarray}
%%%%%%%%%%%%%%%%%%%%%%%%%%%%%%%%%%%%%%%%%%%%%%%%%%%%%%%%%%%%%
is a BRST-closed primary operator of conformal weight $2$.
{}From now on, it is convenient to define 
%**   3.13 %%%%%%%%%%%%%%%%%%%%%%%%%%%%%%%%%%%%%%%%%%%%%%%%%%%%%
\begin{eqnarray}
\hat G^\alpha = G^\alpha +\frac{1}{2} \partial^2 \theta^\alpha,
\label{3.13}
\end{eqnarray}
%%%%%%%%%%%%%%%%%%%%%%%%%%%%%%%%%%%%%%%%%%%%%%%%%%%%%%%%%%%%%%%%%%%%%%
so that (\ref{3.10}) becomes
%**   3.14 %%%%%%%%%%%%%%%%%%%%%%%%%%%%%%%%%%%%%%%%%%%%%%%%%%%%%%%%%%%
\begin{eqnarray}
\{ Q,\hat G^{\alpha} \} = \lambda^{\alpha} T.
\label{3.14}
\end{eqnarray}
%%%%%%%%%%%%%%%%%%%%%%%%%%%%%%%%%%%%%%%%%%%%%%%%%%%%%%%%%%%%%%%%%%%%%%

Now we would like to study the operator  $\lambda^\alpha G^\beta$, that is expected 
to arise in the quantum counterpart of the second recursive relations 
in (\ref{3.2}).
As before, $\lambda^\alpha G^\beta$ is not primary since $< \lambda^\alpha(y)
 G^\beta(z) >$ is different from zero. Indeed,
%%%%%%%%%%%%%%%%%%%%%%%%%%%%%%%%%%%%%%%%%%%%%%%%%%%%%%%%%%%%%%%%%%%%%%%%%
%**   3.15 %%%%%%%%%%%%%%%%%%%%%%%%%%%%%%%%%%%%%%%%%%%%%%%%%%%%%%%%%%%%%%%%%
\begin{eqnarray}
< \lambda^\alpha(y) G^\beta(z) > \equiv \frac{R^{\alpha\beta}_2(z)}{y-z},
\label{3.15}
\end{eqnarray}
%%%%%%%%%%%%%%%%%%%%%%%%%%%%%%%%%%%%%%%%%%%%%%%%%%%%%%%%%%%%%%%%%%%%%%%%%%
where
%**   3.16 %%%%%%%%%%%%%%%%%%%%%%%%%%%%%%%%%%%%%%%%%%%%%%%%%%%%%%%%%%%%%%%%
\begin{eqnarray}
R^{\alpha\beta}_2 
&=& - \partial \theta^\alpha \lambda^\beta + \frac{1}{2} \Gamma_a^
{\alpha\beta} 
(\partial \theta \Gamma^a \lambda).
\label{3.16}
\end{eqnarray}
%%%%%%%%%%%%%%%%%%%%%%%%%%%%%%%%%%%%%%%%%%%%%%%%%%%%%%%%%%%%%%%%%%%%%%%%%%%%
Note that since  $\partial \lambda^\alpha \partial \theta^\beta $ is also primary,
there is an ambiguity in defining a primary operator, say $ B_{2}^{\alpha\beta} $,
 associated to $\lambda^\alpha G^\beta $. Given (\ref{3.16}),
for the symmetric one, one has 
%**   3.17 %%%%%%%%%%%%%%%%%%%%%%%%%%%%%%%%%%%%%%%%%%%%%%%%%%%%%%%%%%%%%%
\begin{eqnarray}
B_{2}^{(\alpha\beta)} = \frac{1}{16}\Gamma_{a}^{\alpha\beta}[(\lambda\Gamma^a
G) + \frac{7}{2}\partial(\lambda\Gamma^a \partial \theta) + c_{+} (\partial
\lambda \Gamma^a \partial \theta)],
\label{3.17}
\end{eqnarray}
%%%%%%%%%%%%%%%%%%%%%%%%%%%%%%%%%%%%%%%%%%%%%%%%%%%%%%%%%%%%%%%%%%%%%%%%%%%
while, for the antisymmetric one, one has
%**   3.18 %%%%%%%%%%%%%%%%%%%%%%%%%%%%%%%%%%%%%%%%%%%%%%%%%%%%%%%%%%%%%%%%
\begin{eqnarray}
B_{2}^{[\alpha\beta]} = \lambda^{[\alpha} G^{\beta]}
 + \frac{1}{2}\partial(\lambda^{[\alpha}\partial \theta^{\beta]}) + c_{-} 
\partial \lambda^{[\alpha} \partial \theta^{\beta]}.
\label{3.18}
\end{eqnarray}
%%%%%%%%%%%%%%%%%%%%%%%%%%%%%%%%%%%%%%%%%%%%%%%%%%%%%%%%%%%%%%%%%%%%%%%%%%%
Let us remark that $\lambda^\alpha G^\beta $ is not BRST-closed. Indeed
$ \{ Q,\lambda^\alpha G^\beta \} = \lambda^\alpha \lambda^\beta T - 
\frac{1}{2} \lambda^\alpha \partial^2 \lambda^\beta $. Whereas $ 
(\lambda^{[\alpha} (\lambda^{\beta]} T)) = (T(\lambda^{[\alpha} \lambda^
{\beta]}) )= 0 $, one has $ ((\lambda^\alpha\Gamma^{a}_{\alpha\beta} (\lambda^
\beta T)) = (T (\lambda\Gamma^a \lambda)) - 2 (\lambda \Gamma^a \partial^2
\lambda) $. Therefore  
the requirement that $B_{2}^{(\alpha\beta)}$ and $ B_{2}^{[\alpha\beta]} $
are BRST-closed implies $c_{+} = -\frac{5}{2} $ and $c_{-}= - \frac{1}{2}$ 
so that
%**   3.19 %%%%%%%%%%%%%%%%%%%%%%%%%%%%%%%%%%%%%%%%%%%%%%%%%%%%%%%%%%%%%%%%%%%
\begin{eqnarray}
B_{2}^{(\alpha\beta)} = \frac{1}{16}\Gamma_{a}^{\alpha\beta}[\lambda\Gamma^a
G + \frac{5}{2} \lambda\Gamma^a \partial^{2} \theta 
+ \partial (\lambda \Gamma^a \partial \theta)],
\label{3.19}
\end{eqnarray}
%%%%%%%%%%%%%%%%%%%%%%%%%%%%%%%%%%%%%%%%%%%%%%%%%%%%%%%%%%%%%%%%%%%%%%%%%%%%%%
%**   3.20 %%%%%%%%%%%%%%%%%%%%%%%%%%%%%%%%%%%%%%%%%%%%%%%%%%%%%%%%%%%%%%%%%%%
\begin{eqnarray}
B_{2}^{[\alpha\beta]} = \lambda^{[\alpha} G^{\beta]}
 + \frac{1}{2}\lambda^{[\alpha}\partial^2 \theta^{\beta]} =
\lambda^{[\alpha} \hat G^{\beta]}.
\label{3.20}
\end{eqnarray}
%%%%%%%%%%%%%%%%%%%%%%%%%%%%%%%%%%%%%%%%%%%%%%%%%%%%%%%%%%%%%%%%%%%%%%%%%%%%%%

\subsection{$H^{\alpha\beta}$}

A minimal choice for  $H^{\alpha\beta}$ is  
%**   3.21  %%%%%%%%%%%%%%%%%%%%%%%%%%%%%%%%%%%%%%%%%%%%%%%%%%%%%
\begin{eqnarray}
H^{\alpha\beta} = H^{(\alpha\beta)} + H^{[\alpha\beta]},
\label{3.21}
\end{eqnarray}
%%%%%%%%%%%%%%%%%%%%%%%%%%%%%%%%%%%%%%%%%%%%%%%%%%%%%%%%%%%%%%%%%%%%%%
where 
%**   3.22 %%%%%%%%%%%%%%%%%%%%%%%%%%%%%%%%%%%%%%%%%%%%%%%%%%%%%%%
\begin{eqnarray}
H^{(\alpha\beta)} = \frac{1}{16} \Gamma^{\alpha\beta}_a ( N^{ab} \Pi_b
- \frac{1}{2} J \Pi^a + c_2 \partial \Pi^a), 
\label{3.22}
\end{eqnarray}
%%%%%%%%%%%%%%%%%%%%%%%%%%%%%%%%%%%%%%%%%%%%%%%%%%%%%%%%%%%%%%%%%%%%%%
%**   3.23  %%%%%%%%%%%%%%%%%%%%%%%%%%%%%%%%%%%%%%%%%%%%%%%%%
\begin{eqnarray}
H^{[\alpha\beta]} = 
\frac{1}{96} \Gamma^{\alpha\beta}_{abc} (\frac{1}{4} d \Gamma^{abc} d 
+ 6 N^{ab} \Pi^c ). 
\label{3.23}
\end{eqnarray}
%%%%%%%%%%%%%%%%%%%%%%%%%%%%%%%%%%%%%%%%%%%%%%%%%%%%%%%%%%%%%%%%%%%%%%
First, we shall evaluate $<T(y) H^{\alpha\beta}(z)>$ in order to fix the 
normal-ordering term.  We can easily show that $H^{[\alpha\beta]}$ and the 
first term in $H^{(\alpha\beta)}$  are primary fields whereas $- \frac{1}
{32}\Gamma_a^{\alpha\beta} J \Pi^a$  and $\frac{c_2}{16} \Gamma_a^{\alpha
\beta} \partial \Pi^a$ have a triple pole with residua 
$- 4 \frac{1}{16} \Gamma^{\alpha\beta}_a \Pi^a$ and $2 c_2 \frac{1}{16} 
\Gamma^{\alpha\beta}_a \Pi^a$, respectively.
Thus, we obtain
%**   3.24  %%%%%%%%%%%%%%%%%%%%%%%%%%%%%%%%%%%%%%%%%%%%%%%%%%%%%%%
\begin{eqnarray}
<T(y) H^{\alpha\beta}(z)> =  \frac{- 4 + 2 c_2}{(y-z)^3} \frac{1}{16} 
\Gamma^{\alpha\beta}_a \Pi^a(z) + \frac{2}{(y-z)^2} H^{\alpha\beta}(z)
+ \frac{1}{y-z} \partial H^{\alpha\beta}(z),
\label{3.24}
\end{eqnarray}
%%%%%%%%%%%%%%%%%%%%%%%%%%%%%%%%%%%%%%%%%%%%%%%%%%%%%%%%%%%%%%%%%%%%%%
thereby taking $c_2 = 2$ makes $H^{\alpha\beta}$ a primary field of 
conformal weight $2$.  This value  agrees with the value in 
the Berkovits' paper \cite{Ber10}.
Next, we wish to evaluate $[ Q, H^{\alpha\beta} ]$:
%**   3.25  %%%%%%%%%%%%%%%%%%%%%%%%%%%%%%%%%%%%%%%%%%%%%%%%%%%%%%%%%%%
\begin{eqnarray}
[ Q, H^{(\alpha\beta)} ] 
&=& \frac{1}{16} \Gamma^{\alpha\beta}_a \Big[ \frac{1}{2}  (\lambda \Gamma^
{ab} d) \Pi_b
+  N^{ab} (\lambda \Gamma_b \partial \theta)  
+ \frac{1}{2}  (\lambda d) \Pi^a \nonumber\\
&-& \frac{1}{2} J (\lambda \Gamma^a \partial \theta)  
+ c_2 \partial (\lambda \Gamma^a \partial \theta)  \Big],
\label{3.25}
\end{eqnarray} 
%%%%%%%%%%%%%%%%%%%%%%%%%%%%%%%%%%%%%%%%%%%%%%%%%%%%%%%%%%%%%%%%%%%%%%%%%
and
%**   3.26 %%%%%%%%%%%%%%%%%%%%%%%%%%%%%%%%%%%%%%%%%%%%%%%%%%%%%%%%%%%%%%
\begin{eqnarray} 
[ Q, H^{[\alpha\beta]}] 
&=& \frac{1}{96} \Gamma^{\alpha\beta}_{abc} \Big[ -\frac{1}{4}( (\Gamma^d 
\lambda)_{\rho} \Pi_d)(\Gamma^{abc} d)^\rho -\frac{1}{4} (\Gamma^{abc} d)^\rho
(\Gamma^d \lambda)_{\rho} \Pi_d \nonumber\\
&+& 3 ( \lambda \Gamma^{ab} d) \Pi^c + 6  N^{ab} (\lambda \Gamma^c 
\partial \theta) +2 c_3 (\partial \lambda \Gamma^{abc}\partial \theta)   \Big].
\label{3.26}
\end{eqnarray}
%%%%%%%%%%%%%%%%%%%%%%%%%%%%%%%%%%%%%%%%%%%%%%%%%%%%%%%%%%%%%%%%%%%%%%%%%
Then, after some algebra and taking into account the normal-ordering terms by the 
rearrangement formula, we get for the symmetric part of $H^{\alpha\beta}$
%**   3.27 %%%%%%%%%%%%%%%%%%%%%%%%%%%%%%%%%%%%%%%%%%%%%%%%%%%%%%%%%%%%%%%    
\begin{eqnarray}
[ Q, H^{(\alpha\beta)} ] = \frac{1}{16} \Gamma^{\alpha\beta}_a [ \lambda \Gamma^a G  
+ \frac{5}{2} \lambda\Gamma^a \partial^2 \theta
+ \partial(\lambda\Gamma^a\partial \theta)],  
\label{3.27}
\end{eqnarray}
%%%%%%%%%%%%%%%%%%%%%%%%%%%%%%%%%%%%%%%%%%%%%%%%%%%%%%%%%%%%%%%%%%%%%%%%%%
and for the more interesting  antisymmetric part $H^{[\alpha\beta]}$ 
%**   3.28 %%%%%%%%%%%%%%%%%%%%%%%%%%%%%%%%%%%%%%%%%%%%%%%%%%%%%%%%%%%%%%%
\begin{eqnarray}
[ Q, H^{[\alpha\beta]} ]= \lambda^{[\alpha} G^{\beta]} 
+ \frac{1}{2} \lambda^{[\alpha} \partial^2 \theta^{\beta]} = \lambda^{[\alpha}
 \hat G^{\beta]},   
\label{3.28}
\end{eqnarray}
%%%%%%%%%%%%%%%%%%%%%%%%%%%%%%%%%%%%%%%%%%%%%%%%%%%%%%%%%%%%%%%%%%%%%%%%%%
in agreement with (\ref{3.19}) and (\ref{3.20}). Notice that the $Y$-dependent 
contributions coming from rearrangement theorem cancel exactly those coming from the 
definitions (\ref{2.17}) and (\ref{2.18}) of $N^{ab}$ and $J$
(For details see Appendix C).

Since the term $ +\partial(\lambda\Gamma^a\partial \theta)$ in (\ref{3.27})
is the BRST variation of $\partial \Pi^a$, (\ref{3.27}) can be rewritten as  
%**   3.29 %%%%%%%%%%%%%%%%%%%%%%%%%%%%%%%%%%%%%%%%%%%%%%%%%%%%%%%%%%%%%%%%    
\begin{eqnarray}
[ Q,\hat H^{(\alpha\beta)} ] = \frac{1}{16} \Gamma^{\alpha\beta}_a [ \lambda
\Gamma^a G  + \frac{5}{2} \lambda\Gamma^a \partial^2 \theta ],
\label{3.29}
\end{eqnarray}
%%%%%%%%%%%%%%%%%%%%%%%%%%%%%%%%%%%%%%%%%%%%%%%%%%%%%%%%%%%%%%%%%%%%%%%%%%%
where we have defined as $ \hat H^{(\alpha\beta)} = H^{(\alpha\beta)} - \frac{1}{16}\Gamma_{a}^{\alpha\beta} \partial \Pi^a $. 

Now let us consider the composite operator $\lambda^\alpha H^{\beta\gamma}$. 
Since $H^{\alpha\beta}$ has conformal weight $2$ but its contraction 
with $\lambda^\alpha$ does not vanish, one can expect that  $\lambda^\alpha 
H^{\beta\gamma}$ is not primary. Actually, using the fact
%**   3.30 %%%%%%%%%%%%%%%%%%%%%%%%%%%%%%%%%%%%%%%%%%%%%%%%%%%%%%%%%%
\begin{eqnarray}
< \lambda^\alpha(y) H^{\beta\gamma}(z) > \equiv \frac{R^{\alpha\beta\gamma}_3
(z)}{y-z},
\label{3.30}
\end{eqnarray}
%%%%%%%%%%%%%%%%%%%%%%%%%%%%%%%%%%%%%%%%%%%%%%%%%%%%%%%%%%%%%%%%%%%%
with $R^{\alpha\beta\gamma}_3$  being given by 
%**   3.31 %%%%%%%%%%%%%%%%%%%%%%%%%%%%%%%%%%%%%%%%%%%%%%%%%%%%%%%%%%%%
\begin{eqnarray}
R^{\alpha\beta\gamma}_3 = - \frac{1}{32} \Gamma^{\beta\gamma}_a 
[ (\Gamma^{ab} \lambda)^\alpha \Pi_b - \lambda^\alpha \Pi^a]
- \frac{1}{32} \Gamma^{\beta\gamma}_{abc} 
(\Gamma^{ab} \lambda)^\alpha \Pi^c,
\label{3.31}
\end{eqnarray}
%%%%%%%%%%%%%%%%%%%%%%%%%%%%%%%%%%%%%%%%%%%%%%%%%%%%%%%%%%%%%%%%%%%%%%%
it turns out that a primary field of conformal weight $2$ related to $\lambda^\alpha 
H^{\beta\gamma}$ is 
%**   3.32 %%%%%%%%%%%%%%%%%%%%%%%%%%%%%%%%%%%%%%%%%%%%%%%%%%%%%%%
\begin{eqnarray}
B^{\alpha\beta\gamma}_3 \equiv \lambda^\alpha H^{\beta\gamma}
+ \frac{1}{2} \partial R^{\alpha\beta\gamma}_3.
\label{3.32}
\end{eqnarray}
%%%%%%%%%%%%%%%%%%%%%%%%%%%%%%%%%%%%%%%%%%%%%%%%%%%%%%%%%%%%%%%%%%%%%%%
Again there is an arbitrariness in choosing the primary field related to 
$\lambda^\alpha H^{\beta\gamma}$ since $ \partial\lambda^\alpha \Pi^a$ is 
primary.

As in previous cases we are especially interested in the antisymmetric part 
$B^{[\alpha\beta\gamma]}_3$ of $B^{\alpha \beta\gamma}_3$. 
Since, in $D=10$, a field which is totally antisymmetric in its three, 
spinor-like indices contains only the $SO(10)$ irreducible representation (irrep.) $560$ and 
$R^{\alpha \beta \gamma}_3$ in Eq. (\ref{3.31}) does not contain such an 
irrep., it follows that
%**   3.33 %%%%%%%%%%%%%%%%%%%%%%%%%%%%%%%%%%%%%%%%%%%%%%%%%%%%%%%%%%%%
\begin{eqnarray}
R^{[\alpha\beta\gamma]}_3  = 0,
\label{3.33}
\end{eqnarray}
%%%%%%%%%%%%%%%%%%%%%%%%%%%%%%%%%%%%%%%%%%%%%%%%%%%%%%%%%%%%%%%%%%%%%
so that $B^{[\alpha\beta\gamma]}_3$ simply becomes 
%**   3.34 %%%%%%%%%%%%%%%%%%%%%%%%%%%%%%%%%%%%%%%%%%%%%%%%%%%%%%%%%%%%
\begin{eqnarray}
B^{[\alpha\beta\gamma]}_3 = \lambda^{[\alpha} H^{\beta\gamma]}.
\label{3.34}
\end{eqnarray}
%%%%%%%%%%%%%%%%%%%%%%%%%%%%%%%%%%%%%%%%%%%%%%%%%%%%%%%%%%%%%%%%%%%
{}From  Eqs. (\ref{3.28}), (\ref{3.15}) and (\ref{3.16}), it is then easy to show that $\lambda^{[\alpha} H^{\beta\gamma]}$ is  BRST-closed. Indeed, 
one finds 
%**   3.35 %%%%%%%%%%%%%%%%%%%%%%%%%%%%%%%%%%%%%%%%%%%%%%%%%%%%
\begin{eqnarray}
[ Q, \lambda^{[\alpha} H^{\beta\gamma]} ] &=& \lambda^{[\alpha}(\lambda^{\beta}
\hat G^{\gamma]})  \nonumber\\ 
&=& \hat G^{[\gamma}\lambda^{\alpha} \lambda^{\beta]} 
+  \lambda^{[\alpha}\partial(\lambda^{\beta}\partial \theta^{\gamma]}) 
+ \partial(\lambda^{[\alpha}\partial \theta^{\gamma})\lambda^{\beta]} 
\nonumber\\
&=& 0.
\label{3.35}
\end{eqnarray}
%%%%%%%%%%%%%%%%%%%%%%%%%%%%%%%%%%%%%%%%%%%%%%%%%%%%%%%%%%%%%%%%%%

\subsection{$K^{\alpha\beta\gamma}$}

A covariant expression of $K^{\alpha\beta\gamma}$ is
%**   3.36 %%%%%%%%%%%%%%%%%%%%%%%%%%%%%%%%%%%%%%%%%%%%%%%%%%%%%%%%
\begin{eqnarray}
K^{\alpha\beta\gamma} &=&  - \frac{1}{48} \Gamma^{\alpha\beta}_a 
( \Gamma_b d )^\gamma N^{ab} - \frac{1}{192} \Gamma^{\alpha\beta}_{abc} 
( \Gamma^a d )^\gamma N^{bc} \nonumber\\
&+& \frac{1}{192} \Gamma^{\beta\gamma}_a \Big[ ( \Gamma_b d )^\alpha N^{ab}
+ \frac{3}{2} ( \Gamma^a d )^\alpha J 
+ c_3 (\Gamma^a \partial d)^\alpha \Big]   \nonumber\\
&-& \frac{1}{192} \Gamma^{\beta\gamma}_{abc} ( \Gamma^a d )^\alpha N^{bc} 
\nonumber\\
&\equiv& K^{\alpha\beta\gamma}_1 + K^{\alpha\beta\gamma}_2 + 
K^{\alpha\beta\gamma}_3 + K^{\alpha\beta\gamma}_4 + K^{\alpha\beta\gamma}_5 
+ K^{\alpha\beta\gamma}_6, 
\label{3.36}
\end{eqnarray}
%%%%%%%%%%%%%%%%%%%%%%%%%%%%%%%%%%%%%%%%%%%%%%%%%%%%%%%%%%%%%%%%%%%%
whereas the totally antisymmetric part is given by
%**   3.37 %%%%%%%%%%%%%%%%%%%%%%%%%%%%%%%%%%%%%%%%%%%%%%%%%%%%%%%%%%
\begin{eqnarray} 
K^{[\alpha\beta\gamma]} = - \frac{1}{96} \Gamma^{[\alpha\beta}_{abc} 
( \Gamma^a d )^{\gamma]} N^{bc}.
\label{3.37}
\end{eqnarray}
%%%%%%%%%%%%%%%%%%%%%%%%%%%%%%%%%%%%%%%%%%%%%%%%%%%%%%%%%%
The term including a constant $c_3$ describes the normal-ordering contribution.
As before, we will calculate $<T(y) K^{\alpha\beta\gamma}(z)>$  in order
to fix the normal-ordering term.  One finds that all the terms 
$K^{\alpha\beta\gamma}_i$  are primary with conformal weight $2$, 
except $K^{\alpha\beta\gamma}_4 \equiv {\frac{1}{192}} \Gamma^{\beta\gamma}_a
 {\frac{3}{2}} ( \Gamma^a d )^\alpha J$ and $K^{\alpha\beta\gamma}_5 
\equiv c_3 \frac{1}{192} \Gamma^{\beta\gamma}_a (\Gamma^a \partial d)^\alpha$ 
which have triple poles in their OPE's with  $T$. In fact, 
%**   3.38 %%%%%%%%%%%%%%%%%%%%%%%%%%%%%%%%%%%%%%%%%%%%%%%%%%%%%%
\begin{eqnarray}
<T(y) K^{\alpha\beta\gamma}_4(z)> &=&  \frac{1}{192} \Gamma^{\beta\gamma}_a 
\frac{12}{(y-z)^3} (\Gamma^a d)^\alpha(z) + \frac{2}{(y-z)^2} 
K^{\alpha\beta\gamma}_4(z)
+ \frac{1}{y-z} \partial K^{\alpha\beta\gamma}_4(z),  \nonumber\\
<T(y) K^{\alpha\beta\gamma}_5(z)> &=&  \frac{1}{192} \Gamma^{\beta\gamma}_a 
\frac{2 c_3}{(y-z)^3} (\Gamma^a d)^\alpha(z) + \frac{2}{(y-z)^2} 
K^{\alpha\beta\gamma}_5(z)
+ \frac{1}{y-z} \partial K^{\alpha\beta\gamma}_5(z).   \nonumber\\
\label{3.38}
\end{eqnarray}
%%%%%%%%%%%%%%%%%%%%%%%%%%%%%%%%%%%%%%%%%%%%%%%%%%%%%%%%%%%%%%%%%%%%
Therefore, one obtains
%**   3.39 %%%%%%%%%%%%%%%%%%%%%%%%%%%%%%%%%%%%%%%%%%%%%%
\begin{eqnarray}
<T(y) K^{\alpha\beta\gamma}(z)> =  \frac{1}{192} \Gamma^{\beta\gamma}_a
\frac{12 + 2 c_3}{(y-z)^3} (\Gamma^a d)^\alpha(z) 
+ \frac{2}{(y-z)^2} K^{\alpha\beta\gamma}(z)
+ \frac{1}{y-z} \partial K^{\alpha\beta\gamma}(z),
\label{3.39}
\end{eqnarray}
%%%%%%%%%%%%%%%%%%%%%%%%%%%%%%%%%%%%%%%%%%%%%%%%%%%%%%%%%%%%%
so that the condition of a primary operator of conformal weight $2$ requires
us to take $c_3 = -6$, which is a new result.

As for $\{ Q, K^{\alpha\beta\gamma} \}$,
we will limit ourselves to considering only the antisymmetric part 
$K^{[\alpha\beta\gamma]}$ of $K^{\alpha\beta\gamma}$
%**   3.40 %%%%%%%%%%%%%%%%%%%%%%%%%%%%%%%%%%%%%%%%%%%%%%%%%%%%%%%
\begin{eqnarray}
\{ Q, K^{[\alpha\beta\gamma]} \} =
 \frac{1}{96} \Gamma^{[\alpha\beta}_{abc} 
\Big[ ( ( \Gamma^a \Gamma^d \lambda)^{\gamma]} \Pi_d) N^{bc}  
+ \frac{1}{2}  ( \Gamma^{a} d )^{\gamma]} (\lambda \Gamma^{bc} d ) \Big].
\label{3.40}
\end{eqnarray}
%%%%%%%%%%%%%%%%%%%%%%%%%%%%%%%%%%%%%%%%%%%%%%%%%%%%%%%%%%%%%%%%%%%
As before, the $Y$-dependent contributions coming from rearrangement
theorem are exactly cancelled by those coming from the 
definition (\ref{2.17}) of $N^{ab}$, as expected from the covariance of the
l.h.s of (\ref{3.40}). Then from the rearrangement theorem and 
with a few algebra one gets 
%**   3.41 %%%%%%%%%%%%%%%%%%%%%%%%%%%%%%%%%%%%%%%%%%%%%%%%%%%%%%%%
\begin{eqnarray}
\{ Q, K^{[\alpha\beta\gamma]} \} =  \lambda^{[\alpha} H^{\beta\gamma]}.
\label{3.41}
\end{eqnarray}
%%%%%%%%%%%%%%%%%%%%%%%%%%%%%%%%%%%%%%%%%%%%%%%%%%%%%%%%%%%%%%%%%%%
Given that the $Y$-dependent terms are absent, (\ref{3.41}) can also been
argued as follows: cohomology arguments based on Eq. (\ref{3.35})
and the classical equivalence between $\{ Q, K^{[\alpha\beta\gamma]} \}$ and 
$\lambda^{[\alpha} H^{\beta\gamma]}$ imply
$ \{ Q, K^{[\alpha\beta\gamma]} \} =  \lambda^{[\alpha} H^{\beta\gamma]}
+ \Lambda^{[\alpha\beta\gamma]}_3,$
where $\Lambda^{[\alpha\beta\gamma]}_3$ is a primary field of conformal 
weight $2$ satisfying $[ Q, \Lambda^{[\alpha\beta\gamma]}_3 ] = 0$.
Then, notice that $\Lambda^{[\alpha\beta\gamma]}_3$ has ghost number $+1$ and
involves $\partial \lambda^\alpha$ and $\Pi^a$ or $\partial \Pi^a$ and 
$\lambda^\alpha$. However, using these fields, it is impossible to construct 
a $560$ irrep. 
of $SO(10)$, so $\Lambda^{[\alpha\beta\gamma]}_3$ vanishes identically.

As before, let us construct a primary field of conformal weight $2$ from
$\lambda^\alpha K^{\beta\gamma\delta}$. We define $R^{\alpha\beta\gamma
\delta}_4$ by
%**   3.42 %%%%%%%%%%%%%%%%%%%%%%%%%%%%%%%%%%%%%%%%%%%%%%%%%%%%%%%
\begin{eqnarray}
< \lambda^\alpha(y) K^{\beta\gamma\delta}(z) > \equiv 
\frac{R^{\alpha\beta\gamma\delta}_4(z)}{y-z},
\label{3.42}
\end{eqnarray}
%%%%%%%%%%%%%%%%%%%%%%%%%%%%%%%%%%%%%%%%%%%%%%%%%%%%%%%%%%%%%%%%%%%
where $R^{\alpha\beta\gamma\delta}_4$ takes the form  
%**   3.43 %%%%%%%%%%%%%%%%%%%%%%%%%%%%%%%%%%%%%%%%%%%%%%%%%%%%%%%%
\begin{eqnarray}
R^{\alpha\beta\gamma\delta}_4 &=& \frac{1}{96} (\Gamma^{ab} \lambda)^\alpha 
\Gamma^{\beta\gamma}_a (\Gamma_b d)^\delta + \frac{1}{384} (\Gamma^{ab} 
\lambda)^\alpha 
\Gamma^{\beta\gamma}_{abc} (\Gamma^c d)^\delta \nonumber\\
&+& \frac{1}{384} [ (\Gamma^{ab} \lambda)^\alpha (\Gamma_b d)^\beta
+ 3 \lambda^\alpha (\Gamma^a d)^\beta ] \Gamma^{\gamma\delta}_a
+ \frac{1}{384} (\Gamma^{ab} \lambda)^\alpha (\Gamma^c d)^\beta
\Gamma^{\gamma\delta}_{abc}.
\label{3.43}
\end{eqnarray}
%%%%%%%%%%%%%%%%%%%%%%%%%%%%%%%%%%%%%%%%%%%%%%%%%%%%%%%%%%%%%%%%%%
Provided that we define $B^{\alpha\beta\gamma\delta}_4$ as
%**   3.44 %%%%%%%%%%%%%%%%%%%%%%%%%%%%%%%%%%%%%%%%%%%%%%%%%%%%%%%%
\begin{eqnarray}
B^{\alpha\beta\gamma\delta}_4 \equiv \lambda^\alpha K^{\beta\gamma\delta}
+ \frac{1}{2} \partial R^{\alpha\beta\gamma\delta}_4,
\label{3.44}
\end{eqnarray}
%%%%%%%%%%%%%%%%%%%%%%%%%%%%%%%%%%%%%%%%%%%%%%%%%%%%%%%%%%%%%%%%%%%%%
it is easy to get
%**   3.45 %%%%%%%%%%%%%%%%%%%%%%%%%%%%%%%%%%%%%%%%%%%%%%%%%%%%%%%%%
\begin{eqnarray}
< T(y) B^{\alpha\beta\gamma\delta}_4(z) > = \frac{2}{(y-z)^2} 
B^{\alpha\beta\gamma\delta}_4(z)
+ \frac{1}{y-z} \partial B^{\alpha\beta\gamma\delta}_4(z),
\label{3.45}
\end{eqnarray}
%%%%%%%%%%%%%%%%%%%%%%%%%%%%%%%%%%%%%%%%%%%%%%%%%%%%%%%%%%%%%%%%%%%%%%
which means that $B^{\alpha\beta\gamma\delta}_4$ is a primary field of 
conformal weight $ 2$ as expected. As before, there is an arbitrariness
in the choice of the primary field related to $\lambda^\alpha 
K^{\beta\gamma\delta}$ since the field $\partial\lambda^\alpha d_\beta$ 
is primary. 

If one considers the completely antisymmetric component 
$B^{[\alpha\beta\gamma\delta]}_4$,
one can notice that, in $D=10$, a field antisymmetric in its four, 
spinor-like indices 
contains only the irreps. $770$ and $1050$ which are absent in the 
expression (\ref{3.43}) 
of $R^{[\alpha\beta\gamma\delta]}_4$ so that one obtains
%**   3.46 %%%%%%%%%%%%%%%%%%%%%%%%%%%%%%%%%%%%%%%%%%%%%%%%%%%%%%%
\begin{eqnarray}
R^{[\alpha\beta\gamma\delta]}_4 = 0.
\label{3.46}
\end{eqnarray}
%%%%%%%%%%%%%%%%%%%%%%%%%%%%%%%%%%%%%%%%%%%%%%%%%%%%%%%%%%%%%%%%%%%%%%%
Consequently, we have
%**   3.47 %%%%%%%%%%%%%%%%%%%%%%%%%%%%%%%%%%%%%%%%%%%%%%%%%%
\begin{eqnarray}
B^{[\alpha\beta\gamma\delta]}_4 = \lambda^{[\alpha} K^{\beta\gamma\delta]}.
\label{3.47}
\end{eqnarray}
%%%%%%%%%%%%%%%%%%%%%%%%%%%%%%%%%%%%%%%%%%%%%%%%%%%%%%%%%%%%%%%%%%%
{}Furthermore, Eq. (\ref{3.41}) together with (\ref{3.33}) gives us the 
equation
%**   3.48 %%%%%%%%%%%%%%%%%%%%%%%%%%%%%%%%%%%%%%%%%%%%%%%%%%%
\begin{eqnarray}
\{ Q, \lambda^{[\alpha} K^{\beta\gamma\delta]} \} = 0.
\label{3.48}
\end{eqnarray}
%%%%%%%%%%%%%%%%%%%%%%%%%%%%%%%%%%%%%%%%%%%%%%%%%%%%%%%%%%

\subsection{$L^{\alpha\beta\gamma\delta}$}

In this final subsection, we wish to consider $L^{\alpha\beta\gamma\delta}$.
In our previous paper \cite{Oda4}, at the classical level, the form of 
$L^{\alpha\beta\gamma\delta}$ was fixed to be
%**   3.49 %%%%%%%%%%%%%%%%%%%%%%%%%%%%%%%%%%%%%%%%%%%%%%%%%%%%%%%%%%%%%%
\begin{eqnarray}
L_{0}^{\alpha\beta\gamma\delta} = - \frac{1}{24} \lambda^\alpha (\tilde\omega 
\Gamma^a)^\beta[ \lambda^\gamma (\tilde\omega \Gamma_a)^\delta - \frac{1}{4}
(\Gamma_{b}\Gamma_a\lambda)^\gamma (\tilde\omega \Gamma^b)^\delta ],
\label{3.49}
\end{eqnarray}
%%%%%%%%%%%%%%%%%%%%%%%%%%%%%%%%%%%%%%%%%%%%%%%%%%%%%%%%%%%%%%%%%
where $\tilde \omega$ is defined in (\ref{2.15}).
One subtle point associated with this expression is that
$L_{0}^{\alpha\beta\gamma\delta}$ cannot be entirely expressed in terms of
$N_{0}^{ab}$
and $J_{0}$. However, we have found that the dangerous terms involving
$\tilde \omega \Gamma^{a_1 a_2 a_3 a_4} \lambda$ cancel exactly in 
constructing the picture raised $b$ ghost. 

On the other hand,  when we consider the 
totally antisymmetrized 
part of $L_{0}^{\alpha\beta\gamma\delta}$,  these dangerous terms never 
appear. In order to show that, let us  notice that, given  Eq. (\ref{3.49}), 
one can write:
%**   3.50 %%%%%%%%%%%%%%%%%%%%%%%%%%%%%%%%%%%%%%%%%%%%%%%
\begin{eqnarray}
L_{0}^{\alpha\beta\gamma\delta} +  L_{0}^{\gamma\delta\alpha\beta} =
 - \frac{1}{24} (\tilde\omega\Lambda_{c}^ {\alpha\beta}\lambda)
(\tilde\omega\Lambda^{\gamma\delta c}\lambda) - \frac{1}{24} \lambda^\alpha 
(\tilde\omega \Gamma^a)^\beta \lambda^\gamma (\tilde\omega \Gamma_a)^\delta, 
\label{3.50}
\end{eqnarray}
%%%%%%%%%%%%%%%%%%%%%%%%%%%%%%%%%%%%%%%%%%%%%%%%%%%%%%%%%%%%%%%%%%%%%
where we have defined 
%**   3.51 %%%%%%%%%%%%%%%%%%%%%%%%%%%%%%%%%%%%%%%%%%%%%%%%%%%%%%
\begin{eqnarray}
 \tilde\omega\Lambda_{c}^ {\alpha\beta}\lambda = (\tilde\omega \Gamma_c)
^\alpha \lambda^\beta - \frac{1}{4} (\tilde\omega \Gamma_b)^\alpha 
(\Gamma^b \Gamma_c \lambda)^\beta.
\label{3.51}
\end{eqnarray}
%%%%%%%%%%%%%%%%%%%%%%%%%%%%%%%%%%%%%%%%%%%%%%%%%%%%%%%%%%
Then, taking the totally antisymmetrized part of Eq. (\ref{3.50}) one gets
%**   3.52 %%%%%%%%%%%%%%%%%%%%%%%%%%%%%%%%%%%%%%%%%%%%%%%%%%
\begin{eqnarray}
L_{0}^{[\alpha\beta\gamma\delta]} = - \frac{1}{48}(\tilde\omega\Lambda_{c}^ 
{[\alpha\beta}\lambda)
(\tilde\omega\Lambda^{\gamma\delta] c}\lambda).
\label{3.52}
\end{eqnarray}
%%%%%%%%%%%%%%%%%%%%%%%%%%%%%%%%%%%%%%%%%%%%%%%%%%%%%%%%%%
Using (\ref{3.51}) and (A.3),  we can rewrite $\tilde\omega
\Lambda_{c}^ {[\alpha\beta]}\lambda$ as
%**   3.53 %%%%%%%%%%%%%%%%%%%%%%%%%%%%%%%%%%%%%%%
\begin{eqnarray}
\tilde\omega\Lambda_{c}^{[\alpha\beta]}\lambda 
&=& \frac{1}{16} \Gamma^{\alpha\beta}_{abc} 
(\omega \Gamma^{ab} \lambda) \nonumber\\
&=& \frac{1}{8} \Gamma^{\alpha\beta}_{abc} N^{bc}_0. 
\label{3.53}
\end{eqnarray}
%%%%%%%%%%%%%%%%%%%%%%%%%%%%%%%%%%%%%%%%%%%%%%%%%%%%%%%%%%%%%%%
Hence, we have shown that $L_{0}^{[\alpha\beta\gamma\delta]}$ is in fact 
expressed by $N^{ab}_0$.

In order to have a covariant expression for $L^{[\alpha\beta\gamma\delta]}$,
at the quantum level, the classical Lorentz generator $N^{bc}_0$ must be 
replaced with  $N^{bc}$ as given in (\ref{2.17}) so that
$L^{[\alpha\beta\gamma\delta]}$ is
%** 3.54 %%%%%%%%%%%%%%%%%%%%%%%%%%%%%%%%%%%%%%%%%%%%%%%%%%%%%%%
\begin{eqnarray}
L^{[\alpha\beta\gamma\delta]} = - \frac{1}{3072} (\Gamma_{abc})^{[\alpha\beta}
(\Gamma^{ade})^{\gamma\delta]} N^{bc} N_{de}.
\label{3.54}
\end{eqnarray}
%%%%%%%%%%%%%%%%%%%%%%%%%%%%%%%%%%%%%%%%%%%%%%%%%%%%%%%%%%%%%%%%%
{}From the OPE's $<T(y)N^{ab}(z)>$ and $<N^{ab}(y)N^{cd}(z)>$, one can 
easily verify that  
$L^{[\alpha\beta\gamma\delta]}$ is a covariant, primary field of conformal 
weight $2$.

At the classical level one has the identities 
%**   3.55  %%%%%%%%%%%%%%%%%%%%%%%%%%%%%%%%%%%%%%%%%%%%%%%%%%%
\begin{eqnarray}
[ Q, L_0^{[\alpha\beta\gamma\delta]} ] &=& \lambda^{[\alpha}
K_0^{\beta\gamma\delta]}, \nonumber\\
\lambda^{[\alpha} L_0^{\beta\gamma\delta\rho]} &=& 0,
\label{3.55}
\end{eqnarray}       
%%%%%%%%%%%%%%%%%%%%%%%%%%%%%%%%%%%%%%%%%%%%%%%%%%%%%%%%%%%%%%%%%%%
where the last identity follows by noting that $L^{[\alpha\beta
\gamma\delta]}_0$ is proportional to $\lambda^{[\alpha} (\omega \Gamma_a)
^\beta (\omega\Gamma_b)^\gamma (\Gamma^{ab} \lambda)^{\delta]}$. 
Since $L^{[\alpha\beta\gamma\delta]}$ and 
$\lambda^{[\epsilon} L^{\alpha\beta\gamma\delta]}$ are covariant tensors and 
a possible quantum failure of these identities would involve $Y_\alpha$, 
thereby inducing violation of Lorentz covariance, one should expect that 
these identities hold at the quantum level as well. It is worthwhile to 
verify this result directly as a nice check of the consistency of the Y-formalism. 

The quantum counterpart of the former equation in Eq. (\ref{3.55}) reads
%**   3.56 %%%%%%%%%%%%%%%%%%%%%%%%%%%%%%%%%%%%%%%%%%%%%
\begin{eqnarray}
[ Q, L^{[\alpha\beta\gamma\delta]} ] = \lambda^{[\alpha} 
K^{\beta\gamma\delta]}.
\label{3.56}
\end{eqnarray}  
%%%%%%%%%%%%%%%%%%%%%%%%%%%%%%%%%%%%%%%%%%%%%%%%%%%%%%%%%%%%%%%%%%%%
In this case there are no contributions from the rearrangement theorem
and using (\ref{3.37}) and (\ref{3.54}) one finds that both sides of 
Eq. (\ref{3.56}) are equal to 
$\frac{1}{768} (\Gamma_{abc})^{[\alpha\beta}(\Gamma^{ade})^{\gamma\delta]} 
(d \Gamma^{bc}\lambda) N_{de}$, thus showing that (\ref{3.56}) is true. 
It is a little more cumbersome to verify the quantum analog of the latter equation 
in Eq. (\ref{3.55}), which is given by 
%**   3.57 %%%%%%%%%%%%%%%%%%%%%%%%%%%%%%%%%%%%%%%%%%%%%%%%%%%%%%%%%%%%%%%%
\begin{eqnarray}
\lambda^{[\epsilon} L^{\alpha\beta\gamma\delta]} = 0.
\label{3.57}
\end{eqnarray}
%%%%%%%%%%%%%%%%%%%%%%%%%%%%%%%%%%%%%%%%%%%%%%%%%%%%%%%%%%%%%%%%%%%
To do that it is convenient to introduce the following notation that extends
that in Eq. (\ref{3.51}):
if $\Psi_\alpha$ and $\Phi^\beta$ are two spinors that (by the conventions 
which we adopt) belong to the $\bar{16}$ and the $16$ of $SO(10)$, 
respectively, we define 
%**  3.58 %%%%%%%%%%%%%%%%%%%%%%%%%%%%%%%%%%%%%%%%%%%%%%%%%%%%%%%%
\begin{eqnarray}
\Psi{\Lambda_{c}}^{[\alpha\beta]}\Phi = (\Psi\Gamma_c)^{[\alpha} \Phi^{
\beta]} -
\frac{1}{4} (\Psi \Gamma_b)^{[\alpha} (\Gamma^b \Gamma_c \Phi)^{\beta]}. 
\label{3.58}
\end{eqnarray}
%%%%%%%%%%%%%%%%%%%%%%%%%%%%%%%%%%%%%%%%%%%%%%%%%%%%%%%%%%%%%
Then, from Eqs. (\ref{2.17}) and (\ref{3.54}), $L^{[\alpha\beta\gamma
\delta]}$
can be rewritten as 
%**   3.59 %%%%%%%%%%%%%%%%%%%%%%%%%%%%%%%%%%%%%%%%%%%%%%%%%%%%%%%%%
\begin{eqnarray}
L^{[\alpha\beta\gamma\delta]} = - \frac{1}{48} N^{c[\alpha\beta}
{N_{c}}^{\gamma\delta]},
\label{3.59}
\end{eqnarray}
%%%%%%%%%%%%%%%%%%%%%%%%%%%%%%%%%%%%%%%%%%%%%%%%%%%%%%%%%%%%%%%%%%%%%
where
%**  3.60  %%%%%%%%%%%%%%%%%%%%%%%%%%%%%%%%%%%%%%%%%%%%%%%%%%%%%%%%%
\begin{eqnarray}
N_{c}^{[\alpha\beta]} \equiv \frac{1}{8}\Gamma_{abc}^{\alpha\beta}N^{ab}
= \Omega\Lambda_{c}^{[\alpha\beta]}\lambda  - 2 Y\Lambda_{c}^{[\alpha
\beta]}\partial\lambda,
\label{3.60}
\end{eqnarray} 
%%%%%%%%%%%%%%%%%%%%%%%%%%%%%%%%%%%%%%%%%%%%%%%%%%%%%%%%%%%%%%%%%%%%%
and $\Omega$ is defined in (\ref{2.28}). 

Using Eqs. (\ref{3.59}) and (\ref{3.60}), the l.h.s. of Eq. (\ref{3.57}) 
splits in three parts: 
%**   3.61 %%%%%%%%%%%%%%%%%%%%%%%%%%%%%%%%%%%%%%%%%%%%%%%%%%%%%%%%%%%
\begin{eqnarray}
 \lambda^{[\epsilon} L^{\alpha\beta\gamma\delta]}= -\frac{1}{48} [ \lambda
^{[\epsilon} L_{1}^{\alpha\beta\gamma\delta]} +\lambda^{[\epsilon} 
L_{2}^{\alpha\beta\gamma\delta]}+ \lambda^{[\epsilon} L_{3}^
{\alpha\beta\gamma\delta]}],
\label{3.61}
\end{eqnarray}
%%%%%%%%%%%%%%%%%%%%%%%%%%%%%%%%%%%%%%%%%%%%%%%%%%%%%%%%%%%%%%%%%%%%%
where we have defined
%**   3.62 %%%%%%%%%%%%%%%%%%%%%%%%%%%%%%%%%%%%%%%%%%%%%%%%%%%%%%%%%%%%
\begin{eqnarray}
\lambda^{[\epsilon} L_{1}^{\alpha\beta\gamma\delta]}= 
\lambda^{[\epsilon}(\Omega\Lambda_{c}^{\alpha\beta}\lambda)(\Omega
\Lambda^{\gamma\delta]c}\lambda), 
\label{3.62}
\end{eqnarray}
%%%%%%%%%%%%%%%%%%%%%%%%%%%%%%%%%%%%%%%%%%%%%%%%%%%%%%%%%%%%%%%%%%%%%%
%**   3.63 %%%%%%%%%%%%%%%%%%%%%%%%%%%%%%%%%%%%%%%%%%%%%%%%%%%%%%%%%%%
\begin{eqnarray}
 \lambda^{[\epsilon} L_{2}^{\alpha\beta\gamma\delta]}= -2 \Big[
\lambda^{[\epsilon}(\Omega\Lambda_{c}^{\alpha\beta}\lambda)(Y
\Lambda^{\gamma\delta]c}\partial\lambda)+\lambda^{[\epsilon}
(Y\Lambda_{c}^{\alpha\beta}\partial\lambda)(\Omega
\Lambda^{\gamma\delta]c}\lambda) \Big],
\label{3.63}
\end{eqnarray}
%%%%%%%%%%%%%%%%%%%%%%%%%%%%%%%%%%%%%%%%%%%%%%%%%%%%%%%%%%%%%%%%%%%%%%
%**   3.64 %%%%%%%%%%%%%%%%%%%%%%%%%%%%%%%%%%%%%%%%%%%%%%%%%%%%%%%%%%%
\begin{eqnarray}
\lambda^{[\epsilon} L_{3}^{\alpha\beta\gamma\delta]}= 
4\lambda^{[\epsilon}(Y\Lambda_{c}^{\alpha\beta}\partial\lambda)
(Y\Lambda^{\gamma\delta]c}\partial\lambda). 
\label{3.64}
\end{eqnarray}
%%%%%%%%%%%%%%%%%%%%%%%%%%%%%%%%%%%%%%%%%%%%%%%%%%%%%%%%%%%%%%%%%%%%%%
To compute the l.h.s. of Eq. (\ref{3.57}), one must shift the
fields $\Omega$ to the left using the rearrangement formula. Then
%**   3.65 %%%%%%%%%%%%%%%%%%%%%%%%%%%%%%%%%%%%%%%%%%%%%%%%%%%%%%%%%%%
\begin{eqnarray}
\lambda^{[\epsilon} L_{1}^{\alpha\beta\gamma\delta]}= \Omega_\sigma(\Omega_
\tau (\lambda^{[\epsilon}(\Lambda_{c}^{\alpha\beta}\lambda)^{\sigma}
(\Lambda^{\gamma\delta]c}\lambda)^{\tau})) + \Omega_\sigma A_{1}^{[
\epsilon\alpha\beta\gamma\delta]\sigma} + A_{0}^{[
\epsilon\alpha\beta\gamma\delta]}, 
\label{3.65}
\end{eqnarray}
%%%%%%%%%%%%%%%%%%%%%%%%%%%%%%%%%%%%%%%%%%%%%%%%%%%%%%%%%%%%%%%%%%%%%%%
%**  3.66  %%%%%%%%%%%%%%%%%%%%%%%%%%%%%%%%%%%%%%%%%%%%%%%%%%%%%%%%%%%%
\begin{eqnarray}
\lambda^{[\epsilon} L_{2}^{\alpha\beta\gamma\delta]} = 
\Omega_\sigma B_{1}^{[\epsilon\alpha\beta\gamma\delta]\sigma} 
+ B_{0}^{[\epsilon\alpha\beta\gamma\delta]}, 
\label{3.66}
\end{eqnarray}
%%%%%%%%%%%%%%%%%%%%%%%%%%%%%%%%%%%%%%%%%%%%%%%%%%%%%%%%%%%%%%%%%%%%%%%
where $A_{1}$, $A_{0}$, $B_{1}$ and $B_{0}$ are $\Omega$-independent,
$Y$-dependent fields.

The term quadratic in $\Omega$ in the r.h.s. of Eq. (\ref{3.65}) 
vanishes since it contains the factor $ \lambda^{[\epsilon}
\lambda^{\beta}(\Gamma_{bc}\lambda)^{\delta]}$. An explicit calculation shows 
that the terms linear in $\Omega$ in (\ref{3.65}) and
(\ref{3.66}) cancel each other:
%**  3.67 %%%%%%%%%%%%%%%%%%%%%%%%%%%%%%%%%%%%%%%%%%%%%%%%%%%%%%%%%%%%%
\begin{eqnarray}
\Omega_\sigma A_{1}^{[\epsilon\alpha\beta\gamma\delta]\sigma} +
\Omega_\sigma B_{1}^{[\epsilon\alpha\beta\gamma\delta]\sigma} = 0,
\label{3.67}
\end{eqnarray}
%%%%%%%%%%%%%%%%%%%%%%%%%%%%%%%%%%%%%%%%%%%%%%%%%%%%%%%%%%%%%%%%%%%%%%%
and that the sum of the terms of zero-order in $\Omega$ in (\ref{3.65}),
(\ref{3.66}) and (\ref{3.64}) vanishes
%**  3.68 %%%%%%%%%%%%%%%%%%%%%%%%%%%%%%%%%%%%%%%%%%%%%%%%%%%%%%%%%%%%%
\begin{eqnarray}
A_{0}^{[\epsilon\alpha\beta\gamma\delta]} +
B_{0}^{[\epsilon\alpha\beta\gamma\delta]} + \lambda^{[\epsilon} L_{3}^
{\alpha\beta\gamma\delta]} = 0,
\label{3.68}
\end{eqnarray}
%%%%%%%%%%%%%%%%%%%%%%%%%%%%%%%%%%%%%%%%%%%%%%%%%%%%%%%%%%%%%%%%%%%%%%% 
so that (\ref{3.57}) is proved.

The details of this calculation are given in Appendix C.

%%%%%%%%%%%%%%%%%%%%%%%%%%%%%%   SEC  4    %%%%%%%%%%%%%%%%%%%%%%%%%%%
%%%%%%%%%%%%%%%%%%%%%%%%%%%%%%%%%%%%%%%%%%%%%%%%%%%%%%%%%%%%
\section{Y-formalism for the non-minimal pure spinor formalism}

In this section, we would like to construct the Y-formalism for the 
${\it non-minimal}$ pure spinor formalism which has been recently 
proposed by Berkovits \cite{Ber12}.
Before doing that, we will first review the non-minimal pure spinor 
formalism briefly.
The main idea is to add to the fields involved 
in the ${\it minimal}$ formalism a BRST quartet of fields 
$\bar \lambda_\alpha, 
\bar \omega^\alpha, r_\alpha$ and $s^\alpha$ in such a way that their BRST 
variations 
are $\delta \bar \lambda_\alpha = r_\alpha$, $\delta r_\alpha = 0$, 
$\delta s^\alpha = \bar \omega^\alpha$ and $\delta \bar \omega^\alpha = 0$.  
Here, $\bar \lambda_\alpha$ 
is a bosonic field, $r_\alpha$ is a fermionic field, and 
$\bar \omega^\alpha$ and $s^\alpha$ are the conjugate momenta 
of $\bar \lambda_\alpha$ and $r_\alpha$, respectively.  These fields are 
required to satisfy the pure spinor conditions
%**   4.1  %%%%%%%%%%%%%%%%%%%%%%%%%%%%%%%%%%%%%%%%%%%%%%%%%%%%%%%%%%%%%%
\begin{eqnarray}
\bar \lambda \Gamma^a \bar \lambda &=& 0,  \nonumber\\  
\bar \lambda \Gamma^a r &=& 0.
\label{4.1}
\end{eqnarray}       
%%%%%%%%%%%%%%%%%%%%%%%%%%%%%%%%%%%%%%%%%%%%%%%%%%%%%%%%%%%%%%%%%%%%%%
The action is then obtained by adding to the conventional pure spinor 
action $I$ in 
Eq. (\ref{2.2}), $\bar I$ given by the BRST variation of the "gauge fermion"
$ F = - \int (s \bar \partial \bar \lambda)$ so that  
%**   4.2  %%%%%%%%%%%%%%%%%%%%%%%%%%%%%%%%%%%%%%%%%%%%%%%%%%%
\begin{eqnarray}
I_{nm} \equiv I + \bar I  = \int d^2z ( \frac{1}{2} \partial X^a  \bar 
\partial X_a + p_\alpha \bar \partial \theta^\alpha - \omega_\alpha \bar 
\partial \lambda^\alpha + s^\alpha \bar \partial r_\alpha - 
\bar \omega^\alpha \bar \partial \bar \lambda_\alpha).
\label{4.2}
\end{eqnarray}       
%%%%%%%%%%%%%%%%%%%%%%%%%%%%%%%%%%%%%%%%%%%%%%%%%%%%%%%%%%%%%%%%%%%%%%
In addition to the $\omega$-symmetry Eq. (\ref{2.4}), due to the conditions 
Eq. (\ref{4.1}), this action is invariant under new gauge symmetries 
involving $\bar \omega$ and $s$,
%**   4.3 %%%%%%%%%%%%%%%%%%%%%%%%%%%%%%%%%%%%%%%%%%%%%%%%%%%%%%%%
\begin{eqnarray}
\delta \bar \omega^\alpha &=& \Lambda^{(1)}_a (\Gamma^a \bar \lambda)^\alpha
- \Lambda^{(2)}_a (\Gamma^a r)^\alpha, \nonumber\\
\delta s^\alpha &=& \Lambda^{(2)}_a (\Gamma^a \bar \lambda)^\alpha,
\label{4.3}
\end{eqnarray}
%%%%%%%%%%%%%%%%%%%%%%%%%%%%%%%%%%%%%%%%%%%%%%%%%%%%%%%%%%%%%%%%%%%
where $\Lambda^{(1)}_a$ and $\Lambda^{(2)}_a$ are local gauge parameters.
Let us note that the conditions Eq. (\ref{4.1}) and these symmetries
reduce the independent components of each field in the quartet to 
eleven components.  
It is easy to show that the action $I_{nm}$ is invariant under the 
new BRST transformation
with BRST charge 
%**   4.4  %%%%%%%%%%%%%%%%%%%%%%%%%%%%%%%%%%%%%%%%%%%%%%%%%%%%
\begin{eqnarray}
Q_{nm}= \oint dz (\lambda^\alpha d_\alpha + \bar \omega^\alpha r_\alpha).
\label{4.4}
\end{eqnarray}       
%%%%%%%%%%%%%%%%%%%%%%%%%%%%%%%%%%%%%%%%%%%%%%%%%%%%%%%%%%%%%%%%
Of course the quartet does not contribute to the central charge and has 
trivial cohomology with
respect to the (new) BRST charge.

 As a final remark, it is worthwhile to 
recall that this new formalism can be interpreted \cite{Ber12} as a critical 
topological string with $\hat c = 3$ and (twisted)
$N = 2$ supersymmetry. Then it is possible to apply topological methods to 
the computation of multiloop amplitudes where a suitable 
regularization factor replaces picture-changing operators to soak up 
zero modes. The covariant $b$ field and the regulator proposed 
in \cite{Ber12} allow to compute loop amplitudes up to  $g = 2$. 
A more 
powerful regularization of $b$ that allows to compute loop amplitudes at any 
$g$ loop has been presented in \cite{BerNek}. This regularization is gauge 
invariant but Lorentz non-covariant since it involves the $Y$-field. However,
this non-covariance is harmless since the regularized $b$ field differs from
the covariant one by BRST-exact terms.     
 
Now we are ready to present the Y-formalism for the non-minimal pure spinor 
quantization. As in Eqs. (\ref{2.10}) and (\ref{2.12}), 
we first introduce
the non-covariant object 
%**   4.5 %%%%%%%%%%%%%%%%%%%%%%%%%%%%%%%%%%%%%%%%%%%%%%%%%%%%%%%%%
\begin{eqnarray}
\bar Y^\alpha = \frac{\bar v^\alpha}{\bar v \bar \lambda},
\label{4.5}
\end{eqnarray}
%%%%%%%%%%%%%%%%%%%%%%%%%%%%%%%%%%%%%%%%%%%%%%%%%%%%%%%%%%%%%%%%%%%
and the projector
%**   4.6 %%%%%%%%%%%%%%%%%%%%%%%%%%%%%%%%%%%%%%%%%%%%%%%%%%%%%%%%%
\begin{eqnarray}
\bar K^\alpha \ _\beta= \frac{1}{2}(\Gamma^a \bar \lambda)^\alpha
(\bar Y \Gamma_a)_\beta,
\label{4.6}
\end{eqnarray}
%%%%%%%%%%%%%%%%%%%%%%%%%%%%%%%%%%%%%%%%%%%%%%%%%%%%%%%%%%%%%%%%%%%
where $\bar v^\alpha$ is a constant pure spinor so that we have 
%**   4.7 %%%%%%%%%%%%%%%%%%%%%%%%%%%%%%%%%%%%%%%%%%%%%%%%%%%%%%
\begin{eqnarray}
\bar Y \Gamma^a \bar Y = 0.
\label{4.7}
\end{eqnarray}
%%%%%%%%%%%%%%%%%%%%%%%%%%%%%%%%%%%%%%%%%%%%%%%%%%%%%%%%%%%%%%%%%%%
Let us note that the conditions (\ref{4.1}) lead to relations
$\bar \lambda_\alpha \bar K^\alpha \ _\beta = r_\alpha \bar K^\alpha \ _\beta
= 0$, which imply that $\bar \lambda_\alpha$ and $r_\alpha$ have
respectively eleven independent components. 

Next we postulate the following OPE's among $\bar \omega^\alpha$, 
$\bar \lambda_\alpha$, $s^\alpha$ and $r_\alpha$:
%**   4.8  %%%%%%%%%%%%%%%%%%%%%%%%%%%%%%%%%%%%%%%%%%%%%%%%%%%%%%%%%
\begin{eqnarray}
<\bar \omega^\alpha (y) \bar \lambda_\beta (z)> = \frac{1}{y-z} 
(\delta_\beta^\alpha - \bar K^\alpha \ _\beta (z)),
\label{4.8}
\end{eqnarray}
%%%%%%%%%%%%%%%%%%%%%%%%%%%%%%%%%%%%%%%%%%%%%%%%%%%%%%%%%%%%%%%%%%%%
%**  4.9   %%%%%%%%%%%%%%%%%%%%%%%%%%%%%%%%%%%%%%%%%%%%%%%%%%%%%%%%
\begin{eqnarray}
<s^\alpha (y) r_\beta (z)> = \frac{1}{y-z} 
(\delta_\beta^\alpha - \bar K^\alpha \ _\beta (z)), 
\label{4.9}
\end{eqnarray}
%%%%%%%%%%%%%%%%%%%%%%%%%%%%%%%%%%%%%%%%%%%%%%%%%%%%%%%%%%%%%%%%%%%%
%**  4.10 %%%%%%%%%%%%%%%%%%%%%%%%%%%%%%%%%%%%%%%%%%%%%%%%%%%%%%%%%%
\begin{eqnarray}
<\bar \omega^\alpha (y) r_\beta (z)> = \frac{1}{y-z} 
[\bar K^\alpha \ _\beta(z) (\bar Y r)(z) - \frac{1}{2} (\Gamma^a r)^\alpha(z)
(\bar Y \Gamma_a)_\beta(z)], 
\label{4.10}
\end{eqnarray}
%%%%%%%%%%%%%%%%%%%%%%%%%%%%%%%%%%%%%%%%%%%%%%%%%%%%%%%%%%%%%%%%%%%%%
%**   4.11 %%%%%%%%%%%%%%%%%%%%%%%%%%%%%%%%%%%%%%%%%%%%%%%%%%%%%%%%%%
\begin{eqnarray}
<s^\alpha (y) \bar \lambda_\beta (z)> = 0.
\label{4.11}
\end{eqnarray}
%%%%%%%%%%%%%%%%%%%%%%%%%%%%%%%%%%%%%%%%%%%%%%%%%%%%%%%%%%%%%%%%%%%
Then, with these OPE's it is easy to check that the OPE's between 
the conjugate
momenta $\bar \omega^\alpha$ and $s^\alpha$, and the conditions (\ref{4.1})
vanish identically:
%**   4.12 %%%%%%%%%%%%%%%%%%%%%%%%%%%%%%%%%%%%%%%%%%%%%%%%%%%%%%%%%
\begin{eqnarray}
<\bar \omega^\alpha (y) (\bar \lambda \Gamma^a \bar \lambda) (z)> &=& 0,
\nonumber\\
<\bar \omega^\alpha (y) (\bar \lambda \Gamma^a r) (z)> &=& 0,
\nonumber\\
<s^\alpha (y) (\bar \lambda \Gamma^a \bar \lambda) (z)> &=& 0,
\nonumber\\
<s^\alpha (y) (\bar \lambda \Gamma^a r) (z)> &=& 0.
\label{4.12}
\end{eqnarray}
%%%%%%%%%%%%%%%%%%%%%%%%%%%%%%%%%%%%%%%%%%%%%%%%%%%%%%%%%%%%%%%%%%
Notice that (\ref{4.10}) follows for consistency by acting with the BRST 
charge $Q_{nm}$ on (\ref{4.8}) (or (\ref{4.9})). 

{}Following \cite{Ber12}, the only holomorphic currents involving 
$\bar\omega$ and $s$ and gauge invariant under (\ref{4.3}) are:
\begin{itemize}
\item{i)} the bosonic currents 
%**   4.13 %%%%%%%%%%%%%%%%%%%%%%%%%%%%%%%%%%%%%%%%%%%%%%%%%%%%%%
\begin{eqnarray}
\bar N_{ab} &=& \frac{1}{2} (\bar \omega \Gamma_{ab} \bar \lambda
- s \Gamma_{ab} r), \nonumber\\
\bar J_{\bar \lambda} &=& \bar \omega \bar \lambda - s r, \nonumber\\
T_{\bar \lambda} &=& \bar \omega \partial \bar \lambda - s \partial r,  
\label{4.13}
\end{eqnarray}
%%%%%%%%%%%%%%%%%%%%%%%%%%%%%%%%%%%%%%%%%%%%%%%%%%%%%%%%%%%%%%%%%%
those are, the Lorentz current, the ghost current and the stress energy tensor 
of the non-minimal fields, respectively.
\item{ii)} the fermionic currents
%**  4.14  %%%%%%%%%%%%%%%%%%%%%%%%%%%%%%%%%%%%%%%%%%%%%%%%%%%%%%%%
\begin{eqnarray}
S_{ab} &=& \frac{1}{2} s \Gamma_{ab} \bar \lambda, \nonumber\\
S &=& s \bar \lambda, \nonumber\\
S_{(b)} &=& s \partial \bar \lambda. 
\label{4.14}
\end{eqnarray}
%%%%%%%%%%%%%%%%%%%%%%%%%%%%%%%%%%%%%%%%%%%%%%%%%%%%%%%%%%%%%%%%%%%
\item{iii)} the doublet 
%**  4.15  %%%%%%%%%%%%%%%%%%%%%%%%%%%%%%%%%%%%%%%%%%%%%%%%%%%%%%%%
\begin{eqnarray}
J_0 &=& rs, \nonumber\\
\Phi_0 &=& \bar \omega r. 
\label{4.15}
\end{eqnarray}
%%%%%%%%%%%%%%%%%%%%%%%%%%%%%%%%%%%%%%%%%%%%%%%%%%%%%%%%%%%%%%%%%%%
\end{itemize}
Using the fundamental OPE's (\ref{4.8})-(\ref{4.11}), one can compute  the 
OPE's among these operators. The OPE's of $\bar N^{ab}$, $T_{\bar \lambda}$ and
$\bar J_{\bar \lambda}$ with $\bar \lambda$ and $r$ and the ones 
among themselves are canonical, namely
%**   4.16 %%%%%%%%%%%%%%%%%%%%%%%%%%%%%%%%%%%%%%%%%%%%%%%%%%%%%%%%
\begin{eqnarray}
<\bar N_{ab} (y) \bar \lambda_\alpha (z)> 
&=& \frac{1}{2} \frac{1}{y-z} (\Gamma_{ab} \bar \lambda)_\alpha (z), \
<\bar N_{ab} (y) r_\alpha (z)> = \frac{1}{2}
\frac{1}{y-z} (\Gamma_{ab} r)_\alpha (z), \nonumber\\
<\bar J_{\bar \lambda} (y) \bar \lambda_\alpha (z)> 
&=& \frac{1}{y-z} \bar \lambda_\alpha (z), \
<\bar J_{\bar \lambda} (y) r_\alpha (z)> = 
\frac{1}{y-z} r_\alpha (z), \nonumber\\
<T_{\bar \lambda} (y) \bar \lambda_\alpha (z)> 
&=& \frac{1}{y-z} \partial \bar \lambda_\alpha (z), \
<T_{\bar \lambda} (y) r_\alpha (z)> = 
\frac{1}{y-z} \partial r_\alpha (z), 
\label{4.16}
\end{eqnarray}
%%%%%%%%%%%%%%%%%%%%%%%%%%%%%%%%%%%%%%%%%%%%%%%%%%%%%%%%%%%%%%%%%%%%%
and
%**   4.17 %%%%%%%%%%%%%%%%%%%%%%%%%%%%%%%%%%%%%%%%%%%%%%%%%%%%%%%%%%%
\begin{eqnarray}
<\bar N_{ab} (y) \bar N_{cd} (z)> 
&=& - \frac{1}{y-z} (\eta_{c[b} \bar N_{a]d} - \eta_{d[b} \bar N_{a]c})(z),  \nonumber\\
<\bar N_{ab} (y) \bar J_{\bar \lambda} (z)> &=& 0, \
<\bar N_{ab} (y) T_{\bar \lambda} (z)> = \frac{1}{(y-z)^2} \bar N_{ab} (z), 
\nonumber\\
<\bar J_{\bar \lambda} (y) \bar J_{\bar \lambda} (z)> &=& 0, \
<\bar J_{\bar \lambda} (y) T_{\bar \lambda} (z)> = 
\frac{1}{(y-z)^2} \bar J_{\bar \lambda}(z), \nonumber\\
<T_{\bar \lambda} (y) T_{\bar \lambda} (z)> &=& 
\frac{2}{(y-z)^2} T_{\bar \lambda} (z) + \frac{1}{y-z} \partial 
T_{\bar \lambda} (z).
\label{4.17}
\end{eqnarray}
%%%%%%%%%%%%%%%%%%%%%%%%%%%%%%%%%%%%%%%%%%%%%%%%%%%%%%%%%%%%%%%%%%%%%%%%

Notice that in contrast with the operators $T$, $N^{ab}$ and $J$ in (\ref{2.16})-(\ref{2.18}), 
the operators  $\bar N_{ab}$, $T_{\bar \lambda}$ and
$\bar J_{\bar \lambda}$  do not involve $\bar Y$-correction terms
since the $\bar Y$-dependent terms which arise in their OPE's  are absent or 
cancel in the combinations (\ref{4.13}).
It is instructive to see explicitly how this cancellation arises. Let us write 
$\bar N_{ab}^{(\bar \omega \bar \lambda)} = \frac{1}{2} \bar \omega \Gamma_{ab} 
\bar \lambda$ and $\bar N_{ab}^{(sr)} = \frac{1}{2} s \Gamma_{ab} r$ 
and consider for instance the OPE between $\bar N_{ab} = 
\bar N_{ab}^{(\bar\omega\bar\lambda)} - \bar N_{ab}^{(sr)}$ and $r_\alpha$. {}From Eq. (\ref{4.9}), one obtains 
%**   4.18 %%%%%%%%%%%%%%%%%%%%%%%%%%%%%%%%%%%%%%%%%%%%%%%%%%%%%%%%%%%%%%
\begin{eqnarray}
< \bar N_{ab}^{(sr)}(y) r_{\alpha}(z)> = \frac{1}{2} \frac{1}{y-z} 
(\Gamma_{ab} r)_\alpha 
+ \frac{1}{4} \frac{1}{y-z} (\Gamma_f \bar Y)_\alpha (\bar \lambda \Gamma^f \Gamma_{ab} r).
\label{4.18}
\end{eqnarray}
%%%%%%%%%%%%%%%%%%%%%%%%%%%%%%%%%%%%%%%%%%%%%%%%%%%%%%%%%%%%%%%%%%%%%%%%%
Then, the second term in the r.h.s. of (\ref{4.18}) is exactly cancelled by the 
contribution of the OPE $< \bar N_{ab}^{(\bar \omega\bar\lambda)}(y) 
r_{\alpha}(z)>$ in terms of Eq. (\ref{4.10}). As a second example, consider 
the OPE $<\bar N_{ab} (y) \bar N_{cd} (z)>$. The double poles coming from
$<\bar N_{ab}^{(\bar \omega \bar\lambda)}(y) \bar N_{ab}^{(\bar \omega\bar
\lambda)}(z)>$ are cancelled by those coming from $<\bar N_{ab}^{(sr)}(y)
\bar N_{ab}^{(sr)}(z)>$. As for the simple poles, one has
%**   4.19 %%%%%%%%%%%%%%%%%%%%%%%%%%%%%%%%%%%%%%%%%%%%%%%%%%%%%%%%%%%%%%%
\begin{eqnarray}
&{}& <\bar N_{ab}^{(\bar \omega \bar \lambda)}(y) 
\bar N_{cd}^{(\bar \omega \bar \lambda)}(z)>     
+ <\bar N_{ab}^{(sr)}(y)\bar N_{cd}^{(sr)}(z)>   \nonumber\\
&=& - \frac{1}{y-z} (\eta_{c[b} \bar N_{a]d} - \eta_{d[b} \bar N_{a]c}) 
+ \frac{1}{8} [ (s \Gamma_{ab} \Gamma_f \bar Y) 
(r \Gamma^f \Gamma_{cd} \bar \lambda) 
+ (s \Gamma_{cd} \Gamma_f \bar Y)(r \Gamma^f \Gamma_{ab} \bar \lambda)],
\label{4.19}
\end{eqnarray}
%%%%%%%%%%%%%%%%%%%%%%%%%%%%%%%%%%%%%%%%%%%%%%%%%%%%%%%%%%%%%%%%%%%%%%%%%%%
but the terms, which are independent of $\bar N_{ab}$ in the r.h.s. of (\ref{4.19}), 
are just cancelled by the contributions stemming from 
$ - ( < \bar N_{ab}^{( \bar \omega \bar\lambda)}(y)
\bar N_{cd}^{(sr)}(z)> + <\bar N_{ab}^{(sr)}(y)\bar N_{cd}^{(\bar\omega
\bar\lambda)}(z)>) $.  For all the remaining OPE's in both (\ref{4.16}) and (\ref{4.17}), 
the spurious, $\bar Y$-dependent terms are absent or cancelled in a similar way.
Moreover, the OPE's among $S^{ab}$, $S$ and $S_{(b)}$ are regular and those of  
$\bar N_{ab}$, $\bar J_{\bar \lambda}$ and $T_{\bar \lambda}$ with $S^{ab}$, $S$
and $S_{(b)}$ are canonical so that $S^{ab}$, $S$ and $S_{(b)}$ are covariant primary 
fields with weight $1$ and ghost number $2$ with respect to the ghost current 
$\bar J_{\bar \lambda}$.
Thus, as for $\bar N_{ab}$, $\bar J_{\bar \lambda}$ and  $T_{\bar \lambda}$, 
they do not have to include $\bar Y$-dependent corrections.

The story is completely different for the currents $J_r$ and $\Phi$.
Indeed, using the OPE's (\ref{4.8})-(\ref{4.11}), one finds 
%**   4.20 %%%%%%%%%%%%%%%%%%%%%%%%%%%%%%%%%%%%%%%%%%%%%%%%%%%%%%%%%%%%%%%%%
\begin{eqnarray}
<(rs)(y) \bar N^{ab}(z)> = \frac{3}{2} \frac{1}{(y-z)^2}
\bar Y \Gamma^{ab} \bar \lambda. 
\label{4.20}
\end{eqnarray}
%%%%%%%%%%%%%%%%%%%%%%%%%%%%%%%%%%%%%%%%%%%%%%%%%%%%%%%%%%%%%%%%%%%%%%%%%%%%%
And since one has
%**   4.21 %%%%%%%%%%%%%%%%%%%%%%%%%%%%%%%%%%%%%%%%%%%%%%%%%%%%%%%%%%%%%%%%%%
\begin{eqnarray}
<(\bar Y \partial \bar \lambda )(y) \bar N^{ab}(z)> = 
\frac{1}{2} \frac{1}{(y-z)^2} \bar Y \Gamma^{ab}\bar \lambda, 
\label{4.21}
\end{eqnarray}
%%%%%%%%%%%%%%%%%%%%%%%%%%%%%%%%%%%%%%%%%%%%%%%%%%%%%%%%%%%%%%%%%%%%%%%%%%%%%
the $\bar Y$-dependent term in $<J_{r}(y) \bar N^{ab}(z)>$ disappears if one 
assumes, as definition of $J_r$ at quantum level, 
%**   4.22 %%%%%%%%%%%%%%%%%%%%%%%%%%%%%%%%%%%%%%%%%%%%%%%%%%%%%%%%%%%%%%%%%%%
\begin{eqnarray}
J_r = rs - 3 \bar Y \partial \bar \lambda.
\label{4.22}
\end{eqnarray}
%%%%%%%%%%%%%%%%%%%%%%%%%%%%%%%%%%%%%%%%%%%%%%%%%%%%%%%%%%%%%%%%%%%%%%%%%%%%%%
With this definition, the OPE's of $J_r$ with $\bar N_{ab}$, $\bar J_{\bar 
\lambda}$,  $T_{\bar \lambda}$,  $S^{ab}$, $S$ and $S_{(b)}$ read 
%**   4.23 %%%%%%%%%%%%%%%%%%%%%%%%%%%%%%%%%%%%%%%%%%%%%%%%%%%%%%%%%%%%%%%%%%%
\begin{eqnarray} 
<J_r (y) J_r (z)> &=& \frac{11}{(y-z)^2}, \nonumber\\
<J_r (y) \bar N^{ab}(z)> &=& 0, \nonumber\\
<\bar J_{\bar \lambda} (y) J_r (z)> &=& \frac{8}{(y-z)^2}, \nonumber\\
<J_r (y) T_{\bar \lambda} (z)> &=&
\frac{11}{(y-z)^3} + \frac{1}{(y-z)^2} J_r, \nonumber\\
<J_r (y)  S^{ab}(z)> &=& \frac{1}{y-z}S^{ab}, \nonumber\\
<J_r (y)  S(z)> &=&  \frac{1}{y-z}S, \nonumber\\ 
 <J_r (y)  S_{(b)}(z)> &=& \frac{1}{y-z}S_{(b)}.  
\label{4.23}
\end{eqnarray}
%%%%%%%%%%%%%%%%%%%%%%%%%%%%%%%%%%%%%%%%%%%%%%%%%%%%%%%%%%%%%%%%%%%%%%%%%%%%%%%
In particular, note that the coefficient $8$ of the double pole in the
contraction $<\bar J_{\bar \lambda} (y) J_r (z)>$ emerges from the arithmetic 
$8 = 11 - 3$ where $11$ comes from the first term and $-3$ from the second term in (\ref{4.22}).

In a similar manner, for $\Phi$ one has
%**   4.24 %%%%%%%%%%%%%%%%%%%%%%%%%%%%%%%%%%%%%%%%%%%%%%%%%%%%%%%%%%%%%%%%%%%%
\begin{eqnarray}
<(\bar \omega r)(y) \bar N^{ab}(z)> &=& -\frac{3}{2} \frac{1}{(y-z)^2}
[\bar Y \Gamma^{ab} r - (\bar Y r)(\bar Y \Gamma^{ab} \bar \lambda)], 
\nonumber\\
<(\bar \omega r)(y) S^{ab}(z)> &=& -\frac{3}{2} \frac{1}{(y-z)^2}
\bar Y \Gamma^{ab} \bar \lambda. 
\label{4.24}
\end{eqnarray}
%%%%%%%%%%%%%%%%%%%%%%%%%%%%%%%%%%%%%%%%%%%%%%%%%%%%%%%%%%%%%%%%%%%%%%%%%%%%%%%
Therefore, at quantum level $\Phi$ must be defined as
%**    4.25 %%%%%%%%%%%%%%%%%%%%%%%%%%%%%%%%%%%%%%%%%%%%%%%%%%%%%%%%%%%%%%%%%%%
\begin{eqnarray}
\Phi = \bar \omega r  + 3 [ \bar Y \partial r -  (\bar Y r)(\bar Y \partial 
\bar \lambda )] = \bar\omega r  + 3 \partial (\bar Y r),
\label{4.25}
\end{eqnarray}
%%%%%%%%%%%%%%%%%%%%%%%%%%%%%%%%%%%%%%%%%%%%%%%%%%%%%%%%%%%%%%%%%%%%%%%%%%%%%%%
in order to avoid spurious $\bar Y$-dependent terms.
With this new definition, one can also derive
%**   4.26 %%%%%%%%%%%%%%%%%%%%%%%%%%%%%%%%%%%%%%%%%%%%%%%%%%%%%%%%%%%%%%%%%%%%
\begin{eqnarray}
<\Phi(y) \bar N^{ab}(z)> &=& 0, \nonumber\\
<\Phi(y) \bar J_{\bar \lambda} (z)> &=& 0, \nonumber\\
<\Phi(y) T_{\bar \lambda} (z)> &=& \frac{1}{(y-z)^2} \Phi, \nonumber\\
<\Phi(y) S^{ab}(z)> &=& \frac{1}{y-z} \bar N^{ab}, \nonumber\\
<\Phi(y) S(z)> &=& \frac{8}{(y-z)^2} + \frac{1}{y-z} \bar J_{\bar \lambda},
\nonumber\\
<\Phi(y) S_{(b)}(z)> &=& \frac{11}{(y-z)^{3}} + \frac{1}{(y-z)^2}J_{r} +
\frac{1}{y-z}T_{\bar\lambda},  \nonumber\\
< \Phi(y) J_r (z) > &=& \frac{1}{y-z} \Phi.
\label{4.26}
\end{eqnarray}
%%%%%%%%%%%%%%%%%%%%%%%%%%%%%%%%%%%%%%%%%%%%%%%%%%%%%%%%%%%%%%%%%%%%%%%%%%%%%%%
The operator $\Phi$ is part of the BRST current and $S_{(b)}$ is a 
contribution of the $b$ ghost as will be seen in the next section.
 
{}From the definitions (\ref{4.13}) and (\ref{4.14}), one finds that the operators 
$\bar N_{ab}$, $\bar J_{\bar \lambda}$ and $T_{\bar \lambda}$ are the BRST variations 
of the operators $S^{ab}$, $S$, 
and $S_{(b)}$, respectively.  Moreover, contrary to what happens 
for the operators in (\ref{2.16})-(\ref{2.18}), the correction term of $ J_r$ in 
(\ref{4.22}) is not BRST-exact but its BRST variation is just the correction term 
for $ -\Phi$ in (\ref{4.25}), so that $\Phi$ is just the BRST variation of $ - J_r$. 
These properties are fully consistent with the OPE's we have computed
thus far. $\footnote {Apart from a 
difference in the OPE $<\Phi S> $ where we find a double pole with residuum 8, 
not present in \cite{Ber12} (perhaps a misprint in \cite{Ber12}), our results 
agree  with those computed in \cite{Ber12} by using the $U(5)$-formalism.}$ 

As a final remark, let us note that in all the derivations of this section
(but the second equality of (\ref{4.25})) we
have never used the fact that $\bar v$ in (\ref{4.5}) is constant and therefore
all the equations in this section remain true even if one replaces $\bar Y^\alpha$ 
with $\tilde Y^\alpha \equiv \frac{\lambda^\alpha}{\lambda \bar \lambda}$.

%%%%%%%%%%%%%%%%%%%%%%%%%%%%%%%%%%%%%%%%%%%%%%%%%%%%%%%%%%%%%%%%%%%%%
%%%%%%%%%%%%%%%%%%%%%%%%%%%%%%   SEC  5    %%%%%%%%%%%%%%%%%%%%%%%%%%
%%%%%%%%%%%%%%%%%%%%%%%%%%%%%%%%%%%%%%%%%%%%%%%%%%%%%%%%%%%%%%%%%%%%%
\section{A quantum $b$ ghost in the non-minimal pure spinor formalism}

In Ref. \cite{Ber12}, Berkovits has obtained an expression for a covariant 
$b$ ghost in the framework of ${\it non-minimal}$ formalism. 
%$\footnote{In our previous report \cite{Oda5}, the cohomological relation 
%for this $b$ ghost was clarified at the classical level.}$  
His idea was triggered by the 
observation that in this  formalism the non-covariant $Y_\alpha$ field 
can be replaced by a covariant field $\tilde \lambda_\alpha$ (which will be 
defined soon)
and then one can look for a new, covariant $b$ ghost satisfying the defining 
equation 
%**   5.1 %%%%%%%%%%%%%%%%%%%%%%%%%%%%%%%%%%%%%%%%%%%%%%%%%%%%%%%%
\begin{eqnarray}
\{ Q_{nm}, b_{nm}(z) \} = T(z) + T_{\bar \lambda}(z) \equiv T_{nm}(z),
\label{5.1}
\end{eqnarray}
%%%%%%%%%%%%%%%%%%%%%%%%%%%%%%%%%%%%%%%%%%%%%%%%%%%%%%%%%%%%%%%%%%%
by starting with $b_{nm} = \tilde\lambda_\alpha G^\alpha + s^\alpha \partial
\bar \lambda_\alpha + \cdots$. The result, given in \cite{Ber12}, is
%**  5.2  %%%%%%%%%%%%%%%%%%%%%%%%%%%%%%%%%%%%%%%%%%%%%%%%%%%%%%%%%%
\begin{eqnarray}
b_{nm} &=& s^\alpha \partial \bar \lambda_\alpha + \tilde \lambda_\alpha 
G^\alpha
-2 \tilde \lambda_\beta \tilde r_\alpha H^{[\alpha\beta]} \nonumber\\
&+& 6 \tilde \lambda_\gamma \tilde r_\beta \tilde r_\alpha K^{[\alpha\beta
\gamma]}
- 24 \tilde \lambda_\delta \tilde r_\gamma \tilde r_\beta \tilde r_\alpha 
L^{[\alpha\beta\gamma\delta]},
\label{5.2}
\end{eqnarray}       
%%%%%%%%%%%%%%%%%%%%%%%%%%%%%%%%%%%%%%%%%%%%%%%%%%%%%%%%%%%%%%%%%%%%
where we have defined
%**  5.3  %%%%%%%%%%%%%%%%%%%%%%%%%%%%%%%%%%%%%%%%%%%%%%%%%%%%%%%%%%
\begin{eqnarray}
\tilde \lambda_\alpha &=& \frac{\bar \lambda_\alpha}{(\bar \lambda \lambda)},
\nonumber\\
\tilde r_\alpha &=& \frac{r_\alpha}{(\bar \lambda \lambda)}.
\label{5.3}
\end{eqnarray}       
%%%%%%%%%%%%%%%%%%%%%%%%%%%%%%%%%%%%%%%%%%%%%%%%%%%%%%%%%%%%%%%%%%%%
Note that $\tilde \lambda_\alpha$ and $\tilde r_\alpha$ are primary fields 
of conformal weight $0$ with respect to $T_{nm}$.

In this section, we will construct a covariant, quantum-mechanical $b$ ghost
in the ${\it non-minimal}$ pure spinor formalism on the basis of our 
Y-formalism, taking care of normal-ordering effects.
Furthermore, we shall show that this covariant $b$ ghost is cohomologically 
equivalent to the non-covariant $\tilde b_Y $ ghost, improved by 
the non-minimal term $s^\alpha \partial \bar \lambda_\alpha$ which takes the
form at the classical level
%**  5.4 %%%%%%%%%%%%%%%%%%%%%%%%%%%%%%%%%%%%%%%%%%%%%%%%%%%%%%%%%%%%
\begin{eqnarray}
\tilde b_{0Y} = Y_\alpha G_{0}^\alpha + s^\alpha \partial \bar \lambda_\alpha.
\label {5.4}
\end{eqnarray}
%%%%%%%%%%%%%%%%%%%%%%%%%%%%%%%%%%%%%%%%%%%%%%%%%%%%%%%%%%%%%%%%%%%%%% 
It is now convenient to consider the following operators:
%**   5.5 %%%%%%%%%%%%%%%%%%%%%%%%%%%%%%%%%%%%%%%%%%%%%%%%%%%%%%%%%%%%%
\begin{eqnarray}
\frac{1}{2}\rho_{[\alpha\beta]} &\equiv& \frac{1}{2}( \tilde r_\alpha \tilde 
\lambda_\beta -  \tilde r_\beta \tilde \lambda_\alpha ) \equiv \tilde 
r_{[\alpha} \tilde \lambda_{\beta]},  \nonumber\\
\frac{1}{3!}\rho_{[\alpha\beta\gamma]} &\equiv& 
- \tilde r_{[\alpha} \tilde r_\beta \tilde \lambda_{\gamma]},  \nonumber\\
\frac{1}{4!}\rho_{[\alpha\beta\gamma\delta]} &\equiv& 
- \tilde r_{[\alpha} \tilde r_\beta \tilde r_{\gamma} \tilde 
\lambda_{\delta]},\nonumber\\
\frac{1}{5!}\rho_{[\alpha\beta\gamma\delta\epsilon]} &\equiv&  
\tilde r_{[\alpha} \tilde r_\beta \tilde r_\gamma \tilde r_\delta \tilde 
\lambda_{\epsilon]},
\label{5.5}
\end{eqnarray}
%%%%%%%%%%%%%%%%%%%%%%%%%%%%%%%%%%%%%%%%%%%%%%%%%%%%%%%%%%%%%%%%%%%%%%%
that satisfy the recursive relations
%**  5.6 %%%%%%%%%%%%%%%%%%%%%%%%%%%%%%%%%%%%%%%%%%%%%%%%%%%%%%%%%%%
\begin{eqnarray}
\Big[ Q_{nm}, \tilde \lambda_\alpha \Big] &=& \lambda^\beta \rho_{[\alpha 
\beta]}, \nonumber\\ 
\{ Q_{nm},\rho_{[\alpha\beta]} \} &=& \lambda^\gamma \rho_{[\alpha
\beta\gamma]},  \nonumber\\
\Big[ Q_{nm},\rho_{[\alpha\beta\gamma]} \Big] &=&  \lambda^
\delta \rho_{[\alpha\beta\gamma\delta]}, \nonumber\\
 \{ Q_{nm},\rho_{[\alpha\beta
\gamma\delta]} \} &=& \lambda^\epsilon \rho_{[\alpha\beta\gamma\delta
\epsilon]}.
\label{5.6}
\end{eqnarray}
%%%%%%%%%%%%%%%%%%%%%%%%%%%%%%%%%%%%%%%%%%%%%%%%%%%%%%%%%%%%
Next, let us also recall the results which were obtained in section 3
and hold at the quantum level:
%**   5.7 %%%%%%%%%%%%%%%%%%%%%%%%%%%%%%%%%%%%%%%%%%%%%%%%%%%%%%%%%%
\begin{eqnarray}
\{ Q,\hat G^\alpha \} &=& \lambda^\alpha T,    \nonumber\\
\Big[ Q, H^{[\alpha\beta]} \Big]  &=& \lambda^{[\alpha}\hat G^{\beta]},   
\nonumber\\
\{ Q, K^{[\alpha\beta\gamma]} \} &=& \lambda^{[\alpha} H^{\beta\gamma]},
\nonumber\\
\Big[ Q, L^{[\alpha\beta\gamma\delta]} \Big] &=& \lambda^{[\alpha} K^{\beta
\gamma\delta]}, \nonumber\\
\lambda^{[\alpha} L^{\beta\gamma\delta\rho]} &=& 0,
\label{5.7}
\end{eqnarray}
%%%%%%%%%%%%%%%%%%%%%%%%%%%%%%%%%%%%%%%%%%%%%%%%%%%%%%%%%%%%%%%%%%%%
where $\hat G^\alpha $ is defined in (\ref{3.13}).

It is also useful to compute the contractions:
%**   5.8 %%%%%%%%%%%%%%%%%%%%%%%%%%%%%%%%%%%%%%%%%%%%%%%%%%%%%%%%%%
\begin{eqnarray}
<\hat G^\beta(y) \rho_{[\beta\alpha]}(z)> &=& \frac{R_{1\alpha}}{y-z},
\nonumber\\ 
<H^{[\beta\gamma]}(y) \rho_{[\gamma\beta\alpha]}(z)> &=& \frac{R_{2\alpha}}
{y-z}, 
\nonumber\\ 
<K^{[\beta\gamma\delta]}(y) \rho_{[\delta\gamma\beta\alpha]}(z)> &=& 
\frac{R_{3\alpha}}{y-z}, 
\nonumber\\ 
<L^{[\beta\gamma\delta\epsilon]}(y) \rho_{[\epsilon\delta\gamma\beta
\alpha]}(z)> 
&=& \frac{\tilde R_{4\alpha}}{(y-z)^2} +  \frac{R_{4\alpha}}{y-z}.
\label{5.8}
\end{eqnarray}
%%%%%%%%%%%%%%%%%%%%%%%%%%%%%%%%%%%%%%%%%%%%%%%%%%%%%%%%%%%%%%%%%%%%
After a simple calculation, it turns out that $R_{1\alpha}$ is given by 
%**   5.9 %%%%%%%%%%%%%%%%%%%%%%%%%%%%%%%%%%%%%%%%%%%%%%%%%%%%%%%%%
\begin{eqnarray}
R_{1\alpha} = - 2 \rho_{[\alpha\beta]}[ \lambda^\beta(\tilde\lambda 
\partial\theta) - \frac{1}{2} (\Gamma^a \tilde \lambda)^\beta (\lambda \Gamma_a \partial \theta)],
\label{5.9}
\end{eqnarray}
%%%%%%%%%%%%%%%%%%%%%%%%%%%%%%%%%%%%%%%%%%%%%%%%%%%%%%%%%%%%%%%%%%%%%
but the second term in the square bracket vanishes when contracted with 
$\rho_{[\alpha\beta]}$ due to the conditions (\ref{4.1}).
As for $R_{2\alpha}$, $ R_{3\alpha}$, $R_{4\alpha}$ and $\tilde R_{4\alpha}$, 
they all contain (at least) a factor $\frac{1}{16}\Gamma_{abc}^{\alpha\beta}
(\tilde \lambda \Gamma^{ab}\lambda) \equiv \tilde \lambda \Lambda_c^{[\alpha\beta]} \lambda$ 
and therefore vanish when contracted with $\rho_{[\beta\gamma \cdots]}$ 
by taking into account (\ref{5.5}), (\ref{3.58}) and (\ref{4.1}). 
$\footnote{In $\tilde R_{4\alpha}$, there is also a term proportional to 
$\rho_{[\alpha\beta\gamma\delta]} (\Gamma_{abc})^{\alpha\beta} 
(\Gamma^{dec})^{\gamma\delta} (\tilde \lambda \Gamma^{ab} \Gamma_{de} \lambda)$ 
that vanishes for the same reason.}$  To summarize, we have the following results:
%**   5.10 %%%%%%%%%%%%%%%%%%%%%%%%%%%%%%%%%%%%%%%%%%%%%%%%%%%%%%%%
\begin{eqnarray}   
R_{1\alpha} &=& -2\rho_{[\alpha\beta]}\lambda^\beta(\tilde \lambda 
\partial\theta), \nonumber\\ 
R_{2\alpha} &=& R_{3\alpha} = R_{4\alpha} = \tilde R_{4\alpha} = 0. 
\label{5.10}
\end{eqnarray}
%%%%%%%%%%%%%%%%%%%%%%%%%%%%%%%%%%%%%%%%%%%%%%%%%%%%%%%%%%%%%%% 

As already noted, the non-minimal $b$ field is expected to be of the form:
%**   5.11 %%%%%%%%%%%%%%%%%%%%%%%%%%%%%%%%%%%%%%%%%%%%%%%%%%%%%%%%
\begin{eqnarray}
b_{nm} = S_{(b)} + \tilde \lambda G + \cdots.
\label{5.11}
\end{eqnarray}
%%%%%%%%%%%%%%%%%%%%%%%%%%%%%%%%%%%%%%%%%%%%%%%%%%%%%%%%%%%%%%%%%%%
The anticommutator of $Q_{nm}$ with $S_{(b)} = s \partial \bar \lambda$ is
%**   5.12 %%%%%%%%%%%%%%%%%%%%%%%%%%%%%%%%%%%%%%%%%%%%%%%%%%%%%%%%
\begin{eqnarray}
\{ Q_{nm},S_{(b)} \} = T_{\bar\lambda}.
\label{5.12}
\end{eqnarray}
%%%%%%%%%%%%%%%%%%%%%%%%%%%%%%%%%%%%%%%%%%%%%%%%%%%%%%%%%%%%%%%%%%%
Now let us compute the anticommutator 
$\{ Q_{nm}, (\tilde \lambda_\alpha \hat G^\alpha) \}$
%**   5.13  %%%%%%%%%%%%%%%%%%%%%%%%%%%%%%%%%%%%%%%%%%%%%%%%%%%%%%%%
\begin{eqnarray}
 \{ Q_{nm}, (\tilde \lambda_\alpha \hat G^\alpha) \} 
 = \tilde \lambda_\alpha (\lambda^\alpha T)  
 + (\lambda^\beta \rho_{[\alpha\beta]}) \hat G^\alpha.
\label{5.13}
\end{eqnarray}
%%%%%%%%%%%%%%%%%%%%%%%%%%%%%%%%%%%%%%%%%%%%%%%%%%%%%%%%%%%%%%%%%%%%%
Using the rearrangement theorem and some algebra,
(\ref{5.13}) can be rewritten as
%**   5.14 %%%%%%%%%%%%%%%%%%%%%%%%%%%%%%%%%%%%%%%%%%%%%%%%%%%%%%%%%
\begin{eqnarray}
\{ Q_{nm}, (\tilde \lambda_\alpha \hat G^\alpha) \} = T + \rho_{[\alpha\beta]}
(\lambda^\beta \hat G^\alpha) + \{ Q_{nm}, \partial \tilde \lambda \partial \theta
- (\tilde \lambda \partial \lambda)(\tilde \lambda \partial \theta) \}.
\label{5.14}
\end{eqnarray}
%%%%%%%%%%%%%%%%%%%%%%%%%%%%%%%%%%%%%%%%%%%%%%%%%%%%%%%%%%%%%%%%%%% 

Here it is of interest to remark that the term $\partial \tilde 
\lambda \partial \theta - (\tilde \lambda \partial \lambda)(\tilde \lambda \partial \theta)$ 
that arises in the r.h.s. of (\ref{5.14}) is just the difference between 
the generalized normal ordering $( \cdots )$ in \cite{Francesco} and 
the improved one $: \cdots :$ of $ \tilde \lambda_\alpha \hat G^\alpha$, 
that is
%**  5.15  %%%%%%%%%%%%%%%%%%%%%%%%%%%%%%%%%%%%%%%%%%%%%%%%%%%%%%%%%%%
\begin {eqnarray}
(\tilde \lambda_\alpha \hat G^\alpha) = : \tilde \lambda_\alpha \hat G^\alpha:
+  \partial \tilde \lambda \partial \theta - (\tilde \lambda \partial \lambda)
(\tilde \lambda \partial \theta),
\label{5.15}
\end{eqnarray}
%%%%%%%%%%%%%%%%%%%%%%%%%%%%%%%%%%%%%%%%%%%%%%%%%%%%%%%%%%%%%%%%%%%%%
so that (\ref{5.14}) becomes
%**  5.16  %%%%%%%%%%%%%%%%%%%%%%%%%%%%%%%%%%%%%%%%%%%%%%%%%%%%%%%%
\begin{eqnarray}
\{ Q_{nm},:\tilde\lambda_\alpha \hat G^\alpha: \} = T + \rho_{[\alpha\beta]}
(\lambda^\beta\hat G^\alpha).
\label {5.16}
\end{eqnarray}
%%%%%%%%%%%%%%%%%%%%%%%%%%%%%%%%%%%%%%%%%%%%%%%%%%%%%%%%%%%%%%%%%%%%%%%%
With the help of the second recursive equations in  (\ref{5.6}) and (\ref{5.7}) 
the last term in the r.h.s. of Eq. (\ref{5.16}) reads
%**  5.17  %%%%%%%%%%%%%%%%%%%%%%%%%%%%%%%%%%%%%%%%%%%%%%%%%%
 \begin {eqnarray}
\rho_{[\alpha\beta]}(\lambda^{[\beta} \hat G^{\alpha]}) 
&=& \rho_{[\alpha\beta]} ([Q_{nm}, H^{[\beta\alpha]}])  \nonumber\\ 
&=&  \{Q_{nm}, \rho_{[\alpha\beta]} H^{[\alpha\beta]}\} 
- \frac{1}{3}(\rho_{[\alpha\beta\gamma]}\lambda^{[\alpha}) H^{\beta\gamma]}.
\label{5.17}
\end{eqnarray}
%%%%%%%%%%%%%%%%%%%%%%%%%%%%%%%%%%%%%%%%%%%%%%%%%%%%%%%%%%%%%%%%%%%%%%%%%%%%%
In this case, the rearrangement theorem does not give extra contributions since 
%**  5.18  %%%%%%%%%%%%%%%%%%%%%%%%%%%%%%%%%%%%%%%%%%%%%%%%%%%%%%%%%%%%%%%%%%
\begin{eqnarray}
(\rho_{[\alpha\beta\gamma]}\lambda^{[\alpha}) H^{\beta\gamma]}
- \rho_{[\alpha\beta\gamma]}(\lambda^{[\alpha} H^{\beta\gamma]}) 
= R_{2\gamma} \partial \lambda^\gamma + \partial \rho_{[\alpha\beta\gamma]} 
R_{3}^{[\alpha\beta \gamma]},  
\label{5.18}
\end{eqnarray}
%%%%%%%%%%%%%%%%%%%%%%%%%%%%%%%%%%%%%%%%%%%%%%%%%%%%%%%%%%%%%%%%%%%%%%%%%%%%%
and the r.h.s. vanishes from (\ref{5.10}) and (\ref{3.33}) .
Therefore,  Eq. (\ref{5.17}) can be rewritten as
%**  5.19  %%%%%%%%%%%%%%%%%%%%%%%%%%%%%%%%%%%%%%%%%%%%%%%%%%%%%%%%
\begin{eqnarray}
\rho_{[\alpha\beta]}(\lambda^{[\beta} \hat G^{\alpha]}) 
= \{Q_{nm}, \rho_{[\alpha\beta]} H^{[\alpha\beta]}\}
- \frac{1}{3}\rho_{[\alpha\beta\gamma]}(\lambda^{[\alpha} H^{\beta\gamma]}).
\label{5.19}
\end{eqnarray}
%%%%%%%%%%%%%%%%%%%%%%%%%%%%%%%%%%%%%%%%%%%%%%%%%%%%
{}For the last term in the r.h.s. of this equation, one can repeat the same 
procedure using the third recursive equations in (\ref{5.6}) and (\ref{5.7}).
Again the contributions from the rearrangement theorem are absent since they 
involve the operators $R_{4}^{[\alpha\beta\gamma\delta]}$ and 
$R_{3\alpha}$ that vanish according to (\ref{3.46}) and (\ref{5.10}). 
As a result, one obtains 
%**  5.20  %%%%%%%%%%%%%%%%%%%%%%%%%%%%%%%%%%%%%%%%%%%
\begin{eqnarray}
\rho_{[\alpha\beta\gamma]}(\lambda^{[\alpha} H^{\beta\gamma]}) 
&=&  \{ Q_{nm}, \rho_{[\alpha\beta\gamma]} K^{[\alpha \beta\gamma ]} \}  
+ \frac{1}{4} (\rho_{[\alpha\beta\gamma\delta]}\lambda^{[\alpha}) 
K^{\beta\gamma\delta]}.
\label{5.20}
\end{eqnarray}
%%%%%%%%%%%%%%%%%%%%%%%%%%%%%%%%%%%%%%%%%%%%%%%%%%%%%%%%%%%%%%%%%%%%%
As a last step, one can express the last term in (\ref{5.20}) 
in terms of $\{ Q_{nm}, \rho_{[\alpha\beta\gamma\delta]}
L^{[\alpha\beta\gamma\delta]} \}$ 
by using the fourth recursive equations in (\ref{5.6}) and (\ref{5.7}). 
Again the contributions from the rearrangement theorem are absent as before, so we have
%**  5.21  %%%%%%%%%%%%%%%%%%%%%%%%%%%%%%%%%%%%%%%%%%%%
\begin{eqnarray}
(\rho_{[\alpha\beta\gamma\delta]}\lambda^{[\alpha}) K^{\beta\gamma\delta]}=
- \{Q_{nm}, (\rho_{[\alpha\beta\gamma\delta]} L^{[\alpha\beta\gamma
\delta]})\},
\label{5.21}
\end{eqnarray}
%%%%%%%%%%%%%%%%%%%%%%%%%%%%%%%%%%%%%%%%%%%%%%%%%%%%%%%%%%%
where we have disregarded the term $\rho_{[\alpha\beta\gamma\delta\epsilon]}
\lambda^{[\epsilon} L^{\alpha\beta\gamma\delta]}$ that vanishes according to
(\ref{3.57}). 

Finally, using (\ref{5.12}) and (\ref{5.16})-(\ref{5.21})
we arrive at the result
%**   5.22 %%%%%%%%%%%%%%%%%%%%%%%%%%%%%%%%%%%%%%%%%%%%%%%%
\begin{eqnarray}
\{ Q_{nm}, b_{nm}\} = T_{nm},
\label{5.22}
\end{eqnarray}
%%%%%%%%%%%%%%%%%%%%%%%%%%%%%%%%%%%%%%%%%%%%%%%%%%%%%%%%%%
where
%**  5.23  %%%%%%%%%%%%%%%%%%%%%%%%%%%%%%%%%%%%%%%%%%%%%
\begin{eqnarray}
b_{nm} &=& s^\alpha \partial \bar \lambda_\alpha
+ :\tilde \lambda_\alpha \hat{G}^\alpha:
- 2(\tilde \lambda_\beta \tilde r_\alpha) H^{[\alpha\beta]} \nonumber\\
&+& 6 (\tilde \lambda_\gamma \tilde r_\beta \tilde r_\alpha) K^{[\alpha\beta
\gamma]} - 24 (\tilde \lambda_\delta \tilde r_\gamma \tilde r_\beta \tilde 
r_\alpha) L^{[\alpha\beta\gamma\delta]}.
\label{5.23}
\end{eqnarray}       
%%%%%%%%%%%%%%%%%%%%%%%%%%%%%%%%%%%%%%%%%%%%%%%%%%%%%%%
In conclusion, we have confirmed  Eq. (\ref{5.2}) provided that one 
interprets the compound field $\tilde \lambda_\alpha G^\alpha$
as the operator $:\tilde \lambda_\alpha \hat G^\alpha:$ which is 
normal-ordered according to the improved prescription (For the other terms in 
(\ref{5.23}) the generalized and the improved normal-ordering 
prescriptions coincide).
Incidentally, we have also checked that this $b_{nm}$ 
possesses conformal weight 2

It might appear from (\ref{5.23}) and the definition of $\tilde \lambda$ and
$\tilde r$ that $b_{nm}$ is singular at $\bar \lambda \lambda 
\rightarrow 0$ with  poles up to fourth order. However, as explained in 
\cite{BerNek}, this singularity is not dangerous. Indeed in this case, the 
analogous of the operator $\xi = Y \theta$ that would trivialize the 
cohomology, is  
%**   5.24 %%%%%%%%%%%%%%%%%%%%%%%%%%%%%%%%%%%%%%%%%%%%%%%%
\begin{eqnarray}
\xi_{nm} = \frac{\tilde \lambda \theta}{\tilde \lambda \lambda 
+ \tilde r \theta} =  \bar \lambda \theta
\sum_{n=1}^{11} \frac{(- r \theta)^{n-1}}{(\lambda \bar\lambda)^n},
\label{5.24}
\end{eqnarray}
%%%%%%%%%%%%%%%%%%%%%%%%%%%%%%%%%%%%%%%%%%%%%%%%%%%%%%%%%%
since $\{Q_{nm}, \xi_{nm}\} = 1$.
However, $\xi_{nm}$ diverges like $(\lambda \bar \lambda)^{-11}$ and to have a 
nontrivial cohomology it is sufficient to exclude from the Hilbert 
space operators that diverge like $\xi_{nm}$ or stronger. Therefore $b_{nm}$
is allowed as insertion to compute higher loop amplitudes.
To do actual calculations at more than two loops \cite{BerNek}, $b_{nm}$ 
must be regularized properly. In fact, in \cite{BerNek} a consistent regularization 
has been proposed.  

Now let us come back to the non-covariant $b$ ghost $\tilde b_{0Y}$ in (\ref{5.4}). 
As a first step, let us derive a quantum counterpart of 
$\tilde b_{0Y}$, which is denoted as $\tilde b_Y$. From the first equation in (\ref{5.7}), 
one has
%**  5.25  %%%%%%%%%%%%%%%%%%%%%%%%%%%%%%%%%%%%%%%%%%%%%%%%
\begin{eqnarray}
\{ Q_{nm}, Y_\alpha \hat G^\alpha \} = Y_\alpha (\lambda^\alpha T).
\label{5.25}
\end{eqnarray}
%%%%%%%%%%%%%%%%%%%%%%%%%%%%%%%%%%%%%%%%%%%%%%%%%%%%%%%%%%%%%%%%%%%
Moreover, since $Y_\alpha(\lambda^\alpha T) - (Y_\alpha\lambda^\alpha) T 
= 2 \partial Y \partial \lambda$ from the rearrangement theorem, one obtains
%**  5.26 %%%%%%%%%%%%%%%%%%%%%%%%%%%%%%%%%%%%%%%%%%%%%%%%%%%%%%%%
\begin{eqnarray}
\{ Q_{nm}, Y_\alpha \hat G^\alpha - 2 \partial Y \partial \theta \} = T.
\label{5.26}
\end{eqnarray}
%%%%%%%%%%%%%%%%%%%%%%%%%%%%%%%%%%%%%%%%%%%%%%%%%%%%%%%%%%%%
As before, the term $2 \partial Y \partial \theta$ is just the difference 
between $(Y_\alpha \hat G^\alpha)$ and $ :Y_\alpha \hat G^\alpha: $ and 
therefore the quantum non-covariant $b$ ghost takes the form
%**  5.27  %%%%%%%%%%%%%%%%%%%%%%%%%%%%%%%%%%%%%%%%%%%%%%%%%
\begin{eqnarray}
\tilde b_{Y} = :Y_\alpha \hat G^\alpha: + (s \partial \bar\lambda),
\label{5.27}
\end{eqnarray}
%%%%%%%%%%%%%%%%%%%%%%%%%%%%%%%%%%%%%%%%%%%%%%%%%%%%%%%%%%%%%%%%%
and it satisfies
%**  5.28 %%%%%%%%%%%%%%%%%%%%%%%%%%%%%%%%%%%%%%%%%%%%%%%%%%%%%%
\begin{eqnarray}
\{ Q_{nm}, \tilde b_{Y} \} = T_{nm}.
\label{5.28}
\end{eqnarray}
%%%%%%%%%%%%%%%%%%%%%%%%%%%%%%%%%%%%%%%%%%%%%%%%%%%%%%%%%%%%%%%%
Even if $ \tilde b_Y$ is non-covariant, its Lorentz variation is BRST-exact. 
Actually, one has
%**   5.29 %%%%%%%%%%%%%%%%%%%%%%%%%%%%%%%%%%%%%%%%%%%%%%%%%
\begin{eqnarray}
\delta_L \tilde b_Y = \Big[ Q_{nm}, 2(L_{\alpha}^{\beta} Y_{\beta}Y_{\gamma}) 
H^{[\gamma \alpha]}\Big],
\label{5.29}
\end{eqnarray} 
%%%%%%%%%%%%%%%%%%%%%%%%%%%%%%%%%%
where $L_{\alpha}^\beta$ are (global) Lorentz parameters.

{}From (\ref{5.22}) and (\ref{5.28}), it follows that $\tilde b_{Y} - b_{nm}$ 
is BRST-closed and then it is plausible that it is also exact. Indeed
in \cite{Oda5}, we have shown that, at the classical level, the covariant 
non-minimal $b$ ghost (\ref{5.2}) and the non-covariant one (\ref{5.4}) are 
cohomologically equivalent.
In this respect, we wish to verify the cohomological equivalence between 
$b_{nm}$ and $\tilde b_{Y}$ even at the quantum level
%**  5.30  %%%%%%%%%%%%%%%%%%%%%%%%%%%%%%%%%%%%%%%%%%%%%%%
\begin{eqnarray}
b_{nm} = \tilde b_{Y} + [ Q_{nm}, W ],
\label{5.30}
\end{eqnarray}       
%%%%%%%%%%%%%%%%%%%%%%%%%%%%%%%%%%%%%%%%%%%%%%%%%%%%%%%%%%%%%%%%
where
%**  5.31  %%%%%%%%%%%%%%%%%%%%%%%%%%%%%%%%%%%%%%%%%%%%%%%%%%
\begin{eqnarray}
W &=& 2 (\tilde \lambda_\beta Y_\alpha) H^{[\alpha\beta]}
+ 3! (\tilde \lambda_\gamma \tilde r_\beta Y_\alpha) K^{[\alpha\beta\gamma]} 
+ 4! (\tilde \lambda_\delta \tilde r_\gamma \tilde r_\beta Y_\alpha) 
L^{[\alpha\beta\gamma\delta]} + W_R,
\label{5.31}
\end{eqnarray}     
%%%%%%%%%%%%%%%%%%%%%%%%%%%%%%%%%%%%%%%%%%%%%%%%%%%%%%%%%%%%%%%%%%
with $W_R$ being a quantum contribution coming from the rearrangement theorem,
which will be determined later.

In order to verify (\ref{5.30}), let us compute the (anti)-commutators of 
$Q_{nm}$ with the first three terms in the r.h.s. of (\ref{5.31}). We have 
%**  5.32  %%%%%%%%%%%%%%%%%%%%%%%%%%%%%%%%%%%%%%%%%%%%%%%%%%%%%%
\begin{eqnarray}
 2[ Q_{nm}, (\tilde \lambda_\beta Y_\alpha) H^{[\alpha\beta]} ]
&=& -  \rho_{\alpha\beta} H^{[\alpha\beta]} 
+  (Y_{[\gamma} \rho_{\alpha\beta]} \lambda^\gamma)H^
{[\alpha\beta]} 
+ (\tilde\lambda_\beta Y_\alpha)(\lambda^{[\alpha} \hat G^{\beta]})  
\nonumber\\ 
&=& - 2(\tilde \lambda_\beta \tilde r_\alpha) H^{[\alpha\beta]} 
- 3!(Y_{[\gamma} \tilde r_\alpha \tilde \lambda_{\beta]})(\lambda^\gamma
H^{\alpha\beta}) + (\tilde \lambda_\alpha \hat G^\alpha) 
- (Y_\alpha \hat G^\alpha)  \nonumber\\
&+& R_H + R_G, 
\label{5.32}
\end{eqnarray}
%%%%%%%%%%%%%%%%%%%%%%%%%%%%%%%%%%%%%%%%%%%%%%%%%%%%%%%%%%%%%%
where $R_H$ and $R_G$ are the contributions coming from the rearrangement 
theorem of the last two terms in the first row of this equation. Then
%**  5.33  %%%%%%%%%%%%%%%%%%%%%%%%%%%%%%%%%%%%%%%%%%%%%%%%%%%%%%
\begin{eqnarray}
3!\{ Q_{nm}, (Y_\alpha \tilde r_\beta \tilde \lambda_\gamma) K^{[\alpha\beta
\gamma]} \} 
&=& 3!(\tilde r_\alpha \tilde r_\beta \tilde \lambda_\gamma) K^{[\alpha\beta
\gamma]}
-4! (Y_\alpha \tilde r_\beta \tilde r_\gamma \tilde \lambda_\delta) 
(\lambda^{[\delta}  K^{\alpha\beta\gamma]}) \nonumber\\
&+& 3! (Y_\alpha \tilde r_\beta \tilde \lambda_\gamma) 
(\lambda^{[\alpha}  H^{\beta\gamma]}) + R_K,
\label{5.33}
\end{eqnarray}
%%%%%%%%%%%%%%%%%%%%%%%%%%%%%%%%%%%%%%%%%%%%%%%%%%%%%%%%%%%%%%%%%%%
where $R_K$ arises from rearrangement theorem.
Finally, we have
%**  5.34 %%%%%%%%%%%%%%%%%%%%%%%%%%%%%%%%%%%%%%%%%%%%%%%%%%%%%%%%%
\begin{eqnarray}
4!\Big[ Q_{nm}, (Y_\alpha \tilde r_\beta \tilde r_\gamma
 \tilde \lambda_\delta L^{[\alpha\beta\gamma\delta]})\Big] = 
4!\tilde r_\alpha \tilde r_\beta \tilde r_\gamma
 \tilde \lambda_\delta L^{[\alpha\beta\gamma\delta]}+4! (Y_\alpha \tilde 
r_\beta \tilde r_\gamma
 \tilde \lambda_\delta) (\lambda^{[\alpha} K^{\beta\gamma\delta]}) + R_L,
\label{5.34}
\end{eqnarray}
%%%%%%%%%%%%%%%%%%%%%%%%%%%%%%%%%%%%%%%%%%%%%%%%%%%%%%%%%%%%%%%%%%
where $R_L$ comes from rearrangement formula. 
The quantum contributions $R_G$, $R_H$, $R_K$ and $R_L$  are explicitly given
by
%**  5.35  %%%%%%%%%%%%%%%%%%%%%%%%%%%%%%%%%%%%%%%%%%%%%%%%%%%%%%%
\begin{eqnarray}
R_G = - [\partial \tilde \lambda \partial \theta - (\tilde \lambda 
\partial\lambda)(\tilde \lambda \partial \theta) 
- 2 \partial Y \partial \theta]
+ 2[Q_{nm}, (Y_{\alpha}\tilde\lambda_{\beta})W_{R1}^{[\alpha\beta]}]
- \frac{1}{2} A_{G}^a \Pi_a,
\label{5.35}
\end{eqnarray}
%%%%%%%%%%%%%%%%%%%%%%%%%%%%%%%%%%%%%%%%%%%%%%%%%%%%%%%%%%%%%%%%%%%%
%**  5.36  %%%%%%%%%%%%%%%%%%%%%%%%%%%%%%%%%%%%%%%%%%%%%%%%%%%%%%%%%
\begin{eqnarray}
R_H = 3![Q_{nm}, (Y_{\alpha}\tilde r_{\beta}\tilde\lambda_
{\gamma})W_{R2}^{[\alpha\beta\gamma]}]
- \frac{1}{4} A^{a}_{H\alpha}(d \Gamma_a )^\alpha
+ \frac{1}{2} A_{G}^a \Pi_a,
\label{5.36}
\end{eqnarray}
%%%%%%%%%%%%%%%%%%%%%%%%%%%%%%%%%%%%%%%%%%%%%%%%%%%%%%%%%%%%%%%%%%%%%%%
%**  5.37  %%%%%%%%%%%%%%%%%%%%%%%%%%%%%%%%%%%%%%%%%%%%%%%%%%%%%%
\begin{eqnarray}
R_K = 4![Q_{nm},  (Y_{\alpha} \tilde r_\beta \tilde r_\gamma \tilde 
\lambda_{\delta}) W_{R3}^{[\alpha\beta\gamma\delta]}]
+ \frac{1}{4} A^{a}_{H\alpha}(d \Gamma_a )^\alpha 
+\frac{1}{12} A^{c}_{K\alpha\beta}N^{\alpha\beta}_c ,
\label{5.37}
\end{eqnarray}
%%%%%%%%%%%%%%%%%%%%%%%%%%%%%%%%%%%%%%%%%%%%%%%%%%%%%%%%%%%%%%%%%%%%%%
%**  5.38  %%%%%%%%%%%%%%%%%%%%%%%%%%%%%%%%%%%%%%%%%%%%%%%%%%%%%%%%%
\begin{eqnarray}
R_L = -\frac{1}{12}  A^{c}_{K\alpha\beta} N^{\alpha\beta}_c + B_{L},
\label{5.38} 
\end{eqnarray}
%%%%%%%%%%%%%%%%%%%%%%%%%%%%%%%%%%%%%%%%%%%%%%%%%%%%%%%%%%%%%%%%%%%%%%
where
%**  5.39 %%%%%%%%%%%%%%%%%%%%%%%%%%%%%%%%%%%%%%%%%%%%%%%%%%%%%%%%%%%
\begin{eqnarray}
W_{R1}^{[\alpha\beta]} &=& \frac{1}{2}((Y+\tilde\lambda)
\Gamma_{a})^{[\alpha} \partial \lambda^{\beta]} \Pi^a,  \nonumber\\
W_{R2}^{[\alpha\beta\gamma]} &=& \frac{1}{8}((Y+2\tilde\lambda)\Gamma_{a})^
{[\alpha}\partial\lambda^{\beta}(\Gamma^a d)^{\gamma ]},   \nonumber\\
W_{R3}^{[\alpha\beta\gamma\delta]} &=& \frac{1}{12} ((Y + 3 \tilde \lambda)
\Gamma_{a})^{[\alpha} \partial \lambda^{\beta}N^{\gamma\delta ] a},
\label{5.39}
\end{eqnarray}
%%%%%%%%%%%%%%%%%%%%%%%%%%%%%%%%%%%%%%%%%%%%%%%%%%%%%%%%%%%%%%%%%%%%%%
and
%**   5.40 %%%%%%%%%%%%%%%%%%%%%%%%%%%%%%%%%%%%%%%%%%%%%%%%%%%%%%%%%%%
\begin{eqnarray}
A^a_G &=& 3! Y_{[\alpha}\tilde r_{\beta}\tilde \lambda_{\gamma]}\lambda^\gamma
((Y + 2 \tilde \lambda) \Gamma^a)^\alpha \partial \lambda^\beta, 
\nonumber\\
A^a_{H \alpha} &=& 4! Y_{[\alpha} \tilde r_\beta \tilde r_\gamma
 \tilde \lambda_{\delta]} \lambda^\delta
 ((Y +3 \tilde \lambda) \Gamma^a)^\beta \partial \lambda^\gamma,
\nonumber\\
A^a_{K \alpha\beta} &=& 5!  Y_{[\alpha}\tilde r_\beta \tilde r_\gamma
\tilde r_\delta \tilde \lambda_{\epsilon]} \lambda^\epsilon
((Y +4 \tilde \lambda) \Gamma^a)^\delta \partial\lambda^\gamma.
\label{5.40}
\end{eqnarray}
%%%%%%%%%%%%%%%%%%%%%%%%%%%%%%%%%%%%%%%%%%%%%%%%%%%%%%%%%%%%%%%%%%%%%%
The $Y$-dependent operators $A_{G}^a$, $A^{a}_{H\alpha}$ and 
$A^{a}_{K\alpha\beta}$ cancel when (\ref{5.35})-(\ref{5.38}) are summed up.

As for $B_L$, it turns out that it is BRST-exact:  
%**   5.41 %%%%%%%%%%%%%%%%%%%%%%%%%%%%%%%%%%%%%%%%%%%%%%%%%%%%%%%%%%%
\begin{eqnarray}
 B_L = 4!\Big[ Q_{nm}, (Y_{\alpha} \tilde r_\beta \tilde r_\gamma \tilde 
\lambda_{\delta})  W_{R4}^{[\alpha\beta\gamma\delta]}\Big] 
+ 4!\Big[ Q_{nm}, \partial( (Y_{\alpha} \tilde r_\beta \tilde r_\gamma \tilde 
\lambda_{\delta})  W_{R5}^{[\alpha\beta\gamma\delta]})\Big],
\label{5.41}
\end{eqnarray}
%%%%%%%%%%%%%%%%%%%%%%%%%%%%%%%%%%%%%%%%%%%%%%%%%%%%%%%%%%%%%%%%%%%%%%%
where
%**   5.42 %%%%%%%%%%%%%%%%%%%%%%%%%%%%%%%%%%%%%%%%%%%%%%%%%%%%%%%%%%%%%
\begin{eqnarray}
W_{R4}^{[\alpha\beta\gamma\delta]} 
&=& \frac{1}{96}[ (\Gamma^c Y)^{[\alpha} (\Gamma^b (Y + 3 \tilde\lambda))^\beta 
(\Gamma_{bc}\partial\lambda)^\gamma 
\partial\lambda^{\delta]} -  (\Gamma^c Y)^{[\alpha} 
(\Gamma^b(Y +3\tilde\lambda))^\beta (\Gamma_{bc}\lambda)^\gamma 
\partial^2 \lambda^{\delta]}   \nonumber\\
&+& 3 (\tilde\lambda\Gamma^c Y)(\Gamma^b 
(Y+2\tilde\lambda))^{[\alpha} \partial\lambda^\beta (\Gamma_c \Gamma^a 
\lambda)^\gamma (\Gamma_b \Gamma_a \partial\lambda)^{\delta]} ],  
\label{5.42}
\end{eqnarray}
%%%%%%%%%%%%%%%%%%%%%%%%%%%%%%%%%%%%%%%%%%%%%%%%%%%%%%%%%%%%%%%%%%%%%%%
and
%**   5.43 %%%%%%%%%%%%%%%%%%%%%%%%%%%%%%%%%%%%%%%%%%%%%%%%%%%%%%%%%%%%
\begin{eqnarray}
W_{R5}^{[\alpha\beta\gamma\delta]}  = \frac{1}{96} (\Gamma^c Y)^{[\alpha} 
(\Gamma^b (Y +3 \tilde\lambda))^\beta [ (\Gamma_{bc}\lambda)^\gamma \partial\lambda^{\delta]} 
- \frac{1}{2} (\Gamma_c \Gamma^a \lambda)^\gamma 
(\Gamma_b \Gamma_a \partial\lambda)^{\delta]}].
\label{5.43}
\end{eqnarray}
%%%%%%%%%%%%%%%%%%%%%%%%%%%%%%%%%%%%%%%%%%%%%%%%%%%%%%%%%%%%%%%%%%%%%%%%
Some details on the derivations of these results will be given in Appendix C.
{}From (\ref{5.15}), one finds that the term $- \partial \tilde \lambda \partial 
\theta + (\tilde \lambda \partial\lambda)(\tilde \lambda \partial \theta) 
- 2 \partial Y \partial \theta$ transforms $(\tilde \lambda_\alpha \hat G^\alpha) 
- (Y_\alpha \hat G^\alpha)$ to $ :\tilde \lambda_\alpha \hat G^\alpha: 
- :Y_\alpha \hat G^\alpha:$. 

Collecting Eqs. (\ref{5.32})-(\ref{5.43}), one recovers (\ref{5.30})
where $b_{nm}$ and $\tilde b_{Y}$ are given in (\ref{5.23}) and (\ref{5.27}), 
respectively and
%**  5.44  %%%%%%%%%%%%%%%%%%%%%%%%%%%%%%%%%%%%%%%%%%%%%%%%%%%%%%%%%%%%%%
\begin{eqnarray}
W &=& 2 (\tilde \lambda_\beta Y_\alpha)( H^{[\alpha\beta]} + W_{R1}^{[\alpha
\beta]}) + 3! (\tilde \lambda_\gamma \tilde r_\beta Y_\alpha)
(K^{[\alpha\beta\gamma]} + W_{R2}^{[\alpha\beta\gamma]}) \nonumber\\
 &+& 4! (\tilde \lambda_\delta \tilde r_\gamma \tilde r_\beta Y_\alpha) 
(L^{[\alpha\beta\gamma\delta]} +W_{R3}^{[\alpha\beta\gamma\delta]} + 
 W_{R4}^{[\alpha\beta\gamma\delta]}) + 4! \partial[ (\tilde \lambda_\delta 
\tilde r_\gamma \tilde r_\beta Y_\alpha) W_{R5}^{[\alpha\beta\gamma\delta]}].
\label{5.44}
\end{eqnarray}     
%%%%%%%%%%%%%%%%%%%%%%%%%%%%%%%%%%%%%%%%%%%%%%%%%%%%%%%%%%%%%%%%%%%%%%%%%%

%%%%%%%%%%%%%%%%%%%%%%%%%%%%%%%%%%%%%%%%%%%%%%%%%%%%%%%%%%%%%%%%%%%%%
%%%%%%%%%%%%%%%%%%%%%%%%%%%%%%   SEC  6    %%%%%%%%%%%%%%%%%%%%%%%%%%
%%%%%%%%%%%%%%%%%%%%%%%%%%%%%%%%%%%%%%%%%%%%%%%%%%%%%%%%%%%%%%%%%%%%%
\section{Conclusion}

In this article, using the Y-formalism \cite{Oda4}, we have calculated the 
normal-ordering contributions existing in various composite 
operators in the pure spinor formalism of superstrings. These operators 
naturally appear when we try to construct a $b$ ghost. Moreover, we have 
constructed the Y-formalism for the  non-minimal sector. 
Using these information, we have presented a quantum-mechanical expression 
of the $b$ ghost, $b_{nm}$, in the non-minimal formulation and we have shown,
 in this case, that the non-covariant $b$ field $b_Y$ and $b_{nm}$, are 
equivalent in cohomology.

The consistent results we have obtained in this article could be regarded 
as a consistency check of the Y-formalism in the both minimal 
and non-minimal pure spinor formulation of superstrings. 

In the case of the non-minimal formulation, due to its field content and 
structure, it is natural to ask if it is possible to reach a  fully covariant 
system of rules for the OPE's in the minimal and non-minimal ghost sectors, 
by replacing the non-covariant fields $Y_\alpha $ and $\bar Y^\alpha $ 
with the covariant ones $\tilde\lambda_\alpha = \frac{\bar \lambda_\alpha}
{\lambda \bar \lambda}$ and $\tilde Y^\alpha = \frac{\lambda^\alpha}
{\lambda \bar \lambda}$, 
respectively. As for the replacement of 
$\bar Y^\alpha$ with $\tilde Y^\alpha $, that is of $\bar v^\alpha$ with
$\lambda^\alpha$ for the non-minimal sector, we do not see any problem, as 
noted at the end of section 4 because $\bar v^\alpha$ and $\lambda^\alpha$ 
are both 
BRST invariant and all the OPE's among the currents of the non-minimal sector 
remain unchanged.

On the contrary, a naive, straightforward replacement of $Y_\alpha$
with $\tilde \lambda_\alpha$ looks problematic.
Indeed, even if the OPE's among the Lorentz current $N_{ab}$, the ghost current $J$, 
and the stress energy tensor $T_{\lambda}$ of the minimal ghost sector are unchanged, 
those among these operators and that of the non-minimal sector become different 
from zero, since the correction terms in (\ref{2.16})-(\ref{2.18}) now acquire a 
dependence from $\bar\lambda$. Therefore the OPE's among the total Lorentz current, ghost 
current and stress energy tensor of the (minimal and non-minimal) ghost sector do not 
close correctly.  Moreover, the BRST variation of (\ref{2.14}) appears to be inconsistent.
We cannot exclude a possibility that these problems could be overcome by a smart modification 
of the basic OPE's, but it is far from obvious that a consistent modification could be found.
Thus, in this paper, we have refrained from exploring this possibility further 
and we hope to come back to this question in future.

%%%%%%%%%%%%%%%%%%%%%%%%%%%%%%%%%%%%%%%%%%%%%%%%%%%%%%%%%%%%%%%%%%
%%%%%%%%%%%%%%%%%%%%%%%% Acknowledgements %%%%%%%%%%%%%%%%%%%%%%%%%%%%%
%%%%%%%%%%%%%%%%%%%%%%%%%%%%%%%%%%%%%%%%%%%%%%%%%%%%%%%%%%%%%%%%%%
\begin{flushleft}
{\bf Acknowledgements}
\end{flushleft}

The work of the first author (I.O.) was partially supported by
the Grant-in-Aid for Scientific Research (C) No.14540277 from 
the Japan Ministry of Education, Science and Culture.
The work of the second author (M.T.) was  supported by the European
Community's Human Potential Programme under contract MRTN-CT-2004-005104 
"Constituents, Fundamental Forces and Symmetries of the Universe".

%%%%%%%%%%%%%%%%%%%%%%%%%%%%%%%%%%%%%%%%%%%%%%%%%%%%%%%%%%%%%%%%%%
%%%%%%%%%%%%%%%%%%%%%%%% Appendices %%%%%%%%%%%%%%%%%%%%%%%%%%%%%%
%%%%%%%%%%%%%%%%%%%%%%%%%%%%%%%%%%%%%%%%%%%%%%%%%%%%%%%%%%%%%%%%%%
\appendix

\section{Notation, Conventions and Useful identities}

In this appendix, we collect our notation, conventions and
some useful formulae employed in this paper.

As usual, in ten space-time dimensions, $\Gamma^a$ are the Dirac matrices
$\gamma^a$ times the charge conjugation matrix $C$, that is, $  (\Gamma^{a})^
{\beta\alpha} = (\gamma^a C)^{\alpha\beta}$ and   $  (\Gamma^{a})_
{\beta\alpha} = ( C^{-1}\gamma^a )_{\alpha\beta}$; they are 16 $\times$ 16 
symmetric matrices with respect to the spinor indices, 
and satisfiy the Clifford algebra $\{ \Gamma^a, \Gamma^b \} = 2 \eta^{ab}$.
Our metric convention is $\eta^{ab} = (-, +, \cdots, +)$. 

The square bracket and the brace respectively denote the antisymmetrization
and the symmetrization of $ p $ indices, normalized with a numerical factor
$\frac{1}{p!}$ so that, for 
instance $A_{[\mu} B_{\nu]} = \frac{1}{2}( A_{\mu}B_{\nu} - A_{\nu}B_{\mu}) $.
As for the products of $\Gamma^a$, $\Gamma^{a_1 \cdots a_p} =
\Gamma^{[a_1 \cdots a_p]}$.
These antisymmetrized products of $\Gamma$ have
definite symmetry properties, which are given by 
$(\Gamma^{ab})^\alpha \ _\beta = - (\Gamma^{ab})_\beta \ ^\alpha$, 
$(\Gamma^{abc})_{\alpha\beta} = - (\Gamma^{abc})_{\beta\alpha}$,
$(\Gamma^{abcd})^\alpha \ _\beta =  (\Gamma^{abcd})_\beta \ ^\alpha$, 
$(\Gamma^{abcde})_{\alpha\beta} = (\Gamma^{abcde})_{\beta\alpha}$,
etc.  

The product of generic spinors $f_\alpha$ and $g_\beta$ can be
expanded in terms of the complete set of gamma matrices as
%**  A.1  %%%%%%%%%%%%%%%%%%%%%%%%%%%%%%%%%%%%%%%%%%%%%%%%%%%%%%%%%%%%%%%%%%%%%
\begin{eqnarray}
f_\alpha g_\beta = \frac{1}{16} \Gamma^a_{\alpha\beta} (f \Gamma_a g)
+  \frac{1}{16 \times 3!} \Gamma^{abc}_{\alpha\beta} (f \Gamma_{abc} g)
+ \frac{1}{16 \times 5!} \Gamma^{abcde}_{\alpha\beta} (f \Gamma_{abcde} g).
\label{A.1}
\end{eqnarray}       
%%%%%%%%%%%%%%%%%%%%%%%%%%%%%%%%%%%%%%%%%%%%%%%%%%%%%%%%%%%%%%%%%%%%%%%%%%%%%%%%%
Similarly, for spinors $f_\alpha$ and $g^\beta$ we have 
%**  A.2  %%%%%%%%%%%%%%%%%%%%%%%%%%%%%%%%%%%%%%%%%%%%%%%%%%%%%%%%%%%%%%%%%%%%%%
\begin{eqnarray}
f_\alpha g^\beta = \frac{1}{16} \delta^\beta_\alpha (f g)
+  \frac{1}{16 \times 2!} (\Gamma^{ab})_\alpha \ ^\beta (f \Gamma_{ab} g)
+ \frac{1}{16 \times 4!} (\Gamma^{abcd})_\alpha \ ^\beta (f \Gamma_{abcd} g).
\label{A.2}
\end{eqnarray}       
%%%%%%%%%%%%%%%%%%%%%%%%%%%%%%%%%%%%%%%%%%%%%%%%%%%%%%%%%
A useful identity, involving three spinor-like operators $A_\alpha$, $B^\beta$
and $C^\gamma$ is
%** A.3  %%%%%%%%%%%%%%%%%%%%%%%%%%%%%%%%%%%%%%%%%%%%%%%%%%%%%%%%%%%%%
\begin{eqnarray}
  - \frac{1}{8} ( B \Gamma^{ab} A) (\Gamma_{ab} C)^\alpha 
 - \frac{1}{4} (B A) C^\alpha
= ( B_\beta A^\alpha) C^\beta 
 - \frac{1}{2} (( \Gamma^a B)^\alpha A^\beta) (\Gamma_a C)_\beta.
\label{A.3}
\end{eqnarray}
%%%%%%%%%%%%%%%%%%%%%%%%%%%%%%%%%%%%%%%%%%%%%%%%%%%%%%%%%%%%%%%%%

\section{Normal ordering, the generalized Wick theorem and rearrangement
theorem}

In this appendix, we explain the prescription of normal ordering,
the generalized Wick theorem and rearrangement theorem, which are
used in this paper. The more detail of them can be seen in the texbook
of conformal field theory \cite{Francesco}.

\subsection{Normal ordering}

In conformal field theory, we usually consider normal ordering for free 
fields where the OPE contains only one singular term with a constant
coefficient. Then, normal ordering is defined as the subtraction
of this singular term. This definition of normal ordering is found to be
equivalent to the conventional normal ordering in the mode expansion 
where the annihilation operators are placed at the rightmost position. 
However, we sometimes meet the case for which the fields are not free in 
this sense. 
One of the well-known examples happens when we try to regularize the OPE 
between 
two stress enery tensors $T(y) T(z)$.
In this case, we have two singular terms where one singular term contains
the quartic pole whose coefficient is proportional to the central charge
while the other singular term contains the quadratic pole whose coefficient
is not a constant but ($2 \times$) stress energy tensor itself. The ususal
normal ordering prescription amounts to subtraction of  the former, most 
singular term, but the latter singular term is still remained. Let us note 
that in the present context, the OPE between $\omega$ and $\lambda$ is not 
free owing to the existence of the projection $K$ reflecting the pure spinor 
constraint.
{}From the physical point of view, we want to subtract $\it{all}$ the 
singular terms 
in the OPE's, so we have to generalize the definition of normal ordering.

To this end, we introduce the generalized normal ordering which is usually 
denoted by parentheses, that is, explicitly, the generalized normal ordering 
of operators $A$ and $B$ is written as $(A B)(z)$.
A definition of the generalized normal ordering is given by the contour 
integration \cite{Francesco} 
%**   B.1 %%%%%%%%%%%%%%%%%%%%%%%%%%%%%%%%%%%%%%%%%%%%%%%%%%%%%%%%%
\begin{eqnarray}
(A B)(z) = \oint_z \frac{dw}{w-z} A(w) B(z).
\label{B.1}
\end{eqnarray}
%%%%%%%%%%%%%%%%%%%%%%%%%%%%%%%%%%%%%%%%%%%%%%%%%%%%%%%%%%%%%%%%%%%
Then the OPE of $A(z)$ and $B(w)$ is described by
%**   B.2 %%%%%%%%%%%%%%%%%%%%%%%%%%%%%%%%%%%%%%%%%%%%%%%%%%%%%%%%%
\begin{eqnarray}
A(z) B(w) = <A(z) B(w)> + (A(z) B(w)),
\label{B.2}
\end{eqnarray}
%%%%%%%%%%%%%%%%%%%%%%%%%%%%%%%%%%%%%%%%%%%%%%%%%%%%%%%%%%%%%%%%%%%
where $<A(z) B(w)>$ denotes the contraction containing $\it{all}$ the 
singular
terms of the OPE and $(A(z) B(w))$ stands for the complete sequence of 
regular terms whose explicit forms can be extracted from the Taylor
expansion of $A(z)$ around $w$:
%**   B.3 %%%%%%%%%%%%%%%%%%%%%%%%%%%%%%%%%%%%%%%%%%%%%%%%%%%%%%%%%
\begin{eqnarray}
(A(z) B(w)) = \sum_{k \geq 0} \frac{(z-w)^k}{k!} (\partial^k A \cdot B)(w).
\label{B.3}
\end{eqnarray}
%%%%%%%%%%%%%%%%%%%%%%%%%%%%%%%%%%%%%%%%%%%%%%%%%%%%%%%%%%%%%%%%%%%

Another definition of the generalized normal ordering is provided by the mode
expansion. If the OPE of $A$ and $B$ is written as
%**   B.4 %%%%%%%%%%%%%%%%%%%%%%%%%%%%%%%%%%%%%%%%%%%%%%%%%%%%%%%%%
\begin{eqnarray}
A(z) B(w) = \sum_{k= - \infty}^N \frac{\{ AB \}_k (w)}{(z-w)^k},
\label{B.4}
\end{eqnarray}
%%%%%%%%%%%%%%%%%%%%%%%%%%%%%%%%%%%%%%%%%%%%%%%%%%%%%%%%%%%%%%%%%%%
where $N$ is some positive integer, the definition of the generalized normal 
ordering reads 
%**   B.5 %%%%%%%%%%%%%%%%%%%%%%%%%%%%%%%%%%%%%%%%%%%%%%%%%%%%%%%%%
\begin{eqnarray}
(A B)(z) = \{ AB \}_0 (z).
\label{B.5}
\end{eqnarray}
%%%%%%%%%%%%%%%%%%%%%%%%%%%%%%%%%%%%%%%%%%%%%%%%%%%%%%%%%%%%%%%%%%%
Incidentally, in this context, the contraction is expressed by
%**   B.6 %%%%%%%%%%%%%%%%%%%%%%%%%%%%%%%%%%%%%%%%%%%%%%%%%%%%%%%%%
\begin{eqnarray}
<A(z) B(w)> = \sum_{k= 1}^N \frac{\{ AB \}_k (w)}{(z-w)^k},
\label{B.6}
\end{eqnarray}
%%%%%%%%%%%%%%%%%%%%%%%%%%%%%%%%%%%%%%%%%%%%%%%%%%%%%%%%%%%%%%%%%%%

In this paper, we adopt the definition of the contour integration (\ref{B.1}).
Moreover, for simplicity, we do not write explicitly the outermost 
parenthesis representing the generalized normal ordering whenever we can 
easily judge from
the context whether some operators are normal-odered or not.

\subsection{The generalized Wick theorem}

Relating to the generalization of the normal-ordering prescription,
we also have to reformulate the Wick theorem for $\it{interacting}$ fields.
In general, the Wick theorem relates the time-ordered product to the 
normal-ordered product of free fields. However, such a relation cannot be
generalized to interacting fields in a straightforward manner. Hence,
the generalized Wick theorem is defined by generalizing a special form
of the Wick theorem for the contraction of free fields. More explicitly,
the generalized Wick theorem is simply defined as
%**   B.7 %%%%%%%%%%%%%%%%%%%%%%%%%%%%%%%%%%%%%%%%%%%%%%%%%%%%%%%%%
\begin{eqnarray}
<A(z)(B C)(w)> = \oint_w \frac{dx}{x-w} [<A(z) B(x)> C(w) + B(x) <A(z) C(w)>].
\label{B.7}
\end{eqnarray}
%%%%%%%%%%%%%%%%%%%%%%%%%%%%%%%%%%%%%%%%%%%%%%%%%%%%%%%%%%%%%%%%%%%
{}From this definition, it is important to notice that the first $\it{regular}$
term of the various OPE's always contributes. 
If we would like to calculate $<(B C)(z)A(w)>$, we first calculate $<A(z)(B C)(w)>$, 
then interchange $w$ and $z$, and finally expand the fields evaluated at $z$ 
in the Taylor series around $w$.

\subsection{Rearrangement theorem}

We often encounter the situation where many of operators are normal-ordered,
e.g., $(A (B C))(z)$. With the generalized normal ordering, some complication
occurs since there is no associativity in such normal-ordered operators
%**   B.8 %%%%%%%%%%%%%%%%%%%%%%%%%%%%%%%%%%%%%%%%%%%%%%%%%%%%%%%%%
\begin{eqnarray}
(A (B C))(z)  \not= ((A B) C)(z).
\label{B.8}
\end{eqnarray}
%%%%%%%%%%%%%%%%%%%%%%%%%%%%%%%%%%%%%%%%%%%%%%%%%%%%%%%%%%%%%%%%%%%
To deal with normal ordering of such composite operators, 
we make use of the rearrangement theorem. The useful formulae are
given by
%**   B.9-11 %%%%%%%%%%%%%%%%%%%%%%%%%%%%%%%%%%%%%%%%%%%%%%%%%%%%%%%%%
\begin{eqnarray}
(A B) &=& (B A) + ([A, B]), \\
(A (B C)) &=& (B (A C)) + (([A, B]) C), \\ 
((A B) C) &=& (A (B C)) + (A ([C, B])) + (([C, A]) B) 
+ ([(A B), C]),
\label{B.9-11}
\end{eqnarray}
%%%%%%%%%%%%%%%%%%%%%%%%%%%%%%%%%%%%%%%%%%%%%%%%%%%%%%%%%%%%%%%%%%%
where $A$, $B$, and $C$ are all the Grassmann-even quantities. Note that if 
the Grassmann-odd quantities are involved, we must change the sign and the commutator
in a suitable manner. For instance, for the Grassmann-even $A$ and
the Grassmann-odd $B$ and $C$, the last rearrangement theorem
is modified as
%**   B.12 %%%%%%%%%%%%%%%%%%%%%%%%%%%%%%%%%%%%%%%%%%%%%%%%%%%%%%%%%
\begin{eqnarray}
((A B) C) =  (A (B C)) - (A (\{B, C\})) - (([C, A]) B) 
+ (\{(A B), C\}).
\label{B.12}
\end{eqnarray}
%%%%%%%%%%%%%%%%%%%%%%%%%%%%%%%%%%%%%%%%%%%%%%%%%%%%%%%%%%%%%%%%%%%
In making use of these rearrangement theorems, we are forced to evaluate
the generalized normal ordering of the (anti-)commutator 
$( [ A, B ] )$. Then, we rely on the useful formula
%**   B.13 %%%%%%%%%%%%%%%%%%%%%%%%%%%%%%%%%%%%%%%%%%%%%%%%%%%%%%%%%
\begin{eqnarray}
( [ A,  B ] )(z) = \sum_{k= 1} \frac{(-1)^{k+1}}{k!} \partial^k \{ AB \}_k(z).
\label{B.13}
\end{eqnarray}
%%%%%%%%%%%%%%%%%%%%%%%%%%%%%%%%%%%%%%%%%%%%%%%%%%%%%%%%%%%%%%%%%%%
Note that field-dependent singular terms contribute to the normal-ordering
(anti-)commutator while the non-singular term $\{ AB \}_0$ does not.
In this paper, we make heavy use of these formulae in evaluating
various normal-ordered products of operators.

\section {Some details about the calculations}

\subsection { BRST variation of $G^\alpha$ }
To compute the BRST variation of  $G^\alpha $ it is 
convenient to use the following notation 
%**  C.1   %%%%%%%%%%%%%%%%%%%%%%%%%%%%%%%%%%%%%%%%%%%%%%%%%%%%%%%%%%%%%
\begin{eqnarray}
g^\alpha (B,A,C) 
&=& - \frac{1}{8} ( B \Gamma^{ab} A) (\Gamma_{ab} C)^\alpha 
 - \frac{1}{4} (B A) C^\alpha \nonumber\\
&=& ( B_\beta A^\alpha) C^\beta 
 - \frac{1}{2} (( \Gamma^a B)^\alpha A^\beta) (\Gamma_a C)_\beta,
\label{C.1}
\end{eqnarray}
%%%%%%%%%%%%%%%%%%%%%%%%%%%%%%%%%%%%%%%%%%%%%%%%%%%%%%%%%%%%%%%%%%%
where $A^\alpha$, $B_\beta$, and $C^\gamma$ are generic spinors and the 
last step is the identity (\ref{A.3}). Then, given (\ref{3.5}), one has
%**   C.2 %%%%%%%%%%%%%%%%%%%%%%%%%%%%%%%%%%%%%%%%%%%%%%%%%%%%%%%%%%
\begin{eqnarray}
 \{ Q, G_1^\alpha \} =
 - \frac{1}{2} \lambda^\alpha (\Pi^a \Pi_a)
+ \frac{1}{2}  (\lambda \Gamma_a \partial \theta) (\Gamma^a d)^\alpha.
\label{C.2}
\end{eqnarray}
%%%%%%%%%%%%%%%%%%%%%%%%%%%%%%%%%%%%%%%%%%%%%%%%%%%%%%%%%%%%%%%%%%%%
Moreover,
%** C.3  %%%%%%%%%%%%%%%%%%%%%%%%%%%%%%%%%%%%%%%%%%%%%%%%%%%%%%%%%%%%
\begin{eqnarray}
\{ Q, G_2^\alpha + G_3^\alpha \}
&=& - g^\alpha(d,\lambda,\partial \theta) + g^\alpha(\Omega,\lambda,\partial 
\lambda) - 2 g^\alpha(Y,\partial\lambda,\partial\lambda) - (Y\partial 
\lambda)\partial \lambda^\alpha .
\label{C.3}
\end{eqnarray}
%%%%%%%%%%%%%%%%%%%%%%%%%%%%%%%%%%%%%%%%%%%%%%%%%%%%%%%%%%%%%%%%%%%
The last three terms come from the
definitions (\ref{2.25}) and (\ref{2.26}) of $N_{ab}$ and $J$.

Using the rearrangement formula (cf. (B.12)), one has 
%**   C.4 %%%%%%%%%%%%%%%%%%%%%%%%%%%%%%%%%%%%%%%%%%%%%%%%%%%%%%%%%
\begin{eqnarray}
g^\alpha(d,\lambda,\partial\theta) 
= \lambda^\alpha (d \partial \theta) + 8 \partial^2 \lambda^\alpha +
 \frac{1}{2} ( \lambda \Gamma_a \partial \theta) ( \Gamma^a d)^\alpha,
\label{C.4}
\end{eqnarray}
%%%%%%%%%%%%%%%%%%%%%%%%%%%%%%%%%%%%%%%%%%%%%%%%%%%%%%%%%%%%%%%%%%%
and
%**   C.5 %%%%%%%%%%%%%%%%%%%%%%%%%%%%%%%%%%%%%%%%%%%%%%%%%%%%%%%%%
\begin{eqnarray}
g^\alpha(\Omega, \lambda,\partial \lambda)
= ( \Omega_\beta \lambda^\alpha) \partial \lambda^\beta - \frac{1}{2} 
(( \Gamma^a \Omega)^\alpha \lambda^\beta) (\Gamma_a \partial \lambda)_\beta.
\label{C.5}
\end{eqnarray}
%%%%%%%%%%%%%%%%%%%%%%%%%%%%%%%%%%%%%%%%%%%%%%%%%%%%%%%%%%%%%%%%%%
Using the rearrangement theorem, the first term in the r.h.s. 
of Eq. (\ref {C.5}) becomes
%**  C.6  %%%%%%%%%%%%%%%%%%%%%%%%%%%%%%%%%%%%%%%%%%%%%%%%%%%%%%%%
\begin{eqnarray}
( \Omega_\beta \lambda^\alpha) \partial \lambda^\beta
= \lambda^\alpha ( \Omega \partial \lambda) 
- \frac{1}{2} (Y \Gamma^a)^\alpha (\partial \lambda \Gamma_a
\partial \lambda) + \frac{11}{2} \partial^2 \lambda^\alpha,
\label{C.6}
\end{eqnarray}
%%%%%%%%%%%%%%%%%%%%%%%%%%%%%%%%%%%%%%%%%%%%%%%%%%%%%%%%%%%%%%%%%%%
whereas the second term can be rewritten as
%**   C.7 %%%%%%%%%%%%%%%%%%%%%%%%%%%%%%%%%%%%%%%%%%%%%%%%%%%%%%%%%
\begin{eqnarray}
- \frac{1}{2} ( ( \Gamma^a \Omega)^\alpha \lambda^\beta) (\Gamma_a 
\partial \lambda)_\beta 
&=& - \frac{3}{2} \partial^2 \lambda^\alpha + \frac{3}{2} (Y \partial^2 
\lambda)\lambda^\alpha +  (Y \partial \lambda) \partial \lambda^\alpha
\nonumber\\
&+& 3 (\partial Y \partial \lambda) \lambda^\alpha
+ 2 g^\alpha (Y, \partial \lambda, \partial \lambda)  + {\frac{1}{2}} 
(Y \Gamma^a)^\alpha (\partial \lambda \Gamma_{a} \partial \lambda),
\label{C.7}
\end{eqnarray}
%%%%%%%%%%%%%%%%%%%%%%%%%%%%%%%%%%%%%%%%%%%%%%%%%%%%%%%%%%%%%%%%%%%%%%
so that from (\ref{C.5})-(\ref{C.7}), one obtains
%**   C.8 %%%%%%%%%%%%%%%%%%%%%%%%%%%%%%%%%%%%%%%%%%%%%%%%%%%%%%%%%%%%
\begin{eqnarray}
g^\alpha (\Omega,\lambda,\partial \lambda)  
&=& \lambda^\alpha ( \Omega \partial \lambda)  + 4 \partial^2 \lambda^\alpha
+ \frac{3}{2} (Y \partial^2 \lambda) \lambda^\alpha  \nonumber\\
&+&  (Y \partial \lambda) \partial \lambda^\alpha 
+ 3 (\partial Y \partial \lambda) \lambda^\alpha 
+ 2 g^\alpha(Y,\partial\lambda, \partial\lambda).
\label{C.8}
\end{eqnarray}
%%%%%%%%%%%%%%%%%%%%%%%%%%%%%%%%%%%%%%%%%%%%%%%%%%%%%%%%%%%%%%%%

Adding Eqs. (\ref{C.2}), (\ref{C.3}) and $\{Q, G_{4}^\alpha \} = c_1 
\partial^2 \lambda^\alpha$ with $c_1 = {\frac{7}{2}}$ ,taking into account 
(\ref{C.4}), (\ref{C.8}) and using the definition (\ref{2.16}) of the stress energy tensor $T$, we finally obtain
%**   C.9   %%%%%%%%%%%%%%%%%%%%%%%%%%%%%%%%%%%%%%%%%%%%%%%%%%%%%%%%
\begin{eqnarray}
\{ Q, G^\alpha \} = \lambda^\alpha T - \frac{1}{2} \partial^2 \lambda^\alpha.
\label{C.9}
\end{eqnarray}
%%%%%%%%%%%%%%%%%%%%%%%%%%%%%%%%%%%%%%%%%%%%%%%%%%%%%%%%%%%%%%%%%%%

\subsection { BRST variation of $H^{\alpha\beta}$ }

Now let us consider the BRST variation of $H^{\alpha\beta}$.
Eq. (\ref{3.25}) can be rewritten as 
%**   C.10 %%%%%%%%%%%%%%%%%%%%%%%%%%%%%%%%%%%%%%%%%%%%%%%%%%%%%%%%%
\begin{eqnarray}
[ Q, H^{(\alpha\beta)} ] = \frac{1}{16} \Gamma^{\alpha\beta}_a h^a,
\label{C.10}
\end{eqnarray} 
%%%%%%%%%%%%%%%%%%%%%%%%%%%%%%%%%%%%%%%%%%%%%%%%%%%%%%%%%%%%%%%%%%%%
where
%**   C.11 %%%%%%%%%%%%%%%%%%%%%%%%%%%%%%%%%%%%%%%%%%%%%%%%%%%%%%%%%
\begin{eqnarray}
 h^a = \frac{1}{2}  (\lambda \Gamma^
{a}\Gamma_{b} d) \Pi^b + N^{ab}(\lambda\Gamma_b\partial\theta) - \frac{1}{2} 
J (\lambda\Gamma^a \partial \theta) 
 + 2 \partial (\lambda \Gamma^a \partial \theta). 
 \label{C11}
\end{eqnarray} 
%%%%%%%%%%%%%%%%%%%%%%%%%%%%%%%%%%%%%%%%%%%%%%%%%%%%%%%%%%%%%%%%%%%% 
The first term in the r.h.s. of this equation can be rewritten as
%%**  C.12 %%%%%%%%%%%%%%%%%%%%%%%%%%%%%%%%%%%%%%%%%%%%%%%%%%%%%%%%%
\begin{eqnarray}
\frac{1}{2} (\lambda \Gamma^a \Gamma_b d)\Pi^b = \frac{1}{2}  (\lambda 
\Gamma^a \Gamma_b \Pi^b d) +5 \partial (\lambda\Gamma^a \partial \theta).
\label{C.12}
\end{eqnarray}
%%%%%%%%%%%%%%%%%%%%%%%%%%%%%%%%%%%%%%%%%%%%%%%%%%%%%%%%%%%%%%%%%%%%
With the notation 
%%**   C.13 %%%%%%%%%%%%%%%%%%%%%%%%%%%%%%%%%%%%%%%%%%%%%%%%%%%%%%%%
\begin{eqnarray}
\Lambda^{\alpha\beta} &\equiv& \frac{1}{2} \partial \lambda^{[\alpha}
\lambda^{\beta]},   \nonumber\\ 
\tilde \Lambda_{[\alpha\beta]} &\equiv& - \frac{1}{4}(\Gamma^c \Lambda 
\Gamma_c)_{[\alpha\beta]},
\label{C.13}
\end{eqnarray}
%%%%%%%%%%%%%%%%%%%%%%%%%%%%%%%%%%%%%%%%%%%%%%%%%%%%%%%%%%%%%%%%%%%%
the vector
%%**   C.14  %%%%%%%%%%%%%%%%%%%%%%%%%%%%%%%%%%%%%%%%%%%%%%%%%%%%%%%
\begin{eqnarray}
V^a = N^{ab}(\lambda\Gamma_b\partial\theta) - \frac{1}{2} J (\lambda  
\Gamma^a \partial \theta),
\label{C.14}
\end{eqnarray}
%%%%%%%%%%%%%%%%%%%%%%%%%%%%%%%%%%%%%%%%%%%%%%%%%%%%%%%%%%%%%%%%%%%%%
becomes
%** C.15  %%%%%%%%%%%%%%%%%%%%%%%%%%%%%%%%%%%%%%%%%%%%%%%%%%%%%%%%%%%
\begin{eqnarray}
V^a = \frac{1}{2} (\Omega\Gamma^a \Gamma^b \lambda)(\lambda\Gamma_b \partial 
\theta)- J (\lambda \Gamma^a \partial\theta) +4(Y\Lambda\Gamma^a 
\partial \theta) + 4(Y\Gamma^a \tilde \Lambda \partial\theta) 
+ 2 (\partial \lambda\Gamma^a \partial \theta).
\label{C.15}
\end{eqnarray}
%%%%%%%%%%%%%%%%%%%%%%%%%%%%%%%%%%%%%%%%%%%%%%%%%%%%%%%%%%%%%%%%%%%%%%
The first term in the r.h.s. of (\ref{C.15}) vanishes modulo a rearrangement 
contribution:
%**  C.16 %%%%%%%%%%%%%%%%%%%%%%%%%%%%%%%%%%%%%%%%%%%%%%%%%%%%%%%%%%%%
\begin{eqnarray}
 \frac{1}{2} (\Omega\Gamma^a \Gamma^b \lambda)(\lambda\Gamma_b \partial 
\theta) = -4(Y\Gamma^a \tilde\Lambda \partial \theta) - 4 (Y\Lambda\Gamma^a 
\partial\theta) + 4(\partial\lambda\Gamma^a\partial \theta),
\label{C.16}
\end{eqnarray}
%%%%%%%%%%%%%%%%%%%%%%%%%%%%%%%%%%%%%%%%%%%%%%%%%%%%%%%%%%%%%%%%%%%%%%
so that $h^a$ becomes
%%**  C.17 %%%%%%%%%%%%%%%%%%%%%%%%%%%%%%%%%%%%%%%%%%%%%%%%%%%%%%%%%%%
\begin{eqnarray}
h^a &=& \frac{1}{2}  (\lambda \Gamma^a \Gamma_b \Pi_b d) 
+ 5 \partial (\lambda \Gamma^a \partial \theta) 
- J (\lambda \Gamma^a \partial \theta) 
 \nonumber\\
&-& 2 (\partial \lambda \Gamma^a \partial 
\theta) + 2 \partial (\lambda \Gamma^a \partial \theta). 
\label{C.17}
\end{eqnarray}
%%%%%%%%%%%%%%%%%%%%%%%%%%%%%%%%%%%%%%%%%%%%%%%%%%%%%%%%%%%%%%%%%%%%%%
On the other hand,
%**  C.18   %%%%%%%%%%%%%%%%%%%%%%%%%%%%%%%%%%%%%%%%%%%%%%%%%%%%%%%%%%
\begin{eqnarray}
\lambda \Gamma^a G = \frac{1}{2}  (\lambda \Gamma^a \Gamma_b \Pi^b d)
+ \frac{7}{2} (\lambda \Gamma^a \partial^2 \theta) + \tilde V^a,
\label{C.18}
\end{eqnarray}
%%%%%%%%%%%%%%%%%%%%%%%%%%%%%%%%%%%%%%%%%%%%%%%%%%%%%%%%%%%%%%%%%%%%%%
where $\tilde V^a = - \frac{1}{4}(\tilde\lambda \Gamma^a N^{bc}\Gamma_{bc} 
\partial \theta) - \frac{1}{4} (\lambda \Gamma^a J \partial \theta) $.
Then, using (\ref{2.25}) and (\ref{2.26})
%**  C.19 %%%%%%%%%%%%%%%%%%%%%%%%%%%%%%%%%%%%%%%%%%%%%%%%%%%%%%%%%%%%
\begin{eqnarray}
\tilde V^a &=& - \frac{1}{2}(\lambda\Gamma_c(\Omega\Gamma^a \Gamma^c\lambda)
\partial \theta)+  \frac{1}{8}(\lambda\Gamma_c \Gamma_b(\Omega\Gamma^b 
\Gamma^c\lambda)\Gamma^a \partial \theta) - (\lambda\Gamma^a (\Omega\lambda)
\partial \theta) \nonumber\\ 
&-& 4(Y\Gamma_a \tilde \Lambda \partial \theta) -
4(Y \Lambda \Gamma^a \partial \theta) + 2(Y \Lambda \Gamma^a \partial 
\theta) + (\partial \lambda\Gamma^a \partial \theta).
\label{C.19}
\end{eqnarray}
%%%%%%%%%%%%%%%%%%%%%%%%%%%%%%%%%%%%%%%%%%%%%%%%%%%%%%%%%%%%%%%%%%%%%%
But the first two terms in the r.h.s. of (\ref{C.19})
vanish modulo the Y-dependent term $ 4 [ (Y\Gamma^a \tilde \Lambda
\partial \theta) +(Y\Lambda\Gamma^a \partial \theta)] - 6(Y\Lambda\Gamma^a 
\partial \theta) $ coming from rearrangement theorem, so that we have
%**  C.20 %%%%%%%%%%%%%%%%%%%%%%%%%%%%%%%%%%%%%%%%%%%%%%%%%%%%%%%%%%
\begin{eqnarray}
\tilde V^a &=& -(\lambda\Gamma^a(\Omega\lambda)\partial \theta) - 
 4 (Y\Lambda\Gamma^a \partial \theta) +(\partial \lambda\Gamma^a \partial 
\theta)\nonumber\\  
&=&  - J(\lambda\Gamma^a \partial \theta)  +4 \partial 
(\lambda\Gamma^a \partial \theta) -4 (\lambda\Gamma^a \partial^2 \theta),
\label{C.20}
\end{eqnarray}
%%%%%%%%%%%%%%%%%%%%%%%%%%%%%%%%%%%%%%%%%%%%%%%%%%%%%%%%%%%%%%%%%%%%%
and therefore 
%**  C.21 %%%%%%%%%%%%%%%%%%%%%%%%%%%%%%%%%%%%%%%%%%%%%%%%%%%%%%%%%%%
\begin{eqnarray}
\lambda\Gamma^a G = \frac{1}{2} (\lambda \Gamma^a \Gamma_b \Pi^b d)
+ \frac{7}{2}(\lambda\Gamma^a \partial^2 \theta) - J (\lambda\Gamma^a \partial 
\theta) - 4 (\lambda \Gamma^a \partial^2 \theta) + 4 \partial (\lambda\Gamma^a 
\partial \theta). 
\label{C.21}
\end{eqnarray}
%%%%%%%%%%%%%%%%%%%%%%%%%%%%%%%%%%%%%%%%%%%%%%%%%%%%%%%%%%%%%%%%%%%%%
Then comparing (\ref{C.17}) with (\ref{C.21}), one gets Eq. (\ref{3.27}). 

Next let us consider the BRST variation of $ H^{[\alpha\beta]}$. Eq.
(\ref{3.26}) can be rewritten as 
%%**  C.22   %%%%%%%%%%%%%%%%%%%%%%%%%%%%%%%%%%%%%%%%%%%%%%%%%%%%%%%%%
\begin{eqnarray}
[ Q, H^{[\alpha\beta]}] 
&=& \frac{1}{96} \Gamma^{\alpha\beta}_{abc} \Big[ \frac{1}{2}(\lambda\Gamma^
{abc}\Gamma^d \Pi_d d) + 6 (\lambda N^{ab} \Gamma^c 
\partial \theta)   
\nonumber\\
&+& 4(\lambda\Gamma^{abc}\partial^2 \theta) + (\partial \lambda \Gamma^{abc}
\partial \theta) \Big],
\label{C.22}
\end{eqnarray}
%%%%%%%%%%%%%%%%%%%%%%%%%%%%%%%%%%%%%%%%%%%%%%%%%%%%%%%%%%%%%%%%%%%%%%%
where the last two terms in the r.h.s. of this equation come from normal
ordering.

On the other hand,
%%**  C.23 %%%%%%%%%%%%%%%%%%%%%%%%%%%%%%%%%%%%%%%%%%%%%%%%%%%%%%%%%%%%%
\begin{eqnarray}
\lambda \Gamma^{abc} \hat G &=& \frac{ 1}{2}(\lambda \Gamma^{abc}\Gamma^d 
\Pi_d d) +4(\lambda\Gamma^{abc}\partial^2 \theta) + 6 (\lambda N^{[ab} \Gamma^
{c]} \partial \theta) \nonumber\\
&+& 3 (\lambda\Gamma_f N^{f[a} \Gamma^{bc]}\partial \theta)  
+ \frac{1}{4} (\lambda \Gamma_f \Gamma_g\Gamma^{abc} N^{fg}\partial \theta) 
-\frac{1}{4} (\lambda \Gamma^{abc} J \partial \theta).
\label{C.23}
\end{eqnarray}
%%%%%%%%%%%%%%%%%%%%%%%%%%%%%%%%%%%%%%%%%%%%%%%%%%%%%%%%%%%%%%%%%%%%%%%%%
Using (\ref{2.30}), (\ref{2.31}) and the notation introduced in (\ref{C.13})
the quantity in the second row of (\ref{C.23}), that is,
%%** C.24 %%%%%%%%%%%%%%%%%%%%%%%%%%%%%%%%%%%%%%%%%%%%%%%%%%%%%%%%%%%%%%% 
\begin{eqnarray}
h^{[abc]} =  +3 (\lambda\Gamma_f N^{f[a} \Gamma^{bc]}\partial \theta)  
+ \frac{1}{4} (\lambda \Gamma_f \Gamma_g\Gamma^{abc} N^{fg}\partial \theta) 
-\frac{1}{4} (\lambda \Gamma^{abc} J \partial \theta),
\label{C.24}
\end{eqnarray}
%%%%%%%%%%%%%%%%%%%%%%%%%%%%%%%%%%%%%%%%%%%%%%%%%%%%%%%%%%%%%%%%%%%%%%%%%
can be rewritten as
%**  C.25 %%%%%%%%%%%%%%%%%%%%%%%%%%%%%%%%%%%%%%%%%%%%%%%%%%%%%%%%%%%%%%%
\begin{eqnarray} 
h^{[abc]} &=& - \frac{3}{2}(\lambda \Gamma_f \Gamma^{[ab}(\Omega \Gamma^{c]}
\Gamma^{f}\lambda)
\partial \theta) + \frac{1}{8}(\lambda \Gamma_f \Gamma_g \Gamma^{abc} (\Omega
\Gamma^f \Gamma^g \lambda)\partial \theta) -12(Y\Gamma^a \tilde \Lambda 
\Gamma^{bc} \partial \theta)  \nonumber\\
&-& 6 (Y \Lambda\Gamma^{abc}\partial \theta) +(\partial \lambda 
\Gamma^{abc}\partial \theta).
\label{C.25}
\end{eqnarray}
%%%%%%%%%%%%%%%%%%%%%%%%%%%%%%%%%%%%%%%%%%%%%%%%%%%%%%%%%%%%%%%%%%%%%%%%%%%
On the other hand, by reordering, the sum of the first two, 
$\Omega$-dependent terms in (\ref{C.25}) yields $ 12(Y\Gamma^a \tilde \Lambda 
\Gamma^{bc} \partial
\theta) + 6 (Y \Lambda\Gamma^{abc}\partial \theta) $ so that
$ h^{[abc]} = \partial \lambda \Gamma^{abc} \partial \theta$ and (\ref{C.23})
becomes
%**  C.26  %%%%%%%%%%%%%%%%%%%%%%%%%%%%%%%%%%%%%%%%%%%%%%%%%%%%%%%%%%%%%%%%
\begin{eqnarray}
\lambda \Gamma^{abc} \hat G = \frac{1}{2}(\lambda \Gamma^{abc}\Gamma^d 
\Pi_d d) +4(\lambda\Gamma^{abc}\partial^2 \theta) + 6 (\lambda N^{[ab} \Gamma^
{c]} \partial \theta)  
+(\partial\lambda\Gamma^{abc}\partial \theta).
\label{C.26}
\end{eqnarray}
%%%%%%%%%%%%%%%%%%%%%%%%%%%%%%%%%%%%%%%
By comparing (\ref{C.22}) with (\ref{C.26}) one gets Eq. (\ref{3.28}).

\subsection { BRST variation of $K^{[\alpha\beta\gamma]}$ }

Now let us  check (\ref{3.41}). Let us rewrite (\ref{3.40}) as
%** C.27 %%%%%%%%%%%%%%%%%%%%%%%%%%%%%%%%%%%%%%%%%%%%%%%%%%%%%%%%%%
\begin{eqnarray}
\{ Q, K^{[\alpha\beta\gamma]} \} = k_1^{[\alpha\beta\gamma]} 
+ k_2^{[\alpha\beta\gamma]}, 
\label{C.27} 
 \end{eqnarray}
%%%%%%%%%%%%%%%%%%%%%%%%%%%%%%%%%%%%%%%%%%%%%%%%%%%%%%%%%%%%%%%%%%%%%
where
%**  C.28 %%%%%%%%%%%%%%%%%%%%%%%%%%%%%%%%%%%%%%%%%%%%%%%%%%%%%%%%%%
\begin{eqnarray}
k_2^{[\alpha\beta\gamma]} &=& - \frac{1}{12}(\Gamma_a d)^{[\alpha} \Big[
\frac{3}{4} (\Gamma^a d)^\beta \lambda^{\gamma]} -\frac{1}{4}(\Gamma_b d)^\beta
\Gamma^{ba}\lambda^{\gamma]} \Big]  \nonumber\\
&=& \frac{1}{384}\lambda^{[\alpha}\Gamma_{abc}^{\beta\gamma]}(d\Gamma^{abc}d),
\label{C.28}
\end{eqnarray}
%%%%%%%%%%%%%%%%%%%%%%%%%%%%%%%%%%%%%%%%%%%%%%%%%%%%%%%%%%%%%%%%%%%%%%
and
%**  C.29 %%%%%%%%%%%%%%%%%%%%%%%%%%%%%%%%%%%%%%%%%%%%%%%%%%%%%%%%%%%
\begin{eqnarray}
 k_1^{[\alpha\beta\gamma]} &=& \frac{1}{12} \Pi_d (\Gamma^a \Gamma^d
\lambda)^{[\alpha} N_a^{\beta\gamma]} \nonumber\\ 
&=& \frac{1}{6} \Pi_d (\Gamma^a \Gamma^d \lambda)^{[\alpha} 
[ \frac{1}{2}(\Omega \Lambda_a^{\beta\gamma]}\lambda) 
- (Y \Lambda_a^{\beta\gamma]} \partial \lambda) ].
\label{C.29} 
\end{eqnarray}
%%%%%%%%%%%%%%%%%%%%%%%%%%%%%%%%%%%%%%%%%%%%%%%%%%%%%%%%%%%%%%%%%%%%%%
 The first term in the r.h.s. of (\ref{C.29}) can be elaborated as follows:
%**  C.30 %%%%%%%%%%%%%%%%%%%%%%%%%%%%%%%%%%%%%%%%%%%%%%%%%%%%%%%%%%%%
\begin{eqnarray}
\frac{1}{12} \Pi_d (\Gamma^a \Gamma^d \lambda)^{[\alpha}(\Omega \Lambda_a^{\beta\gamma]} \lambda) 
&=& \frac{1}{8} \Pi^d (\Omega \Gamma^b)^{[\alpha}\lambda^\beta
(\Gamma_b \Gamma_d \lambda)^{\gamma]} + \Delta^{[\alpha\beta\gamma]} 
\nonumber \\
&=& \frac{1}{2} \Pi^d \lambda^{[\alpha}(\Omega \Lambda_d^{\beta\gamma]}
\lambda) 
+ \Delta^{[\alpha\beta\gamma]}  + \hat \Delta^{[\alpha\beta\gamma]}, 
\label{C.30}
\end{eqnarray}
%%%%%%%%%%%%%%%%%%%%%%%%%%%%%%%%%%%%%%%%%%%%%%%%%%%%%%%%%%%%%%%%%%%%%%
where $\Delta^{[\alpha\beta\gamma]}$ and $\hat\Delta^{[\alpha\beta\gamma]}$
are the contributions of rearrangement theorem and are given by
%**  C.31 %%%%%%%%%%%%%%%%%%%%%%%%%%%%%%%%%%%%%%%%%%%%%%%%%%%%%%%%%%%%
\begin{eqnarray}
\Delta^{[\alpha\beta\gamma]} &=& \frac{1}{192} \Pi^d 
\Gamma_{abc}^{[\alpha\beta}(\Gamma^a \Gamma_d \partial K \Gamma^{bc} \lambda)^{\gamma]} 
\nonumber \\ 
&=& \frac{1}{24} \Pi^f (\Gamma^a Y)^{[\alpha}[\partial \lambda^\beta
(\Gamma_a \Gamma_f \lambda)^{\gamma]} + (\Gamma_a \Gamma_f \partial 
\lambda)^\beta \lambda^{\gamma]} - \frac{1}{2} (\Gamma_f \Gamma_b \partial \lambda)^\beta 
(\Gamma_a \Gamma^b \lambda)^{\gamma]}]
\nonumber\\
&-& \frac{1}{2} \Pi^d (\Gamma_d Y)^{[\alpha} \partial \lambda^\beta \lambda^{\gamma]},
\label{C.31}
\end{eqnarray}
%%%%%%%%%%%%%%%%%%%%%%%%%%%%%%%%%%%%%%%%%%%%%%%%%%%%%%%%%%%%%%%%%%%%%%%%%
and
%** C.32 %%%%%%%%%%%%%%%%%%%%%%%%%%%%%%%%%%%%%%%%%%%%%%%%%%%%%%%%%%%%%%%%
\begin{eqnarray}
\hat\Delta^{[\alpha\beta\gamma]} = \frac{1}{4} \Pi^d (\Gamma^f Y)^{[\alpha}
(\partial \lambda \Gamma_f \Lambda_d^{\beta\gamma]} \lambda).
\label{C.32}
\end{eqnarray} 
%%%%%%%%%%%%%%%%%%%%%%%%%%%%%%%%%%%%%%%%%%%%%%%%%%%%%%%%%%%%%%%%%%%%%%%%%
Therefore, $ k_1^{[\alpha\beta\gamma]}$ becomes
%**   C.33 %%%%%%%%%%%%%%%%%%%%%%%%%%%%%%%%%%%%%%%%%%%%%%%%%%%%%%%%%%%%%%
\begin{eqnarray}
k_1^{[\alpha\beta\gamma]} &=& \frac{1}{2} \Pi^d \lambda^{[\alpha} 
N_d^{\beta\gamma]} 
+ \{ \Pi^d (\lambda^{[\alpha}Y\Lambda_d^{\beta\gamma]} \partial \lambda) 
- \frac{1}{6} \Pi_d (\Gamma^a \Gamma^d \lambda)^{[\alpha}
(Y \Lambda_a^{\beta\gamma]} \partial \lambda) 
\nonumber\\
&+& \hat \Delta^{[\alpha\beta\gamma]} 
+ \Delta^{[\alpha\beta\gamma]} \}.
\label{C.33}
\end{eqnarray}
%%%%%%%%%%%%%%%%%%%%%%%%%%%%%%%%%%%%%%%%%%%%%%%%%%%%%%%%%%%%%%%%%%%%%%%%%%
With a little algebra, it is easy to show that the terms in the curly bracket
in the r.h.s. of (\ref{C.33}) vanish so that (\ref{C.33}) becomes
%**  C.34 %%%%%%%%%%%%%%%%%%%%%%%%%%%%%%%%%%%%%%%%%%%%%%%%%%%%%%%%%%%%%%%%
\begin{eqnarray}
k_1^{[\alpha\beta\gamma]} = \frac{1}{2} \lambda^{[\alpha} \Pi^d 
N_d^{\beta\gamma]}.  
\label{C.34}
\end{eqnarray}
%%%%%%%%%%%%%%%%%%%%%%%%%%%%%%%%%%%%%%%%%%%%%%%%%%%%%%%%%%%%%%%%%%%%%%%%%%
Then, (\ref{C.27}), together with (\ref{C.28}), (\ref{C.34}) and (\ref{3.23}), 
reproduces Eq. (\ref{3.41}).

\subsection {The vanishing of  $ \lambda^{[\epsilon} L^{\alpha\beta\gamma
\delta]} $}

Now let us consider $\lambda^{[\epsilon} L^{\alpha\beta\gamma\delta]}$ in 
order to verify that it vanishes. As discussed at the end of section 3, 
it consists of three terms: 
%**   C.35 %%%%%%%%%%%%%%%%%%%%%%%%%%%%%%%%%%%%%%%%%%%%%%%%%%%%%%%%%%%%%%%%
\begin{eqnarray}
\lambda^{[\epsilon} L_{1}^{\alpha\beta\gamma\delta]}&=&  
 \lambda^{[\epsilon}(\Omega\Lambda_{c}^{\alpha\beta}\lambda)(\Omega
\Lambda^{\gamma\delta]c}\lambda) \nonumber\\ 
&=& \Omega_\sigma A_{1}^{[\epsilon\alpha\beta\gamma\delta]\sigma} 
+ A_{0}^{[\epsilon\alpha\beta\gamma\delta]},
\label{C.35}
\end{eqnarray}
%%%%%%%%%%%%%%%%%%%%%%%%%%%%%%%%%%%%%%%%%%%%%%%%%%%%%%%%%%%%%%%%%%%%%%%%%%%%
%**  C.36 %%%%%%%%%%%%%%%%%%%%%%%%%%%%%%%%%%%%%%%%%%%%%%%%%%%%%%%%%%%%%%%%%%
\begin{eqnarray}
\lambda^{[\epsilon} L_2^{\alpha\beta\gamma\delta]} 
= \Omega_\sigma B_{1}^{[\epsilon\alpha\beta\gamma\delta]\sigma} 
+ B_0^{[\epsilon\alpha\beta\gamma\delta]},
\label{C.36}
\end{eqnarray}
%%%%%%%%%%%%%%%%%%%%%%%%%%%%%%%%%%%%%%%%%%%%%%%%%%%%%%%%%%%%%%%%%%%%%%%%%%%%
and
%**  C.37 %%%%%%%%%%%%%%%%%%%%%%%%%%%%%%%%%%%%%%%%%%%%%%%%%%%%%%%%%%%%%%%%%
\begin{eqnarray}
\lambda^{[\epsilon} L_3^{\alpha\beta\gamma\delta]}= 
4\lambda^{[\epsilon}(Y\Lambda_{c}^{\alpha\beta}\partial\lambda)
(Y\Lambda^{\gamma\delta]c}\partial\lambda) 
\label{C.37}
\end{eqnarray}
%%%%%%%%%%%%%%%%%%%%%%%%%%%%%%%%%%%%%%%%%%%%%%%%%%%%%%%%%%%%%%%%%%%%%%%%%%%%
where $\Omega_\sigma B_{1}^{[\epsilon\alpha\beta\gamma\delta]\sigma}$ is
%**  C.38  %%%%%%%%%%%%%%%%%%%%%%%%%%%%%%%%%%%%%%%%%%%%%%%%%%%%%%%%%%%%%%%%%
\begin{eqnarray}
\Omega_{\sigma} B_{1}^{[\epsilon\alpha\beta\gamma\delta]\sigma} 
= - 4 \Omega_\sigma \lambda^{[\epsilon} (Y\Lambda_{c}^{[\alpha \beta}\partial 
\lambda)(\Lambda^{\gamma\delta]c}\lambda)^{\sigma},
\label{C.38}
\end{eqnarray}
%%%%%%%%%%%%%%%%%%%%%%%%%%%%%%%%%%%%%%%%%%%%%%%%%%%%%%%%%%%%%%%%%%%%%%%%%%%%
and  $\Omega_{\sigma}A_{1}^{\sigma}$, $A_0$ and $B_0$ are rearrangement 
contributions coming when $ \Omega_\sigma $ is shifted to the left. The 
explicit calculation of $ \Omega_{\sigma}A_{1}^{\sigma}$ gives
%**  C.39  %%%%%%%%%%%%%%%%%%%%%%%%%%%%%%%%%%%%%%%%%%%%%%%%%%%%%%%%%%%%%%%%%
\begin{eqnarray}
\Omega_\sigma  A_{1}^{[\epsilon\alpha\beta\gamma\delta]\sigma} =
\Omega_{\sigma} (\Lambda^{c[\alpha\beta}\lambda)^{\sigma} (\partial \lambda
\Gamma_{f}\Lambda _{c}^{\gamma\delta}\lambda)(\Gamma^f Y)^{\epsilon]}  
+ \Omega_{\sigma}(\Lambda^{c[\alpha\beta}\Gamma^{f} Y)^{\sigma}(\partial 
\lambda\Gamma_{f}\Lambda _{c}^{\gamma\delta}\lambda) \lambda^{\epsilon]}.
\label{C.39}
\end{eqnarray}
%%%%%%%%%%%%%%%%%%%%%%%%%%%%%%%%%%%%%%%%%%%%%%%%%%%%%%%%%%%%%%%%%%%%%%%%%%%%
The first term in the r.h.s. of this equation can be rewritten as 
$4 \Omega_\sigma \lambda^{[\epsilon} (Y \Lambda_c^{[\alpha \beta}
\partial \lambda)(\Lambda^{\gamma\delta]c}\lambda)^\sigma - 2 \Omega_{\sigma}
(\Lambda^{c[\alpha\beta}\lambda)^{\sigma}(\Gamma_{c}Y)^{\gamma} \partial 
\lambda^{\delta} \lambda^{\epsilon]}$ and the second one as $ 2 \Omega_
{\sigma}(\Lambda^{c[\alpha\beta}\lambda)^{\sigma}(\Gamma_{c}Y)^{\gamma} 
\partial \lambda^{\delta} \lambda^{\epsilon]}$  so that we have
%**  C.40  %%%%%%%%%%%%%%%%%%%%%%%%%%%%%%%%%%%%%%%%%%%%%%%%%%%%%%%%%%%%%%%%%
\begin{eqnarray}
\Omega_\sigma A_{1}^{[\epsilon\alpha\beta\gamma\delta]\sigma} 
= 4 \Omega_\sigma \lambda^{[\epsilon} (Y\Lambda_{c}^{\alpha \beta}\partial 
\lambda)(\Lambda^{\gamma\delta]c}\lambda)^{\sigma}.
\label{C.40}
\end{eqnarray}
%%%%%%%%%%%%%%%%%%%%%%%%%%%%%%%%%%%%%%%%%%%%%%%%%%%%%%%%%%%%%%%%%%%%%%%%%%%%
Then, using (\ref{C.38}) and (\ref{C.40}), Eq. (\ref{3.67}) can be derived.

As for $A_{0}^{[\epsilon\alpha\beta\gamma\delta]}$ and $B_{0}^{[\epsilon
\alpha\beta\gamma\delta]}$, the explicit calculation gives
%**  C.41 %%%%%%%%%%%%%%%%%%%%%%%%%%%%%%%%%%%%%%%%%%%%%%%%%%%%%%%%%%%%%%%%%%
\begin{eqnarray}
A_{0}^{[\epsilon\alpha\beta\gamma\delta]} &=& \frac{1}{2} (\Gamma^{f} Y)^{[
\epsilon}(Y\Lambda_{c}^{\alpha\beta}\lambda)(\partial \lambda\Gamma_f\Lambda^
{\gamma\delta]c}\partial \lambda) - \frac{3}{2} (\Gamma^{f} Y)^{[\epsilon}
(Y\Lambda_{c}^{\alpha\beta}\partial \lambda)(\partial \lambda\Gamma_f\Lambda^
{\gamma\delta]c}\lambda) \nonumber\\
&+& \frac{1}{2}\partial ( (\Gamma^{f} Y)^{[
\epsilon}(Y\Lambda_{c}^{\alpha\beta}\lambda)(\partial \lambda\Gamma_f\Lambda^
{\gamma\delta]c}\lambda)),
\label{C.41}
\end{eqnarray}
%%%%%%%%%%%%%%%%%%%%%%%%%%%%%%%%%%%%%%%%%%%%%%%%%%%%%%%%%%%%%%%%%%%%%%%%%%%%%
and
%**   C.42 %%%%%%%%%%%%%%%%%%%%%%%%%%%%%%%%%%%%%%%%%%%%%%%%%%%%%%%%%%%%%%%%%%
\begin{eqnarray}
B_{0}^{[\epsilon\alpha\beta\gamma\delta]} &=& -2 (\Gamma^{f} Y)^{[\epsilon}
(Y\Lambda_{c}^{\alpha\beta}\partial \lambda)(\partial \lambda\Gamma_f\Lambda^
{\gamma\delta]c}\lambda) 
+ 4 \lambda^{[\epsilon}(Y\Lambda_{c}^{\alpha\beta}\partial \lambda)
(Y\Lambda^{\gamma\delta] c}\partial \lambda) \nonumber \\ 
&+& 2 \partial \lambda^{[\epsilon}(Y\Lambda_{c}^{\alpha\beta} \lambda)
(Y\Lambda^{\gamma\delta]c}\partial \lambda) 
- 2 \partial (\lambda^{[\epsilon}
(Y\Lambda_{c}^{\alpha\beta}\partial \lambda)
(Y\Lambda^{\gamma\delta]c} \lambda)),
\label{C.42}
\end{eqnarray}
%%%%%%%%%%%%%%%%%%%%%%%%%%%%%%%%%%%%%%%%%%%%%%%%%%%%%%%%%%%%%%%%%%%%%%%%%%%%%%%
so that we obtain
%**   C.43 %%%%%%%%%%%%%%%%%%%%%%%%%%%%%%%%%%%%%%%%%%%%%%%%%%%%%%%%%%%%%%%%%%%%
\begin{eqnarray}
A_{0}^{[\epsilon\alpha\beta\gamma\delta]} +B_{0}^{[\epsilon
\alpha\beta\gamma\delta]} 
&=&  \frac{1}{2}(\Gamma^{f} Y)^{[\epsilon}(Y\Lambda_
{c}^{\alpha\beta}\lambda)(\partial \lambda \Gamma_{f}\Lambda^{\gamma\delta]c}
\partial \lambda) 
- \frac{7}{2} (\Gamma^{f} Y)^{[\epsilon}(Y\Lambda_{c}
^{\alpha\beta}\partial \lambda)(\partial \lambda \Gamma_{f}\Lambda^{\gamma
\delta]c}\lambda)
\nonumber\\
&+& 4\lambda^{[\epsilon}(Y\Lambda_{c}^{\alpha\beta}\partial \lambda)
(Y\Lambda^{\gamma\delta] c}\partial \lambda) 
+ 2 \partial \lambda^{[\epsilon}(Y\Lambda_{c}^{\alpha\beta} \lambda)
(Y\Lambda^{\gamma\delta]c} \partial \lambda) 
\nonumber\\
&+& \frac{1}{2}\partial \{ (\Gamma^{f} Y)^{[
\epsilon}(Y\Lambda_{c}^{\alpha\beta}\lambda)(\partial \lambda\Gamma_f\Lambda^
{\gamma\delta]c}\lambda)) - 4 \lambda^{[\epsilon}(Y\Lambda_{c}^{\alpha\beta}
\partial \lambda)(Y \Lambda^{\gamma\delta]c}\lambda)\}.
\label{C.43}
\end{eqnarray}
%%%%%%%%%%%%%%%%%%%%%%%%%%%%%%%%%%%%%%%%%%%%%%%%%%%%%%%%%%%%%%%%%%%%%%%%%%%%%%%%
In order to verify (\ref{3.68}), one needs three useful identites:
%**  C.44  %%%%%%%%%%%%%%%%%%%%%%%%%%%%%%%%%%%%%%%%%%%%%%%%%%%%%%%%%%%%%%%%%%%%%
\begin{eqnarray}
(\Gamma^{f} Y)^{[\epsilon}(Y\Lambda_{c}
^{\alpha\beta}\lambda)(\partial \lambda \Gamma_{f}\Lambda^{\gamma\delta]c}
 \lambda) =  4 \lambda^{[\epsilon}(Y\Lambda_{c}^{\alpha\beta}
\partial \lambda) (Y \Lambda^{\gamma\delta]c}\lambda),
\label{C.44}
\end{eqnarray}
%%%%%%%%%%%%%%%%%%%%%%%%%%%%%%%%%%%%%%%%%%%%%%%%%%%%%%%%%%%%%%%%%%%%%%%%%%%%%%%%%
%**  C.45 %%%%%%%%%%%%%%%%%%%%%%%%%%%%%%%%%%%%%%%%%%%%%%%%%%%%%%%%%%%%%%%%%%%%%%
\begin{eqnarray}
(\Gamma^{f} Y)^{[\epsilon}(Y\Lambda_{c}^{\alpha\beta} \lambda)(\partial \lambda 
\Gamma_{f}\Lambda^{\gamma\delta]c}
\partial \lambda ) 
&=& 5 \lambda^{[\epsilon}(Y\Lambda_{c}^{\alpha\beta}
\partial \lambda)(Y\Lambda^{\gamma\delta]c}\partial \lambda) 
\nonumber\\
&+& 10 \partial \lambda^{[\epsilon}(Y\Lambda_{c}^{\alpha\beta} \lambda)
(Y\Lambda^{\gamma\delta]c}\partial \lambda),
 \label{C.45}
\end{eqnarray}
%%%%%%%%%%%%%%%%%%%%%%%%%%%%%%%%%%%%%%%%%%%%%%%%%%%%%%%%%%%%%%%%%%%%%%%%%%%%%%%%%
%**   C.46  %%%%%%%%%%%%%%%%%%%%%%%%%%%%%%%%%%%%%%%%%%%%%%%%%%%%%%%%%%%%%%%%%%%%
\begin{eqnarray}
(\Gamma^{f} Y)^{[\epsilon}(Y\Lambda_{c}
^{\alpha\beta}\partial \lambda)(\partial \lambda \Gamma_{f}\Lambda^{\gamma
\delta]c} \lambda) 
= 3 \lambda^{[\epsilon}(Y\Lambda_{c}^{\alpha\beta}\partial \lambda)
(Y\Lambda^{\gamma\delta]c}\partial \lambda) 
+ 2 \partial \lambda^{[\epsilon}(Y\Lambda_{c}^{\alpha\beta} \lambda)
(Y\Lambda^{\gamma\delta]c}\partial \lambda).
\label{C.46}
\end{eqnarray}
%%%%%%%%%%%%%%%%%%%%%%%%%%%%%%%%%%%%%%%%%%%%%%%%%%%%%%%%%%%%%%%%%%%%%%%%%%%%%%%%%
{}From the identity (\ref{C.44}), the derivative term in the last row of the
r.h.s. of (\ref{C.43}) vanishes. Then, removing the first two terms in the 
r.h.s. of (\ref{C.43}) by means of the two identites (\ref{C.45}) and (\ref{C.46}), one gets 
%**   C.47  %%%%%%%%%%%%%%%%%%%%%%%%%%%%%%%%%%%%%%%%%%%%%%%%%%%%%%%%%%%%%%%%%%%%
\begin{eqnarray}
A_{0}^{[\epsilon\alpha\beta\gamma\delta]} +B_{0}^{[\epsilon
\alpha\beta\gamma\delta]} = - 4  \lambda^{[\epsilon}(Y\Lambda_{c}^{\alpha\beta}
\partial \lambda)(Y\Lambda^{\gamma\delta]c}\partial \lambda), 
\label{C.47}
\end{eqnarray}
%%%%%%%%%%%%%%%%%%%%%%%%%%%%%%%%%%%%%%%%%%%%%%%%%%%%%%%%%%%%%%%%%%%%%%%%%%%%%%%%%
which is nothing but Eq. (\ref{3.68}). In this way, we have succeeded in
proving Eq. (\ref{3.57}).

\subsection {Equivalence in cohomology of $ b_Y$ and $b_{nm}$}

As a last remark, let us report briefly about the derivation of the 
rearrangement terms $R_G$, $R_H$, $R_K$, $R_L$ and $B_L$, which appear at the end of section 5. 
In particular we shall show that $B_L$ is BRST-exact.  

{}From the recipe given in Appendix B.3 and using the OPE (\ref{3.15}), 
one can compute $R_G =(\tilde\lambda_\beta Y_\alpha)(\lambda^{[\alpha} \hat G^
{\beta]}) -(\tilde\lambda_\beta Y_\alpha\lambda^{[\alpha}) \hat G^{\beta]}$  
with the result
%**   C.48 %%%%%%%%%%%%%%%%%%%%%%%%%%%%%%%%%%%%%%%%%%%%%%%%%%%%%%%%%%%%%%%%%%%%%
\begin{eqnarray}
R_G = - [ \partial \tilde \lambda \partial \theta 
- (\tilde \lambda \partial\lambda)(\tilde \lambda \partial \theta) 
- 2 \partial Y \partial \theta ] + \hat R_G,
\label{C.48}
\end{eqnarray}
%%%%%%%%%%%%%%%%%%%%%%%%%%%%%%%%%%%%%%%%%%%%%%%%%%%%%%%%%%%%%%%%%%%%%%%%%%%%%%%%
where 
%**   C.49 %%%%%%%%%%%%%%%%%%%%%%%%%%%%%%%%%%%%%%%%%%%%%%%%%%%%%%%%%%%%%%%%%%%%%
\begin{eqnarray}
\hat R_G = Y_{[\alpha}\tilde\lambda_{\beta]}((Y+\tilde\lambda)\Gamma_a)^\alpha
\partial\lambda^{\beta}(\lambda\Gamma^{a}\partial\theta).
\label{C.49}
\end{eqnarray}
%%%%%%%%%%%%%%%%%%%%%%%%%%%%%%%%%%%%%%%%%%%%%%%%%%%%%%%%%%%%%%%%%%%%%%%%%%%%%%%%
With the replacement 
%**   C.50 %%%%%%%%%%%%%%%%%%%%%%%%%%%%%%%%%%%%%%%%%%%%%%%%%%%%%%%%%%%%%%%%%%%%%
\begin{eqnarray}
\lambda \Gamma^a \partial \theta = [ Q_{nm}, \Pi^a ],
\label{C.50}
\end{eqnarray}
%%%%%%%%%%%%%%%%%%%%%%%%%%%%%%%%%%%%%%%%%%%%%%%%%%%%%%%%%%%%%%%%%%%%%%%%%%%%%%%%
and some simple algebra, $\hat R_G$ can be rewritten as
%**   C.51  %%%%%%%%%%%%%%%%%%%%%%%%%%%%%%%%%%%%%%%%%%%%%%%%%%%%%%%%%%%%%%%%%%%%
\begin{eqnarray}
\hat R_G =  2 [Q_{nm}, (Y_{\alpha}\tilde\lambda_{\beta})W_{R1}^{[\alpha\beta]}]
- \frac{1}{2} A_G^a \Pi_a, 
\label{C.51}
\end{eqnarray}
%%%%%%%%%%%%%%%%%%%%%%%%%%%%%%%%%%%%%%%%%%%%%%%%%%%%%%%%%%%%%%%%%%%%%%%%%%%%%%%
where $W_{R1}^{[\alpha\beta]}$ is defined in (\ref{5.39}) and
$A^a_G$, defined in (\ref{5.40}), comes from 
%**   C.52 %%%%%%%%%%%%%%%%%%%%%%%%%%%%%%%%%%%%%%%%%%%%%%%%%%%%%%%%%%%%%%%%%%%%
\begin{eqnarray}
 \frac{1}{2} A_{G}^a \Pi_a = \Big[Q_{nm}, Y_{[\alpha}\tilde\lambda_{\beta]}
((Y+\tilde\lambda)\Gamma_a)^{\alpha}\partial\lambda^{\beta}\Big] \Pi^a,
\label{C.52}
\end{eqnarray}
%%%%%%%%%%%%%%%%%%%%%%%%%%%%%%%%%%%%%%%%%%%%%%%%%%%%%%%%%%%%%%%%%%%%%%%%%%%%%%%%
by using a simple algebra.

In a similar way, $R_H$ is given by 
%**   C.53 %%%%%%%%%%%%%%%%%%%%%%%%%%%%%%%%%%%%%%%%%%%%%%%%%%%%%%%%%%%%%%%%%%%%
\begin{eqnarray}
R_H = - \frac{1}{2} A_G^a \Pi_a + \hat R_H,
\label{C.53}
\end{eqnarray}
%%%%%%%%%%%%%%%%%%%%%%%%%%%%%%%%%%%%%%%%%%%%%%%%%%%%%%%%%%%%%%%%%%%%%%%%%%%%%%%%
where $\hat R_H$ contains the factor $ (\Gamma^{c}\Gamma_{a}
\Pi^{a}\lambda)^{\gamma}$ which can be replaced by $- \{ Q_{nm},(\Gamma^{c}d)^\gamma \}$ 
and then, working as before, one arrives at
%**   C.54 %%%%%%%%%%%%%%%%%%%%%%%%%%%%%%%%%%%%%%%%%%%%%%%%%%%%%%%%%%%%%%%%%%%%
\begin{eqnarray}
\hat R_H  =3! \Big[ Q_{nm},Y_{[\alpha}\tilde r_{\beta}\tilde\lambda_{\gamma]}
W_{R2}^{\alpha\beta\gamma}\Big] - \frac{1}{4} A^{a}_{H\alpha}
(d \Gamma_a )^\alpha,
\label{C.54}
\end{eqnarray}
%%%%%%%%%%%%%%%%%%%%%%%%%%%%%%%%%%%%%%%%%%%%%%%%%%%%%%%%%%%%%%%%%%%%%%%%%%%%%%%%
where $W_{R2}^{[\alpha\beta\gamma]}$ and $ A^{a}_{H\alpha}$
are defined in (\ref{5.39}) and (\ref{5.40}), respectively.
Moreover,
%**   C.55 %%%%%%%%%%%%%%%%%%%%%%%%%%%%%%%%%%%%%%%%%%%%%%%%%%%%%%%%%%%%%%%%%%%%%
\begin{eqnarray}
R_K = \frac{1}{4} A^a_{H \alpha}(d \Gamma_a )^\alpha  + \hat R_K,
\label{C.55}
\end{eqnarray}
%%%%%%%%%%%%%%%%%%%%%%%%%%%%%%%%%%%%%%%%%%%%%%%%%%%%%%%%%%%%%%%%%%%%%%%%%%%%%%%%
where
%**   C.56 %%%%%%%%%%%%%%%%%%%%%%%%%%%%%%%%%%%%%%%%%%%%%%%%%%%%%%%%%%%%%%%%%%%%%
\begin{eqnarray}
\hat R_K &=& \frac{4!}{12}
(Y_{[\alpha} \tilde r_{\beta} \tilde r_{\gamma} \tilde \lambda_{\delta]}
((Y + 3 \tilde\lambda)\Gamma^{c})^\alpha \partial \lambda^{\beta}[Q_{nm},N^{
\gamma\delta}_c ] 
\nonumber\\ 
&=& 4!\Big[Q_{nm},  (Y_{[\alpha} \tilde r_\beta 
\tilde r_\gamma \tilde \lambda_{\delta]})W_{R3}^{[\alpha\beta\gamma\delta]}
\Big] + A^{c}_{K[\alpha\beta}N^{\alpha\beta]}_c,
\label{C.56}
\end{eqnarray}
%%%%%%%%%%%%%%%%%%%%%%%%%%%%%%%%%%%%%%%%%%%%%%%%%%%%%%%%%%%%%%%%%%%%%%%%%%%%%%%%
where again $W_{R3}^{[\alpha\beta\gamma\delta]}$ and 
$A^{c}_{K\alpha\beta}$ are defined in (\ref{5.39}) and (\ref{5.40}), respectively.

Now let us move on to $R_L$ which, according to (\ref{5.34}), is
%**   C.57 %%%%%%%%%%%%%%%%%%%%%%%%%%%%%%%%%%%%%%%%%%%%%%%%%%%%%%%%%%%%%%%%%%%%%
\begin{eqnarray}
R_L = 5! \frac{1}{48} (Y_{[\alpha} \tilde r_{\beta} \tilde 
r_{\gamma} \tilde r_{\delta} \tilde \lambda_{\epsilon]}\lambda^\epsilon)
(N^{c \alpha\beta} N_{c}^{\gamma\delta}).
\label{C.57}
\end{eqnarray}
%%%%%%%%%%%%%%%%%%%%%%%%%%%%%%%%%%%%%%%%%%%%%%%%%%%%%%%%%%%%%%%%%%%%%%%%%%%%%%%%
With rearrangement formula and using (\ref{3.58}), $R_L$ becomes
%**   C.58 %%%%%%%%%%%%%%%%%%%%%%%%%%%%%%%%%%%%%%%%%%%%%%%%%%%%%%%%%%%%%%%%%%%%%
\begin{eqnarray}
R_L &=&  5! \frac{1}{48} \{ (\Big[ K_{[\alpha\beta\gamma\delta\epsilon]}
\lambda^\epsilon, N^{c \alpha\beta} N_{c}^{\gamma\delta} \Big]) 
- (\Big[ K_{[\alpha\beta\gamma\delta\epsilon]}, N_{c}^{\alpha\beta}
N^{c \gamma\delta} \Big]) \lambda^\epsilon \} 
\nonumber\\
&=&  5!\frac{1}{48} (-2) (K_{[\alpha\beta\gamma\delta\epsilon]}
((Y+4\tilde\lambda)\Lambda_{c}^{\gamma\delta} \lambda)
\partial \lambda^\epsilon) N^{c \alpha\beta} + \hat R_L, 
\label{C.58}
\end{eqnarray}
%%%%%%%%%%%%%%%%%%%%%%%%%%%%%%%%%%%%%%%%%%%%%%%%%%%%%%%%%%%%%%%%%%%%%%%%%%%%%%%%
where we have defined $K_{[\alpha\beta\gamma\delta\epsilon]}
= Y_{[\alpha} \tilde r_{\beta} \tilde r_{\gamma} \tilde r_{\delta} \tilde \lambda_{\epsilon]}$. 
(The expression of $\hat R_L$ will be given below.)
To the first term in the last row of Eq. (\ref{C.58}),  adding and subtracting 
the term defined by
%**   C.59 %%%%%%%%%%%%%%%%%%%%%%%%%%%%%%%%%%%%%%%%%%%%%%%%%%%%%%%%%%%%%%%%%%%%%
\begin{eqnarray}
R_0 = \frac{5!}{24} ( (Y_{[\alpha} \tilde r_{\beta} \tilde r_{\gamma} 
\tilde r_{\delta} \tilde \lambda_{\epsilon]})((Y + 4 \tilde\lambda) 
\Sigma_c^{\gamma\delta} \lambda) \partial \lambda^\epsilon ) N^{c \alpha\beta},
\label{C.59}
\end{eqnarray}
%%%%%%%%%%%%%%%%%%%%%%%%%%%%%%%%%%%%%%%%%%%%%%%%%%%%%%%%%%%%%%%%%%%%%%%%%%%%%%%%
where we have also defined
%**   C.60 %%%%%%%%%%%%%%%%%%%%%%%%%%%%%%%%%%%%%%%%%%%%%%%%%%%%%%%%%%%%%%%%%%%%%
\begin{eqnarray}
\tilde Y \Sigma^{[\alpha\beta]}_c\lambda = (\tilde Y \Gamma_c)^{[\alpha}
\lambda^{\beta]} + \frac{1}{4} (\tilde Y \Gamma)^{[\alpha}_b 
(\Gamma^b \Gamma_c\lambda)^{\beta]},
\label{C.60}
\end{eqnarray}
%%%%%%%%%%%%%%%%%%%%%%%%%%%%%%%%%%%%%%%%%%%%%%%%%%%%%%%%%%%%%%%%%%%%%%%%%%%%%%%%
and 
%**   C.61 %%%%%%%%%%%%%%%%%%%%%%%%%%%%%%%%%%%%%%%%%%%%%%%%%%%%%%%%%%%%%%%%%%%%%
\begin{eqnarray}
\tilde Y_\alpha = Y_\alpha + 4 \tilde\lambda_\alpha,
\label{C.61}
\end{eqnarray}
%%%%%%%%%%%%%%%%%%%%%%%%%%%%%%%%%%%%%%%%%%%%%%%%%%%%%%%%%%%%%%%%%%%%%%%%%%%%%%%%
$R_L$ is then reduced to
%**   C.62 %%%%%%%%%%%%%%%%%%%%%%%%%%%%%%%%%%%%%%%%%%%%%%%%%%%%%%%%%%%%%%%%%%%%%
\begin{eqnarray}
R_L = - \frac{1}{12} A^c_{K\alpha\beta} N^{\alpha\beta}_c + R_0 +  \hat R_L.
\label{C.62}
\end{eqnarray}
%%%%%%%%%%%%%%%%%%%%%%%%%%%%%%%%%%%%%%%%%%%%%%%%%%%%%%%%%%%%%%%%%%%%%%%%%%%%%%%%
Here we have introduced the quantity
%**   C.63 %%%%%%%%%%%%%%%%%%%%%%%%%%%%%%%%%%%%%%%%%%%%%%%%%%%%%%%%%%%%%%%%%%%%%
\begin{eqnarray}
\hat R_L = 5!\frac{1}{48}( \partial R_1 + R_2 +R_3),
\label{C.63}
\end{eqnarray}
%%%%%%%%%%%%%%%%%%%%%%%%%%%%%%%%%%%%%%%%%%%%%%%%%%%%%%%%%%%%%%%%%%%%%%%%%%%%%%%%
where $R_1$, $R_2$ and $R_3$ are defined by 
%**   C.64 %%%%%%%%%%%%%%%%%%%%%%%%%%%%%%%%%%%%%%%%%%%%%%%%%%%%%%%%%%%%%%%%%%%%%
\begin{eqnarray}
R_1 &=& \frac{1}{4} Y_{[\alpha} \tilde r_\beta \tilde 
r_\gamma \tilde r_\delta \tilde \lambda_{\epsilon]}\lambda^\epsilon 
(\Gamma^a Y)^\alpha (\Gamma^b \tilde Y)^\beta (\Gamma_b\Gamma^c \lambda)^\gamma
[(\Gamma_a\Gamma_c \partial \lambda)^\delta + 2 \delta_{ac}\partial
\lambda^\delta],
\nonumber\\
R_2 &=& - \frac{1}{2} Y_{[\alpha} \tilde r_\beta \tilde r_\gamma \tilde r_\delta 
\tilde \lambda_{\epsilon]}\lambda^\epsilon (\Gamma^c Y)^\alpha (\Gamma^b \tilde Y)^\beta 
\partial [(\Gamma_{bc} \lambda)^\gamma \partial \lambda^\delta], 
\nonumber\\
R_3 &=& Y_{[\alpha} \tilde r_\beta \tilde r_\gamma \tilde r_\delta 
\tilde \lambda_{\epsilon]} \lambda^\epsilon (\tilde \lambda \Gamma^f Y)
(\Gamma^c (2 Y + 5 \tilde \lambda))^\alpha (\Gamma_c \Gamma^b \lambda)^\beta 
(\Gamma_f \Gamma_b \partial \lambda)^\gamma \partial \lambda^\delta.
\label{C.64}
\end{eqnarray}
%%%%%%%%%%%%%%%%%%%%%%%%%%%%%%%%%%%%%%%%%%%%%%%%%%%%%%%%%%%%%%%%%%%%%%%%%%%%%%%%
It is of importance that $R_1$, $R_2$ and $R_3$ are all BRST-exact:
%**   C.65 %%%%%%%%%%%%%%%%%%%%%%%%%%%%%%%%%%%%%%%%%%%%%%%%%%%%%%%%%%%%%%%%%%%%%
\begin{eqnarray}
R_1 &=& \frac{1}{20} \Big[Q_{nm}, Y_{[\alpha} \tilde r_\beta 
\tilde r_\gamma \tilde \lambda_{\delta]} 
(\Gamma^a Y)^\alpha (\Gamma^b (Y + 3 \tilde \lambda))^\beta (\Gamma_b \Gamma^c 
\lambda)^\gamma [(\Gamma_a \Gamma_c \partial \lambda)^\delta + 2 \delta_{ac}
\partial \lambda^\delta] \Big],
\nonumber\\
R_2 &=& -\frac{1}{10} \Big[Q_{nm}, Y_{[\alpha} \tilde r_\beta 
\tilde r_\gamma \tilde \lambda_{\delta]}
(\Gamma^c Y)^\alpha (\Gamma^b (Y + 3 \tilde \lambda))^\beta 
\partial [(\Gamma_{bc} \lambda)^\gamma \partial \lambda^\delta] \Big],
\nonumber\\
R_3 &=& \frac{3}{10} \Big [Q_{nm}, Y_{[\alpha} \tilde r_\beta 
\tilde r_\gamma \tilde \lambda_{\delta]} (\tilde \lambda \Gamma^f Y)
(\Gamma^c ( Y + 2 \tilde \lambda))^\alpha (\Gamma_c \Gamma^b \lambda)^\beta 
(\Gamma_f \Gamma_b \partial \lambda)^\gamma \partial \lambda^\delta \Big].
\label{C.65} 
\end{eqnarray}
%%%%%%%%%%%%%%%%%%%%%%%%%%%%%%%%%%%%%%%%%%%%%%%%%%%%%%%%%%%%%%%%%%%%%%%%%%%%%%%
On the other hand, by rearrangement theorem, $R_0$ can be rewritten as 
%%**   C.66 %%%%%%%%%%%%%%%%%%%%%%%%%%%%%%%%%%%%%%%%%%%%%%%%%%%%%%%%%%%%%%%%%%%
\begin{eqnarray}
R_0 &=& \frac{5!}{24} \Omega_\sigma (Y_{[\alpha} \tilde r_{\beta} \tilde 
r_{\gamma} \tilde r_{\delta} \tilde \lambda_{\epsilon]} (\tilde Y  \Sigma_
{c}^{\gamma\delta}\lambda)\partial \lambda^{\epsilon}(\Lambda^{\alpha\beta c} \lambda)
^\sigma ) 
\nonumber\\ 
&+& \frac{5!}{24} Y_{[\alpha} \tilde r_{\beta} \tilde 
r_\gamma \tilde r_\delta \tilde \lambda_{\epsilon]}\lambda^\epsilon 
(\Gamma^c \tilde Y)^\alpha \partial \lambda^\beta (\Gamma^b Y)^\gamma
(\Gamma_b \Gamma_c \partial\lambda)^\delta.
 \label{C.66}  
\end{eqnarray}
%%%%%%%%%%%%%%%%%%%%%%%%%%%%%%%%%%%%%%%%%%%%%%%%%%%%%%%%%%%%%%%%%%%%%%%%%%%%%%%
The first term in the r.h.s. of (\ref{C.66}) vanishes and the second one is
BRST-exact. Indeed, one has
%**   C.67 %%%%%%%%%%%%%%%%%%%%%%%%%%%%%%%%%%%%%%%%%%%%%%%%%%%%%%%%%%%%%%%%%%%%
\begin{eqnarray}
R_0 = \frac{1}{4} [Q_{nm},  Y_{[\alpha} \tilde r_{\beta} \tilde 
r_{\gamma} \tilde \lambda_{\delta]} 
(\Gamma^c (Y + 3\tilde \lambda))^\alpha \partial \lambda^\beta (\Gamma^b Y)^
\gamma (\Gamma_b\Gamma_c \partial\lambda)^{\delta}].
\label{C.67}
\end{eqnarray}
%%%%%%%%%%%%%%%%%%%%%%%%%%%%%%%%%%%%%%%%%%%%%%%%%%%%%%%%%%%%%%%%%%%%%%%%%%%%%%%
Eq. (\ref{C.62}) is just Eq. (\ref{5.38}) with $B_L = R_0 + \hat R_L$.
Then, from Eqs. (\ref{C.63}), (\ref{C.65}) and (\ref{C.67}), one can reproduce
Eqs. (\ref{5.41})-(\ref{5.43}).

%%%%%%%%%%%%%%%%%%%%%%%% reference %%%%%%%%%%%%%%%%%%%%%%%%%%%%%%%
%%%%%%%%%%%%%%%%%%%%%%%%%%%%%%%%%%%%%%%%%%%%%%%%%%%%%%%%%%%%%%%%%%

\end{document}